\documentclass[12pt]{iopart}
\usepackage{graphicx}% Include figure files
\usepackage{dcolumn}% Align table columns on decimal point
\usepackage{bm}% bold math
\usepackage{iopams}% bold math
\usepackage{color}% bold math

\newcommand{\beqar}{\begin{eqnarray}}
\newcommand{\eeqar}{\end{eqnarray}}
\newcommand{\bea}{\begin{eqnarray}}
\newcommand{\eea}{\end{eqnarray}}
\newcommand{\bcen}{\begin{center}}
\newcommand{\ecen}{\end{center}}

\newcommand{\bra}[1]{\left< #1 \right|}
\newcommand{\ket}[1]{\left| #1 \right>}

\newcommand{\f}[2]{\frac{#1}{#2}}
\renewcommand{\b}[1]{\left({#1}\right)}
\renewcommand{\v}[1]{\vec{#1}}
\newcommand{\pd}[2]{\frac {\partial #1}{\partial #2}}

\renewcommand{\sb}[1]{\left[{#1}\right]}
\newcommand{\mean}[1]{\langle {#1} \rangle}

\newcommand{\ra}{\rightarrow}

\newcommand{\trr}{\textcolor{black}}

\begin{document}

\title[Unification of the first law of quantum thermodynamics]{Unification of the first law of quantum thermodynamics}

\author{Roie Dann}
\address{The Institute of Chemistry, The Hebrew University of Jerusalem, Jerusalem 9190401, Israel}
\ead{roie.dann@mail.huji.ac.il}
\author{Ronnie Kosloff}
\address{The Institute of Chemistry, The Hebrew University of Jerusalem, Jerusalem 9190401, Israel}
\ead{ronnie@fh.huji.ac.il}
\vspace{10pt}
\begin{indented}
\item[]November 2022
\end{indented}

\begin{abstract}
  Underlying the classical thermodynamic principles are analogous microscopic laws, arising from the fundamental axioms of quantum mechanics. These define quantum thermodynamic variables such as quantum work and heat and characterize the possible transformations of open quantum systems. The foremost quantum thermodynamic law is a simple statement concerning the conservation of energy. Nevertheless, there exist ambiguity and disagreement regarding the precise partition of a quantum system's energy change to work and heat. By treating quantum mechanics as a comprehensive theory, applicable to both the micro and macroscopic domains, and employing dynamical symmetries, we bridge the gaps between five popular thermodynamic approaches to the first law. These include both autonomous and semi-classical formulations, which define work in terms of an ensemble average, as well as the single shot paradigm, where work is defined as a deterministic quantity. 
\end{abstract}

%\maketitle
\break

\section{\label{sec:intro}Introduction}
Thermodynamics is often portrayed as the queen of physical theories, as it supplies a unified framework to portray phenomena from the subatomic to cosmological scale \cite{alicki2018introduction,rossnagel2016single,de2000flat}.
Originally, thermodynamics described the behaviour of bulk processes which incorporate an ensemble of particles. Nevertheless, a recent change in paradigm highlights the fact that thermodynamic principles can be associated to an individual quantum system, such as an ion in a trap \cite{rossnagel2016single}. Despite the generality of the theory, the traditional construction is restricted to systems close to equilibrium and the description of processes of infinite time duration (quasi-static) \cite{1956T,callen1998thermodynamics,reif2009fundamentals,sekimoto1998langevin}. To extend the theory beyond this regime, we harness a dynamical description of physical systems.

With the aim of to construct a fundamental theory of the microscopic, the complete theoretical framework must  comply with the postulates of quantum mechanics.
The quantum theory provides a comprehensive dynamical framework which overlaps with thermodynamics in the description of physical reality.  Combining the two theories allows a complete description, including non-equilibrium scenarios and finite time dynamics. 
This realization has led to the emerging field of quantum thermodynamics \cite{kosloff2013quantum,vinjanampathy2016quantum,binder2018thermodynamics}.

The core idea  of the field is that classical thermodynamics emerges from an underlying quantum mechanical description. In this framework, the four classical thermodynamic laws ($0^\textrm{th}$-III) emerge from a set of quantum analogous constructions.
Similarly to  classical thermodynamics, these fundamental laws are achieved by partitioning the quantum world into thermodynamic constituents, such as a local primary system (working medium) and an environment, and identifying the various thermodynamic variables (energy, entropy and temperature). 

In the present study we focus on the first law of thermodynamics; a statement on the conversion rules between various forms of energy change:
\begin{equation}
    \Delta E = {\cal{W}} + {\cal{Q}}~~.
    \label{eq:1st_law}
\end{equation}
The traditional theory identifies work $\cal W$ as the energy change of the system leaving the entropy invariant, and heat $\cal Q$ is associated with entropy production. It is only natural to conjecture that the basic partition also exists in the quantum regime. The issue of determining  suitable partitions has led to controversies and conflicting definitions \cite{spohn1978irreversible,lenard1978thermodynamical,alicki1979quantum,allahverdyan2005fluctuations,tonner2005autonomous,talkner2007fluctuation,esposito2010entropy,horodecki2013fundamental,aaberg2013truly,roncaglia2014work,allahverdyan2014nonequilibrium,skrzypczyk2014work,binder2015quantum,talkner2016aspects,elouard2017role,miller2017time,sampaio2018quantum,lorch2018optimal,silva2021quantum1,beyer2020work,verteletsky2020revealing,de2020unraveling,pekola2021colloquium,juan2021first,soret2022thermodynamic}. We show that the source of conflict can be traced to different idealizations and approximation schemes. Ultimately, as in any physical theory, the aspiration is a single inclusive definition, encapsulating all the different scenarios. Our current aim is to relate various definitions, unify them and explore their limits.

We tackle the issue utilizing an axiomatic approach, based on a first principle microscopic (autonomous) description of all the thermodynamic constituents. We introduce a set of thermodynamically motivated postulates which enforce a strict partition between the thermodynamic constituents: the primary system, environment and work reservoir. These allow constructing the exact structure of equations of motion for the reduced system and evaluating the energy fluxes between the thermodynamic constituents. The precise structure highlights that the nature of the partition and the associated dynamical symmetry is the source of the difference between various possible decompositions of the first law.  The axiomatic approach allows deducing the exact form of work and heat in the autonomous quantum description.

Building upon the autonomous framework, we consider the behaviour of the thermodynamic variables in the semi-classical limit, leading to semi-classical definitions. The connection between the semi-classical description and  corresponding autonomous framework illuminates the  origins of the difference between definitions of the same thermodynamic variable. Namely, it highlights the fact that semi-classical definitions of work and heat implicitly imply a certain dynamical symmetry in the corresponding underlying autonomous description. 
In addition, the relations between the various thermodynamic approaches and the possible decompositions of the first law demonstrate the fundamental role of dynamical symmetries in the thermodynamic analysis of quantum processes.
%\tb{some redundancy here with the intro, I'm contemplating if to rephrase and remove some stuff}

We begin by briefly describing the five different approaches to the first law of thermodynamics  in Sec. \ref{sec:app_quantum_work}. These approaches are labelled as:  Autonomous (Local or Global), Single Shot, Semi-classical, External and Dynamical Map.  Section \ref{sec:Framework} presents the theoretical framework, highlighting the basic postulates and assumptions considered in the analysis. 
Following, Secs. \ref{sec:autonomous_therm} and \ref{sec:local_interaction_model} introduce the autonomous thermodynamic definitions explicitly and derive the exact reduced dynamics under the local interaction autonomous model. The exact dynamical form leads to the precise explicit expressions for the local power and heat flux in terms of system operators. In Sec. \ref{sec:semi_classical} we introduce the semi-classical approach, and derive the connection to the local autonomous approach in Sec. \ref{sec:autonomous_s_c_connection}. The relation between the approaches is obtained by introducing a procedure that defines the semi-classical limit. This transition is then illustrated by analyzing the semi-classical limit and thermodynamics of the Janyes-Cummings model. The connection between the local autonomous and the single shot approaches is established in Sec. \ref{sec:single_shot}.

Section \ref{sec:global_autonomous} introduces the global interaction model and the associated exact reduced dynamics. Thermodynamics variables are then expressed in terms of the system operators and related to the external approach, Sec. \ref{sec:external approach}, and dynamical map approach definitions, Sec. \ref{sec:dynamical map}. We summarize the relations between the approaches in Sec. \ref{sec:summary} and conclude in Sec. \ref{sec:discussion} with a discussion on the close relation between dynamical symmetries, thermodynamic idealizations, and quantum thermodynamic variables. We emphasis the considered idealizations and point out potential future extensions of the presented framework. For the convenience of the reader, the notations employed  throughout the manuscript are summarized in
Table \ref{table:notations}.

\section{The many faces of quantum work}
\label{sec:app_quantum_work}
A number of subtle issues are confronted when trying to define quantum thermodynamic variables. Primarily, any definition of work and heat is only meaningful if it can be related to physical reality by a measurement procedure, directly or indirectly. This immediately poses a difficulty, since gathering information on a quantum  system may perturb its state. An example, is the definition of work according to the two-point measurement protocol \cite{piechocinska2000information,piechocinska2000information,kurchan2000quantum,talkner2007fluctuation,campisi2011colloquium}. Non-stationary systems may include coherence in the energy basis (from here on denoted just as coherence). Applying a projective measurement of energy will eliminate the coherence, and thus modify the potential work extraction. In turn, if the measurement itself influences the thermodynamic variables, the amount of power required to perform the  measurement should also be included in the thermodynamic accounting \cite{debarba2019work,guryanova2020ideal}. Since the measurement apparatus is classical  (consuming a ``classical" amount of power) this introduces addition complications.  This obstacle can be circumvented by performing a measurement in the eigenbasis of the system's density operator \cite{manzano2018quantum} or by considering only initial states with no coherence, preventing any back-action due to the measurement process. The key idea is that the thermodynamic variables are defined in context to an underlying measurement procedure in mind (sometimes implicitly). In the following analysis, we bypass this complication by implicitly considering an appropriate measurement procedure which does not perturb the thermodynamic analysis. 

Another difficulty in identifying thermodynamic variables in the quantum regime, concerns the emergence of significant fluctuations enhanced in miniaturized systems. As a result, one needs to differentiate between the average work and the work distribution, which is the outcome of many ``single shot'' experiments \cite{piechocinska2000information,piechocinska2000information,kurchan2000quantum,tasaki2000jarzynski,talkner2007fluctuation,talkner2016aspects,perarnau2017no}. A distinction emerges between two approaches: work as an average property or work as the outcome of a single shot experiment. 

The single shot analysis is closely related to the thermodynamics resource theory \cite{janzing2000thermodynamic,horodecki2013fundamental,brandao2013resource,faist2015gibbs,halpern2016beyond,lostaglio2019introductory}, where the quantum evolution is described in terms of operations. The theory identifies a set of allowed (free) operations  performed on an arbitrary initial state a single time. It shares common features with classical thermodynamics, essentially aiming to evaluate the bounds to the possible transformations, given an initial state (resource) and allowed operations. 
Single shot work extraction (similarly for consumption) is considered a deterministic process, which includes charging a quantum battery with certainty. This is commonly formulated as a transition between two energy states of the battery \cite{horodecki2013fundamental,aaberg2013truly}. We label this thermodynamic approach (and associated definitions) by
\begin{itemize}
    \item Single shot
\end{itemize}
It should be noted that in the framework of quantum thermodynamic resource theory, the main interest regards the absolute limits, therefore, the method by which the experimenter performs the free operations is not accounted for and in principle may take infinite time.

Alternatively, the dynamical description is more attuned to analyze the thermodynamics of a certain ensemble of processes. Here, the framework is based on a microscopic description in terms of the Hamiltonian of the universe, which completely determines the dynamics of the ensemble of states. Work is then identified as an average quantity \cite{alicki1979quantum,esposito2010entropy,spohn1978irreversible,kosloff1984quantum,szczygielski2013markovian,alicki2012periodically,kolavr2012quantum,levy2012quantum,pusz1978passive,lobejko2020thermodynamics,gelbwaser2013work,roulet2017autonomous,lenard1978thermodynamical,gelbwaser2013work}. 
This paradigm is commonly employed to study quantum heat devices \cite{alicki1979quantum,kosloff1984quantum,quan2007quantum,rossnagel2016single}, quantum fluctuation relations \cite{kurchan2000quantum,campisi2011colloquium,aaberg2018fully}  and thermometery \cite{correa2015individual,mehboudi2019thermometry,seah2019collisional}. Moreover, in the classical limit this notion of quantum work can be viewed in terms of a process over classical phase space \cite{jarzynski2015quantum}.

Even within the paradigm of thermodynamics in terms of ensemble averages a variety of definitions for work and heat have been proposed.
Specifically, we study four common approaches, they are labeled by:
\begin{itemize}
    \item Semi-classical
    \item External
    \item Dynamical map
        \item Autonomous 
\end{itemize}

The traditional approach identifies work as the change in the total energy $\Delta E$ of the total isolated system, encompassing the primary system and environment \cite{alicki1979quantum,spohn1978irreversible,kosloff1984quantum}. In this framework, an external time-dependent field induces a change of energy and an instantaneous  power current. The work is then expressed as an integral over the power. 
This definition does not include the operational cost of the external drive as a part of the thermodynamical accounting \cite{torrontegui2017energy}. Such setting is classified as non-autonomous, with a semi-classical description of the work-reservoir (battery), we will denote it as the {\emph{semi-classical}} approach. 

Semi-classical driven systems inspire definitions of heat and work which depend solely on the system observables. A number of alternative definitions are presented in \cite{juan2021first,de2020unraveling}. In this framework, once coherence is generated heat and work cannot measured simultaneously. As a result, there is no unique definition of heat and work.

Two other related paradigms first identify the flows into the reservoir and utilize an approximation scheme to obtain the thermodynamic variables in terms of local system observables. The {\emph{external approach}} defines the heat flow in terms of the reservoir Hamiltonian \cite{esposito2010entropy,elouard2020thermodynamics}. While the second proposal primarily identifies the entropy production by building on the mathematical properties of the reduced system dynamics. Heat is then defined in terms of the entropy production \cite{szczygielski2013markovian,alicki2012periodically,kolavr2012quantum,levy2012quantum}. We refer to this paradigm as the \emph{dynamical map} approach. In both approaches heat depends on the specific master equation employed to describe the reduced system dynamics and work is determined by the  first law, Eq. (\ref{eq:1st_law}), as the residual energy change. 
%\tg{Such treatment is often used when the WM is driven periodically, and at sufficiently long times, once the WM reaches a steady state (returning to the same state after each period) \cite{}}.
Similarly to the semi-classical approach, these definitions are based on a non-autonomous and semi-classical framework.

An alternative popular framework models the battery explicitly, describing both the primary system and battery under a full quantum setting \cite{lenard1978thermodynamical,pusz1978passive,lobejko2020thermodynamics,gelbwaser2013work,roulet2017autonomous,maffei2021probing,stevens2022energetics}.
Initially, the primary system and battery (can also be considered as the control system) interact autonomously, leading to both energy and information flow. The potential work stored in the battery can then be extracted via an entropy conserving operation. In the following we refer to this setting as the {\emph{autonomous}} approach. 
The autonomous approach has been experimentally realized in a set up of a quantum engine and a quantum refrigerator. In the quantum engine, an harmonic mode of the vibrational motion of a trapped Calcium ion was employed as a work depository. The charging system consisted of an Otto driven two-level system composed of the internal ion spin \cite{von2019spin}. In the refrigerator setup composed of three ions, generating non-linear coupled harmonic modes. The control mode is initially charged with external energy, subsequently it provides the work to  pump energy from a cool mode to a hotter one \cite{maslennikov2019quantum}.

Within the autonomous framework we identify two optional approaches, local and global. These differ form one another by the nature of the environment and control coupling, and corresponding dynamical symmetry.

The different approaches and associated definitions for the same physical quantities may lead to confusion and hinder the development of the quantum thermodynamics research field. Our current goal is to connect the various approaches, and set up a theoretical framework which illuminates the source of the differences, when these exist.

We approach the issue from first principles, by considering a complete (autonomous) quantum description including the system, work depository and environment. Assuming a number of thermodynamically motivated idealizations an accurate reduced dynamical description is obtained. Such dynamics constitute a quantum dynamical thermodynamic process. Within this framework, there is a natural identification of the thermodynamic variables (in terms of a deterministic or an ensemble average). These identifications are then related to the other approaches in the limit of a large control system, where a corresponding semi-classical description emerges.

%\trr{- Quantum mechanics is independent allowing adding dynamics and treating non-equilibrium situations.}

%\trr{- What the analogous quantum thermodynamic quantities?}

%\trr{- Many definition for different situations, confusion, need for unification.}

%\trr{- We propose autonomous frame allowing address this problem. The different definitions can be rationalized by studying different iso-energetic partitions.  The insight is that different partitions lead to different thermodynamic idealizations.}

%\trr{- We treat 4 common definitions of work, we classify the different definitions, unify them and analysis the relationship between them. }

\section{Framework}
\label{sec:Framework}

We construct a theoretical framework which is founded on a set of thermodynamically motivated quantum postulates.
To set the stage, we consider a composite isolated quantum system, which state is represented by a density operator $\hat{\rho}$ in the composite Hilbert space. Such a description complies with the dogma of the church of the larger Hilbert space \cite{everett2012everett,zurek1991quantum,zurek2013wave,gogioso2019process}. According to this tenet any subsystem dynamics can be embedded in a unitary evolution of a larger Hilbert space. This embedding or purification is not unique, and includes the possibility that sub-dynamics arise from the dynamics of a pure state in the extended Hilbert space \cite{stinespring1955positive,nielsen2002quantum}.

Complementing this Hamiltonian setting, we further assume that the dynamics of the composite system are generated by a stationary Hamiltonian. Under this assumption the complete energy distribution is an invariant of the dynamics, trivially implying that the average energy is a constant of motion. What might be less obvious is that the coherence with respect to the energy basis is also conserved \cite{streltsov2017colloquium}. Such a description is consistent with quantum field theory, where the time-dependence emerges form the non-stationary initial state. In the current  framework, explicit time-dependent Hamiltonians are obtained in the semi-classical limit. In the reverse direction, a realistic explicit time-dependence of the semi-classical Hamiltonian can always be  removed by quantizing the electromagnetic field. Meaning that the Hilbert space is extended to include the field degrees of freedom as well. Overall,
this procedure infers that underlying explicitly time-dependent Hamiltonians are non-stationary field states that include coherence.

In practice any measurement is local, we therefore partition the extended Hilbert space to correspond to subsystems which can be measured locally. 
To perform a thermodynamic analysis this partition should be analogous to the classical thermodynamic partition.
In this context the universe (composite system) is partitioned into subsystems, constituting the studied systems and environments. The specific partition depends on the chosen studied scenario. An ideal partition in classical thermodynamics serves as an interface between subsystems that allows transfer of energy or particles, while maintaining the integrity of the subsystems. In essence, the system properties can be identified uniquely by neglecting the accumulated energy or particles in the interface. This condition is a manifestation of the thermodynamic limit, where the interface properties are negligible relative to the bulk.

In the quantum regime a similar partition can be achieved by imposing the strict energy conservation condition, leading to a constant interface energy. In analogy to the classical theory, this condition constitutes an idealization which is expected to hold in weak coupling limit, or in steady state processes.
Under this condition we can study the thermodynamics of very small quantum systems coupled to large environments.
%\tg{The various decompositions of the energy change, Eq. \eqref{eq:1st_law}, can be related to different setups of partitions between the primary system controller and the environment} \tb{not sure this fits here}. 
In the quantum regime, two autonomous models represent possible thermodynamic partitions which encompass the vast majority of the studies thermodynamic phenomena, Fig. \ref{fig:1-a}:
\begin{itemize}
    \item{{\emph{Local interaction model}} - Primary system interacting with an environment and control in a ``tandem" setup (coupled independently).}
    \item{{\emph{Global interaction model}} - An environment which interacts with a device that is composed of the primary and control systems. }
\end{itemize}

A final  idealization assumes the environment is initially in a stationary state with respect to its bare Hamiltonian. Under the present perspective, the stationarity property is the crucial difference between the environment and the control. A control system can be very large but requires coherence in order to operate, and therefore is non-stationary. %\trr{In the thermodynamic limit } 
%Within this framework we would like to analyse the first law of thermodynamics.

We formalize the described restrictions in terms of a set of thermodynamic quantum postulates:
\begin{enumerate}
    \item{ The total state $\hat{\rho}$ is defined as a density operator on the composite Hilbert space $\cal H$.}
    \item{ The composite dynamics are unitary and generated by the static Hamiltonian $\hat{H}$. The associated unitary map which propagates the composite state is defined by
    \begin{equation}
     \hat{\rho}\b t ={\cal{U}}\sb{\hat{\rho}\b 0}= \hat{U}\b{t,0}\hat{\rho}\b 0 \hat{U}^\dagger\b{t,0}  ~~,
    \end{equation}
    where $\hat{U}\b{t,0}=e^{-i\hat{H}t/\hbar}$.}
    This scenario implies three important global dynamical invariants: the von-Neumann entropy, energy entropy and coherence.
    Under arbitrary unitary dynamics the eigenvalues of the density operator are preserved. As a result, any function of the eigenvalues is preserved as well, in particular, the von-Neumann entropy
    $${\cal S}_{VN} =-\textrm{ tr} \{ \hat \rho \ln \hat \rho \}~~.$$
    In addition, if the dynamics are generated by a time-independent Hamiltonian any function of the Hamiltonian is also preserved:
    $$[f( \hat H ),\hat H]=0~~.$$ An important special case is the energy entropy 
    $${\cal S}_E=-\sum_j p_j \ln p_j~~,$$ 
    where $p_j =\textrm{ tr} \{\hat \Pi_j \hat \rho \} $ and $\hat \Pi_j$ is the projection operator on the $j$'th energy state.
    From these two properties we infer the conservation of coherence. Here, coherence is defined as the quantum relative entropy between the state $\hat{\rho}$ and its diagonal form in the energy representation \cite{streltsov2017colloquium}
    $${\cal C} = {\cal D}(\hat \rho | \hat \rho_d) ={\cal S}_E -{\cal S}_{VN}~~,$$
    where $\hat \rho_d$ is the diagonal part of the density operator in the energy basis. 
    \item{ The composite system is partitioned into subsystems, which are initially in an uncorrelated state: $\hat{\rho}\b 0=\hat{\rho}_S\b 0\otimes\cdots\otimes\hat{\rho}_E\b 0$. This boundary condition along with condition 2 defines a completely-positive trace preserving dynamical map of the reduced primary system \cite{kraus1971general}
    \begin{equation}
      \hat{\rho}_S\b t=\Lambda_{t}\sb{\hat{\rho}_S\b 0}=\textrm{tr}_{E,A}\b{\hat{U}\b{t,0}\hat{\rho}\b 0\hat{U}\b{t,0}^\dagger}~~.
      \label{eq:rho_s_3}
    \end{equation} 
where $A$ represents all the auxiliary systems (not including the environment and primary system). }
    \item{Strict energy conservation coupling between chosen subsystems: 
    Strict energy conservation between systems $A$ and $B$ infers that  
    \begin{equation}
    \sb{\hat{H}_A+\hat{H}_B,\hat{H}_{AB}}=0~~,
    \end{equation}
    where $\hat{H}_{A}$ and $\hat{H}_{B}$ are the bare Hamiltonian of subsystems $A$ and $B$ and $\hat{H}_{AB}$ is the interaction term between the two subsystems.
    The local and global interaction setups correspond to different choices for subsystems $A$ and $B$. Such a choice has a substantial impact on the thermodynamic analysis.  }
    \item{The environment is initially in a stationary state with respect to its bare Hamiltonian $\hat{H}_E$. In the thermodynamic limit, the perturbation to the environment state is negligible, and is therefore approximately stationary. }
\end{enumerate}

These postulates supply a sufficient structure which allow analyzing thermodynamics of quantum systems in a consistent manner. As in the classical thermodynamic theory, the present postulates are not generally satisfied, and serve as idealizations that allow describing the essential thermodynamic properties of quantum systems. A comprehensive critical analysis regarding their physical relevance is given in Sec. II of Ref. \cite{dann2021open}. A connection to the quantum version of the second law of thermodynamics can be established from conditions (ii) and (iii)  and the partition to the global system to subsystems. This implies that the sum of local entropies is always larger than the initial one, alternatively the entropy production is guaranteed to be positive \cite{esposito2010entropy,brandao2015second}.

\emph{Dynamical symmetries} serve as an essential tool in the present analysis. These symmetries are reflected in the form of open system dynamical equations of motion, and emerge from the extension of thermodynamic principles to dynamical processes. Specifically, {\emph{time-translation symmetry}} is a consequence of the principle of isothermal partition between subsystems (Postulate (iv)), and the initially stationary environment state (Postulate (v)) \cite{marvian2014extending}. These conditions imply a relation of the form
\begin{equation}
    \Lambda_t \circ {\cal U} = {\cal U} \circ \Lambda_t~,
    \label{eq:time_transl}
\end{equation}
meaning that the open system dynamical map commutes with a unitary superoperator $\cal U$. This property is also known as phase covariance \cite{holevo1996covariant}.
In the present context, ${\cal U}$ will constitute the free dynamical propagator of a subsystem. The physical context of Eq. (\ref{eq:time_transl}) then essentially implies that the environment cannot be utilized to synchronize the subsystem's time. The thermodynamic consequences of the symmetry condition have been studied extensively, leading to the discovery of thermodynamic constraints that go beyond free energy relations \cite{lostaglio2015quantum,lostaglio2015quantum}.

Importantly, the symmetries of the dynamical map $\Lambda_t$ lead to the exact form of the differential generator ${\cal{L}}_t$. This allows obtaining differential expressions for the power and heat flux. These thermodynamic fluxes serve as formal expressions by which to evaluate the observable total work and heat.

\section{Autonomous thermodynamic definitions}
\label{sec:autonomous_therm}
We consider the most general process within the framework of the thermodynamic postulates. Initially, the composite system is prepared in a state $\hat{\rho}\b{0}=\hat{\rho}_S\b{0}\otimes\hat{\rho}_C\b{0} \otimes\hat{\rho}_E\b{0}$ and evolves autonomously until a final time $t$, while the coupling induces energy transfer and introduces correlations between the system and battery.
In the autonomous setting, work is identified as the negative average energy change of the battery 
\begin{equation}
    {\cal{W}}^a \equiv - \Delta E_C=   - \textrm{tr}\b{\hat{H}_C\sb{\hat{\rho}\b{t}-\hat{\rho}\b{0}}}~~,
    \label{eq:Work}
\end{equation}
where $\hat{H}_C$ is the work depository (battery) Hamiltonian \cite{skrzypczyk2014work}.
Such definition implicitly assumes that the initial and final state of the battery is known, as any lack of knowledge prevents accurately evaluating the work. 
In addition, ${\cal W}^a$ depends only on local battery operators. This is a manifestation of the fact that prior to the charging process, the system and battery are separated and the energy can only be consumed by applying local operations on the battery. This setting resembles the day-to-day scenario, where one acquires a battery without prior knowledge of the charging process.  

Another implicit assumption is that the battery energy can be extracted by applying a local operation on the battery. This holds under the following restrictions (a) No entanglement between the system and battery at initial and final times.  (b) The ability to perform a local unitary entropy preserving transformation, restoring the battery to the initial state while extracting the work for another task.  These conditions are motivated by the fact that when significant global correlations between the battery and primary system are generated, a part of the energy extracted from the system cannot be extracted by means of local operations on the battery. This energy remains trapped in the interface between the primary system and the battery and therefore cannot be regarded as useful work.  This  identification of work matches the traditional association of work with the energy transfer that leaves the entropy invariant. It is analogous to storage of work in the potential energy of a classical weight, assuming no dissipation. An example of an experimental demonstration of a quantum work depository has been realized in resonator coupled to a Josephson qubit \cite{stevens2022energetics}. In this case the change in energy of the work depository, Eq. (\ref{eq:Work}), was directly measured. \trr{In \cite{maffei2021probing} strict energy conservation was utilized in order to perform a thermodynamical analysis of a similar experimental setup.}

By the first law heat includes all the residual energy 
\begin{equation}
    {\cal{Q}}^a = \Delta E_S-{\cal W}^{a}~~,
    \label{eq:heat_pre}
\end{equation}
where the average system energy is given by $E_S=\tr\b{\hat{H}_S\hat{\rho}}$.
%\tg{Strict energy conservation now allows to unambiguously determine the heat from the first law. Alternatively, an equivalent  definition identifies the heat as the change of energy in the environment
%\begin{equation}
%    {\cal{Q}}^a \equiv - \Delta E_E=   - \textrm{tr}\b{\hat{H}_E\sb{\hat{\rho}\b{t_f}-\hat{\rho}\b{t_i}}}~~,
%\end{equation}
%where $\hat{H}_E$ is the environment Hamiltonian.} \tb{discuss this, this is not true in the global setup model where $\sb{\hat{H}_{SC},\hat{H}_{SE}}\neq0$ heat should.}

The advantages of the autonomous thermodynamic definitions, Eqs. (\ref{eq:Work}) and (\ref{eq:heat_pre}), include the fact that the idealization applied are relatively clear, and their properties fit the familiar classical notion of work and heat. However, the simplicity comes with a price, the  disadvantage of Eqs. (\ref{eq:Work}) and (\ref{eq:heat_pre}) is their dependence on the global composite state. In practice, the environment is huge, which prevents obtaining a general accurate solution for the composite state dynamics. Hence, it is customary to express ${\cal W}^a$ and ${\cal Q}^a$ in terms of local system observables. To achieve this we first must specify the form of the interaction between the primary-system and environment. 

Note that despite the similarities there is a fundamental difference between the autonomous quantum definitions and the classical thermodynamic variables. In the classical framework work and heat are path dependent variables, while in the quantum autonomous framework they are state functions which are completely determined  by the initial conditions and total Hamiltonian.

\section{Local interaction autonomous model}
\label{sec:local_interaction_model}
\begin{figure}
    \centering
    \includegraphics[width=8cm]{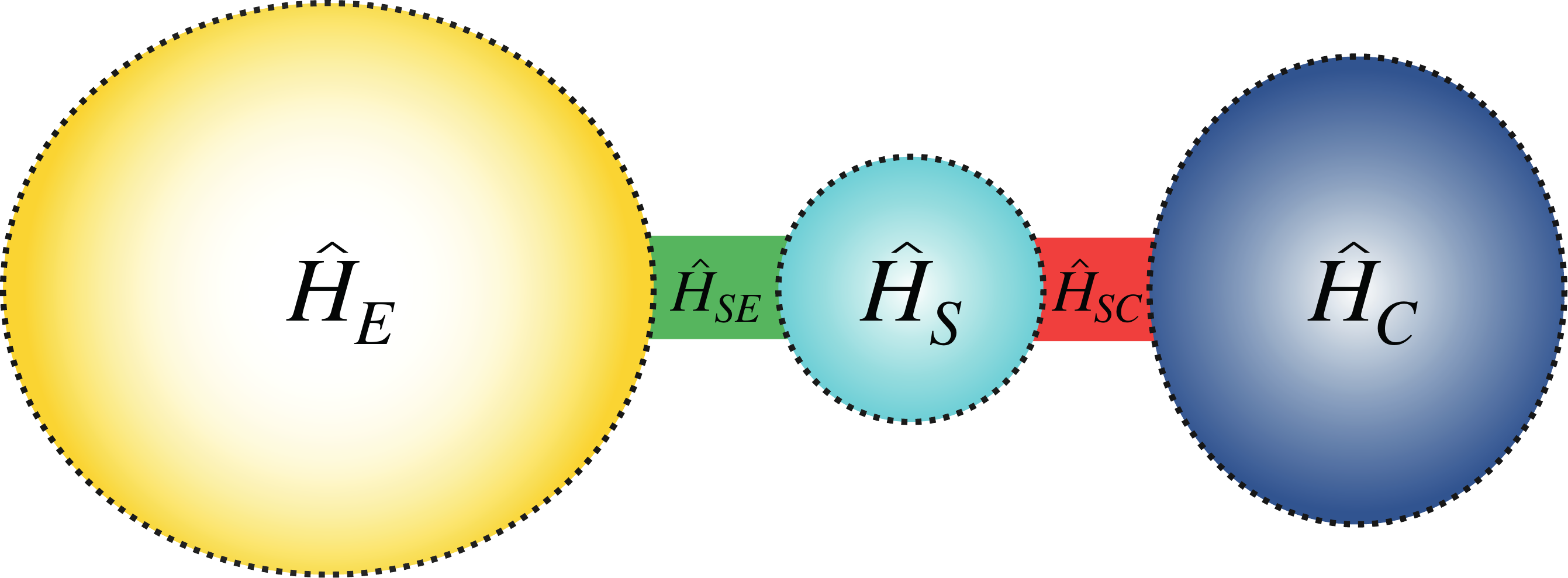}
    \includegraphics[width=8cm]{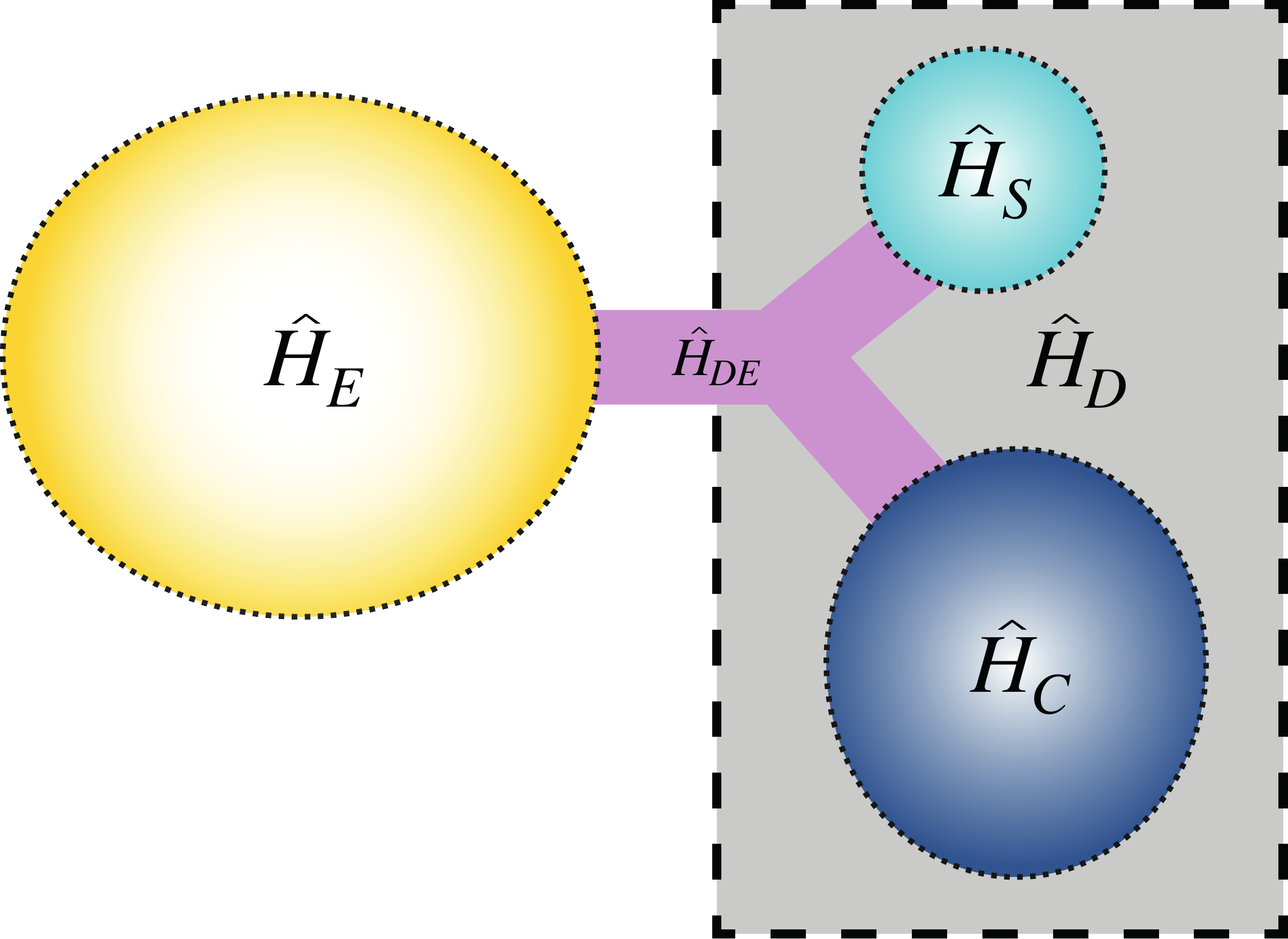}
    \caption{Partitions between the primary-system environment and  control. a) Local interaction model (tandem setup): Primary-system coupled to the environment  and the control. The interface between the primary-system and environment
    and the primary-system and work reservoir (control) obey strict energy conservation Eq. (\ref{eq:SEC_local}). b) Global interaction model: Primary system and control form a composite system called the global device which is embedded in the environment. Strict energy conservation is satisfied between the device and the environment Eq. (\ref{eq:global_sec}).}
    \label{fig:1-a}
\end{figure}

The local interaction model represents a scenario where the direct influence of the environment on the work depository (battery/control) can be discarded. In this framework, the primary system-environment and primary system-work depository coupling are independent, and the three sub-systems are coupled via a `tandem setup', see Fig. \ref{fig:1-a}.
The representing autonomous composite Hamiltonian is of the form
\begin{equation}
    \hat{H}^{\b{L}} = \hat{H}_S+\hat{H}_C+ \hat{H}_{SC}^{\b L} +\hat{H}_{E}+\hat{H}_{SE}^{\b L}~~,
    \label{eq:local_Ham}
\end{equation}
where $\hat{H}_S$ and $\hat{H}_E$, are the bare Hamiltonians of the primary system and environment, with dimensions $N_S$ and $N_E$, where $N_S$ is finite while $N_E$ may be infinite. The third and last terms are the system-work depository and system-environment interaction terms, correspondingly. These operators can generally be expressed as a sum of products of local system, work depository and environment operators $\{\hat{S}\}$, $\{\hat{C}\}$ and $\{\hat{E}\}$
\begin{equation}
    \hat{H}_{SE}^{\b L} = \sum_j \hat{S}_j\otimes \hat{E}_j \label{eq:local_SE_interaction}\\
    \hat{H}_{SC}^{\b L} = \sum_j \hat{S}_j\otimes \hat{C}_j~~.
    \nonumber
\end{equation}
Strict energy conservation (postulate (iv)) implies that the interaction terms obey the following commutation relations
\begin{equation}
\sb{\hat{H}_S+\hat{H}_C,\hat{H}_{SC}^{\b L}}=0 \label{eq:SEC_local}\\
\sb{\hat{H}_S+\hat{H}_E,\hat{H}_{SE}^{\b L}}=0~~.
\nonumber
\end{equation}
In turn, these relations infer that heat coincides with accumulated energy transferred to the environment  
\begin{equation}
    {\cal{Q}}^{a\b{L}} = - \Delta E_E=   - \textrm{tr}\b{\hat{H}_E\sb{\hat{\rho}\b{t_f}-\hat{\rho}\b{t_i}}}~~.
    \label{eq:Heat}
\end{equation}

For the following analysis, it will prove to be convenient to combine the primary-system and control into a single system, labelled as {\emph{device}}. Excluding any environmental influence, the device is represented by the Hamiltonian 
\begin{equation}
    \hat{H}_D^{\b L} = \hat{H}_S+\hat{H}_C+\hat{H}_{SC}^{\b L}~~.
    \label{eq:11_D}
\end{equation}
In the presence of an environment its reduced evolution is governed by the dynamical map
\begin{equation}
\Lambda^{\b{L}}_{D}\b t\sb{\hat{\rho}_{D}\b{0}}=\textrm{tr}_{E}\b{\hat{U}^{\b{L}}\b{t,0}\hat{\rho}\b 0{\hat{U}^{\b{L}}\b{t,0}}}~~,
\label{eq:13_lambda_D}
\end{equation}
where $\hat{U}^{\b{L}}\b{t,0}=e^{-i \hat{H}^{\b{L}} t/\hbar}$.
It is assumed that $\hat{H}_D^{\b{L}}$ has a finite dimension $N_D=N_S\cdot N_C$.

An alternative interpretation of the local model Hamiltonian $\hat{H}^{\b{L}}$ is a system simultaneously coupled to a quantum controller and general environment. This analogous structure, allows us to choose the convenient interpretation within the desired context. Throughout the article we switch between the battery and control descriptions, nevertheless, it should be clear that these titles designate the same subsystem.

At this point the thermodynamics framework is set. But in order to define the thermodynamic quantities in terms of local operators of the system and work depository, we require the exact reduced system dynamics.

\subsection{Local autonomous reduced system dynamics}
\label{subsec:local dynamics}

The common approach to obtain the reduced system dynamics follows a first principle `microscopic' derivation. Such a construction begins with a complete quantum description of the environment Eq. (\ref{eq:local_Ham}), and conducts a series of approximations leading to an equation of motion for the reduced state, i.e., the master equation \cite{davies1974markovian,breuer2002theory}.
The approximation scheme typically casts the dynamics in terms of a master equation of the Gorini-Kossakowski-Lindblad-Sudarshan (GKLS) form \cite{lindblad1976generators,gorini1976completely}. This structure constitutes the most general form of a dynamical semi-group generator \footnote{ A dynamical semi-group is a family of CPTP maps which satisfy the semi-group property}.

The approach is frequently used in a broad range of fields and serves as a faithful description of typical processes in nature \cite{cohen1998atom}. However, the validity of the master equation is guaranteed only within a restricted physical range, which is determined by the approximations conducted. Moreover, a variety of possible microscopic derivation exist \cite{alicki2007quantum,breuer2002theory,nakajima1958quantum,zwanzig1960ensemble,davies1974markovian,diosi1993calderia,lidar2001completely,alicki2012periodically,szczygielski2013markovian,majenz2013coarse,albash2012quantum,dann2018time,mozgunov2020completely}, distinct derivations rely on different approximations and therefore lead to varying validity regimes and definitions for work and heat. 

In the present study we adopt an  alternative methodology. We employ an axiomatic framework \cite{dann2021open,dann2021quantum,dann2021nonmarkovian}, based on the thermodynamic postulates, that determines the structure of the master equation, without conducting any approximation.
The detailed construction is presented in Ref. \cite{dann2021nonmarkovian}, here we summarize the main results and employ the procedure to obtain the reduced dynamics of the device and primary system.  The exact forms of the dynamical equations are then utilized to analyze the thermodynamics of quantum systems.

The construction is based on the relation between a dynamical symmetry of the open system map and Postulates (iv) and (v): strict energy conservation condition and the initial stationary environment state.  Assuming the dynamical map is invertible (or at times when the map is invertible), the symmetry relation, hermitiacy and trace preserving properties, along with the linearity of the map define the operatorial structure of the dynamical generator. This operatorial form is similar to the GKLS form, where the Lindblad operators are determined by the symmetry properties. The obtained form of the master equation is termed the {\emph{dynamical symmetric structure}}. 

Physically, the symmetric structure can be characterized by two ingredients, thermodynamic and kinematic. The dynamical symmetry properties are embodied by the operatorial structure and represent the thermodynamics, while the kinematic properties, concerning the details of the environment, coupling constants and memory effects, are all incorporated within the kinetic coefficients of the symmetric structure. Unlike the standard GKLS form, the coefficients are generally time-dependent and may obtain negative values under non-Markovian dynamics. 
Overall, the dynamical symmetric structure is not restricted to weak coupling, nor requires a large bath and is valid for highly non-Markovian dynamics, cf. Ref \cite{dann2021nonmarkovian} for explicit examples.

\subsubsection{Reduced dynamics of the device in the local interaction setup}

The local interaction commutation relations, Eq. (\ref{eq:SEC_local})
infer the conservation of the interaction energy
 \begin{equation}
     \sb{\hat{H}^{\b{L}},\hat{H}_S+\hat{H}_C+\hat{H}_E}=0~~.
     \label{eq:commu2}
 \end{equation}
In turn, such invariance along with the stationarity of the initial environment state imply a crucial dynamical symmetry \ref{apsec:appendix_A}. This symmetry allows employing a spectral analysis, which infers that the dynamics of the device energy populations and coherences (with respect to $\hat{H}_D$) are decoupled.

In the Heisenberg picture, this symmetry is manifested by the decoupling of the dynamics of the device's non-invariant $\{\hat{G}_\kappa^{\b{L}}\}$ and invariant $\{\hat{R}_j^{\b{L}}\}$ eigenoperators. These eigenoperator obey the eigenvalue-type relation 
${\mathcal{U}}_0
\sb{\hat{G}_\kappa^{\b{L}}}=\lambda_\kappa \hat{G}_\kappa^{\b{L}}$  
and ${\cal{U}}_0\sb{\hat{R}_\kappa^{\b{L}}}= \hat{R}_\kappa^{\b L}$, where 
$\lambda_k\in\mathbb{C}$
and ${\cal U}_{0}\sb{\hat{\rho}_{D}\b 0}=\hat{U}_{0}\b{t,0}\hat{\rho}_{D}\b 0\hat{U}_{0}^{\dagger}\b{t,0}$, with  $\hat{U}_0\b{t,0}=e^{-i\b{\hat{H}_S+\hat{H}_C+\hat{H}_E}t/\hbar}$.
Since the device Hamiltonian has no explicit time-dependence,  $\{\hat{G}_{\kappa}=\ket{\phi_n}\bra{\phi_m}\}$ constitute transition operators between the energy states of $\hat{H}_D^{\b L}$ ($\{\ket{\phi_n}\}$ are the device eigenstates). The invariant operator subspace of ${\cal U}_0$, is spanned by the device's invariant eigenoperators, these are generally linear combinations of the device's energy projection operators ${\ket{\phi_n}\bra{\phi_n}}$.

%In the Heisenberg picture, this symmetry is manifested by a decoupling of the non-invariant eigenoperators $\{\hat{F}_{\alpha}\}$, with $\alpha=1,\dots,N\b{N-1}$ and the invariant eigenoperators  
%\footnote{An operator $\hat{S}$ is an eigenoperator with respect to a map ${\cal{V}}\sb{\bullet}$ if it obeys the eigenvalue-type relation ${\cal{V}}\sb{\hat{S}}=\lam \hat{S}$, where $\lam\in\mathbb{C}$. Invariant eigenoperators or have unity eigenvalues $\lam=1$, while for non-invariant eigenoperators $\lam\neq 1$}.
%For a time-independent system Hamiltonian,  $\{\hat{F}_{\alpha}=\ket{n}\bra{m}\}$ constitute transition operators between the energy states of $\hat{H}_S$. Therefore, $\alpha$ implicitly designates a double index $nm$, referring to the two specific energy levels $\ket{n}$ and $\ket{m}$. The invariant operator subspace of ${\cal U}_0$, is spanned by the invariant eigenoperators, which can be chosen as the energy projection operators $\{\hat{\Pi}_i\}$.

When the spectrum of ${\cal U}_0$ is non-degenerate, the associated non-invariant eigenoperators $\{\hat{G}_{\kappa}\}$ also evolve independently. This condition along with the Hermitiacy and trace preserving properties impose strict constrains on the linear form of the dynamical generator. The outcome is a  dynamical symmetric structure of the following form
\begin{eqnarray}
\begin{array}{ll}
  {\cal L}^{\b{L}}_D\b{t} \sb{\bullet} &= -\f{i}{\hbar}\sb{\bar{H}^{\b{L}}_D\b{t},\bullet}\\& 
   +\sum_{\kappa=1}^{N_{D}\b{N_{D}-1}} g_{\kappa\kappa}\b{t} \left( \hat{G}_{\kappa}^{\b{L}} \bullet\hat{G}_{\kappa}^{\b L\dagger} -\f{1}{2}\{\hat{G}_{\kappa}^{\b L\dagger}\hat{G}_{\kappa}^{\b{L}},\bullet\}\right)\\&
    +\sum_{i,j=1}^{N_D}r_{ij}\b{t}\b{\hat{R}_{i}^{\b{L}}\bullet\hat{R}_{j}^{\b L\dagger}-\f 12\{\hat{R}_{j}^{\b L\dagger}\hat{R}_{i}^{\b{L}},\bullet\}}~~,
   \label{eq:local_master_eq_device}
\end{array}
\end{eqnarray}
where $\bar{H}^{\b{L}}_D\b{t}=\f 1{2i}\b{\hat{R}^{\dagger}\b{t}-\hat{R}\b{t}}$ is a Hermitian operator, with $\hat{R}\b{t}^{\b L}=\b{\f {1}{N_D}}^{1/2}\sum_{i=1}^{N_{D}-1}r_{iN_D}\b{t}\hat{R}_{i}^{\b{L}}$ and $N_D=\textrm{dim}\b{H_D^{\b{L}}}=N_S\cdot N_C$. Here, the set $\{\hat{R}_i^{\b L}\}$ constitutes an operator basis for the invariant subspace, satisfying $\hat{R}_{N_D}^{\b{L}}=\hat{I}/\sqrt{N_D}$, while the rest of the operators are traceless. 
A possible choice for such a set is the diagonal matrices of the $SU\b{N}$ generalized Gell-Mann matrices $\hat{R}_j^{\b L}=\sqrt{\f{2}{j\b{j+1}}}\b{\sum_{k=1}^{j}\ket{\phi_k}\bra{\phi_k}-j\ket{\phi_{j+1}}\bra{\phi_{j+1}}}$, for $j=1,\dots,N_D-1$ \cite{bertlmann2008bloch}. Crucially,
the kinetic coefficients $g_{\kappa \kappa}\b{t}$ must be real and possibly negative, while $r_{ij}\b{t}$ are generally complex and satisfy $r_{ij}\b{t}=r_{ji}^*\b{t}$.

From the thermodynamic point of view it is natural to decompose the local master equation, Eq. (\ref{eq:local_master_eq}), into three significant terms: (i) Isolated unitary dynamics, associated with the device bare Hamiltonian $\hat{H}_D^{\b L}$ (ii) The environment's unitary contribution to the reduced dynamics (a Lamb-shift type term), related to the Hermitian operator $\hat{H}_{LS}$ (iii) Dissipative term, generated by the superoperator  ${\cal D}^{\b{L}}_D$. This decomposition reads
\begin{equation}
    {\cal L}^{\b{L}}_D\b t\sb{\bullet} = -\f{i}{\hbar}\sb{\hat{H}_D^{\b L}+\hat{H}_{D,LS}^{\b L}\b t,\bullet}+{\cal D}^{\b L}_D\b t\sb{\bullet}~~,
    \label{eq:local_me_thermo_device}
\end{equation}
where the Lamb-shift term may be time-dependent $\hat{H}_{D,LS}^{\b L}\b t=\bar{H}^{\b{L}}_D\b{t}-\hat{H}_D^{\b L}$. The definition of $\hat{H}_{D,LS}^{\b L}\b t$ together relations (\ref{eq:local_me_thermo_device}) and (\ref{eq:local_master_eq_device}) define the three significant terms. 
The decomposition simplifies the thermodynamic analysis, as it compresses the complex form Eq. (\ref{eq:local_master_eq_device}) into the basic three thermodynamic ingredients.

\subsubsection{Reduced dynamics of the primary system in the local interaction setup}

We next focus on the reduced dynamics of the primary system. The separability of the free time-evolution $\hat{U}_0\b{t}=e^{-i\hat{H}_S t/\hbar}\otimes e^{-i\hat{H}_C t/\hbar}\otimes e^{-i\hat{H}_E t/\hbar}$, implies that device's eigenoperators can also be expressed as a product of local eigenoperators of the primary-system and control 
\begin{equation}
    \hat{G}_{\kappa}^{\b L} = \hat{F}_{\alpha}^{\b L}\otimes \hat{W}_{\beta}^{\b L}~~~;~~~\hat{R}_j^{\b L} = \hat{P}_i^{\b L} \otimes \hat{W}_\beta^{\b L}~~,
\label{eq:eigenoperators_seper}
\end{equation}
where $\hat{F}_\alpha$ and $\hat{R}_j$ are the non-invariant and invariant eigenoperators of primary system free evolution operator, correspondingly, and $\hat{W}_\beta$ are eigenoperators of the control. The index $\alpha=1,\dots,N_S\b{N_S-1}$ implicitly designates a double index $n,m$, labeling  the transition between the primary system's energy states   $\ket{m}$ and $\ket{n}$ ($\alpha=nm$) and similarly for $\beta=1,\dots,N_C\b{N_C-1}$, where $N_S$ and $N_C$ are the dimensions of the primary system and control Hilbert space.

The dynamical description of the primary system can now be obtained by tracing over the control degrees of freedom $ {\cal L}^{\b L}_S\b t\sb{\hat{\rho}_S}=\tr \b{{\cal L}^{\b L}_D\b t\sb{\hat{\rho}_D}} $. 
This procedure along with the product relation, Eq. (\ref{eq:eigenoperators_seper}), implies that under the time-translation symmetry, the primary system reduced dynamics are governed by 

\begin{eqnarray}
\begin{array}{ll}
{\cal L}^{\b{L}}_S\b{t} \sb{\bullet} &= -\f{i}{\hbar}\sb{\bar{H}^{\b{L}}_S\b{t},\bullet}\\ &
   +\sum_{\alpha=1}^{N\b{N-1}} f_{\alpha\alpha}^{\b{L}}\b{t} \left( \hat{F}_{\alpha}^{\b{L}} \bullet\hat{F}_{\alpha}^{\b{L}\dagger} -\f{1}{2}\{\hat{F}_{\alpha}^{\b L \dagger}\hat{F}_{\alpha}^{\b{L}},\bullet\}\right)\\&
    +\sum_{i,j=1}^{N-1}p_{ij}^{\b{L}}\b{t}\b{\hat{P}_{i}^{\b L}\bullet\hat{P}_{j}^{\b L\dagger}-\f 12\{\hat{P}_{j}^{\b L\dagger}\hat{P}_{i}^{\b{L}},\bullet\}}~~,
   \label{eq:local_master_eq}
\end{array}
\end{eqnarray}
where $\bar{H}^{\b{L}}_S\b{t}$ is an Hermitian operator and the operators and kinetic coefficients satisfy analogous conditions as stated below equation (\ref{eq:local_master_eq_device}), with respect to the primary system Hamiltonian.

The first terms of  Eqs. (\ref{eq:local_master_eq}) includes the unitary dynamics as well as the Lamb-shift term, while the other terms are dissipative. Specifically, the second term is in charge of energy transfer, while the third term leads to change of coherence in the energy basis. When the environment is sufficiently large with short memory time, the later leads to pure dephasing of the device state.

In analogy with the devices reduced dynamics, ${\cal{L}}_S^{\b L}$ can also be decomposed into three thermodynamic ingredients, with analogous phyiscal roles 
\begin{equation}
    {\cal L}^{\b{L}}_S\b t\sb{\bullet} = -\f{i}{\hbar}\sb{\hat{H}_S+\hat{H}_{S,LS}^{\b L}\b t,\bullet}+{\cal D}^{\b L}_S\b t\sb{\bullet}~~,
    \label{eq:local_me_thermo_system}
\end{equation}
where $\hat{H}_{S,LS}^{\b L}\b t=\bar{H}^{\b{L}}_S\b{t}-\hat{H}_S$.

Overall, Eq. (\ref{eq:local_master_eq}), serves as the a general structure of a dynamical generator that complies with the thermodynamic postulates and the local interaction model Hamiltonian. This structure can serve as a benchmark to validate various microscopic derivation. It constitutes a (non-Markovian) generalization of the so-called ``local approach'' to derive the master equation \cite{hofer2017markovian,gonzalez2017testing,scali2020local}.

Note that the non-degeneracy condition concerning the Bohr frequency of the free dynamics (spectrum of $ \hat{\cal U}_0\b t$) can be bypassed by slightly shifting the devices energies so to remove the degeneracy. In addition, it can be shown that a sufficiently small modification of the energies is undetectable, Sec. VIII Ref. \cite{dann2021nonmarkovian}. Hence, the dynamical symmetry structure is also effectively valid for systems where a degeneracy in the spectrum of the free propagator exists. 

The kinetic coefficients of the local symmetric structure, Eq. (\ref{eq:local_master_eq}), are determined by considering a specific environment and coupling. They can then be evaluated utilizing a perturbative treatment, see Sec. V of Ref. \cite{dann2021nonmarkovian}, or  complementing the structure with a microscopic derivation which complies with the thermodynamic postulates. Another option is to employ an exact numerical calculation. In the present analysis, our main interest is in the general form of work and heat. We therefore keep a somewhat general description where the master equation is only determined partially, but the structure is set. 

The form of the local master equation ${\cal L}_S^{\b{L}}$ remains valid in the semi-classical limit. In this limit, the control is initialized in a highly excited state with large coherence. As a consequence, its dynamics is dominated by its initial coherence, and its effect on the primary system be can well approximated by time-dependent scalar functions in the total Hamiltonian, cf. Sec. \ref{sec:semi_classical}. Since the Lindblad jump operators of the local interaction model, $\{\hat{F}_{\alpha}\}$ and $\{\hat{P}_i\}$, do not depend on the control state, they remain unaffected. Thus, the operatorial structure remains intact, while the kinetic coefficients converge in the semi-classical limit to the coefficients as calculated under the semi-classical description.

%\tb{\subsection{Semi-classical limit of the local interaction reduced dynamics}}

%\tb{discuss this, I think here the semi-classical limit coincides with Eq. \eqref{eq:local_me_thermo_system} but separating from $\hat{H}_{S,LS}^{\b L}$}

%first term of Eq. \eqref{eq:D_local_1} generates energy transitions between the energy states of $\hat{H}_S$, while the double commutator is in charge of pure dephasing, terminating coherence in the system's energy basis. Such a dynamical structure leads to independent evolution (uncoupled equations of motion) for the population in the energy basis and the coherence.

\subsection{Autonomous thermodynamic variables in terms of system operators}
\label{subsec:local_thermo}

%\trr{It will serve us later to also analyse a small amount of energy transfer to the battery. For this purpose the accumulated work ${\cal{W}}^a$ can be formally decomposed into infinitesimal energy changes, generally expressed as $ \delta {\cal{W}}^a \b t= -  \textrm{tr}\b{\hat{H}_C\sb{\hat{\rho}\b{t+\Delta t}-\hat{\rho}\b{t}}}$, where $t\in\sb{0,t_f-t_i}$. Expressing $\hat{\rho}\b{t+\Delta t}$ in terms of the evolution operator $\hat{U}\b{t_i+t,t_i}=\exp\b{-i \hat{H} t/\hbar}$, utilizing the cyclic property of the trace and expanding the propagator up to order first in $\Delta t$, leads to
%\begin{equation}
 %   \delta {\cal{W}}^a \b t =\Big{\langle}{\sb{\hat{H}_{C}-e^{i\hat{H}\Delta t/\hbar}\hat{H}_{C}e^{-i\hat{H}\Delta t/\hbar}}}\Big{\rangle}\\\approx \f i{\hbar}\Delta t\,\sum_j\Big{\langle}{\sb{\hat{H}_{C},\hat{W}_j}\otimes\hat{S}_j}\Big{\rangle}~,
  %  \label{eq:delta_C}
%\end{equation}
%where $\mean{\bullet}\equiv\textrm{tr}\b{\bullet\, \hat{\rho}\b t}$. Note that the transition to the last equality of Eq. \eqref{eq:delta_C} relies on the independence of the system-environment interaction on the battery operators, which is a unique property of the local interaction model. 
%In the limit of vanishing $\Delta t$ the associated instantaneous {\emph{ autonomous power}}  becomes
%\begin{equation}
 %{\cal P}^{a}\b t=\b{i/\hbar} \,\sum_j\Big{\langle}{\sb{\hat{H}_{C},\hat{W}_j}\otimes\hat{S}_j}\Big{\rangle}~~.
 %\label{eq:P}
%\end{equation}

%}

The thermodynamic definitions can be recast in term of local system operators by utilizing the accurate form of the reduced system dynamics, Eq. (\ref{eq:local_master_eq}). We begin by formally decomposing the work and heat into infinitesimal changes and study the rates associated with the thermodynamic variables.
The rate of change of the autonomous work is the autonomous power ${\cal P}^a\equiv \dot{\cal W}^a$, which can be expressed as 
\begin{eqnarray}
\begin{array}{ll}
  {\cal P}^{a\b L} &=-\textrm{tr}\b{\hat H_{C}\dot{\hat\rho}} =\f{i}{\hbar}\textrm{tr}\b{\hat H_{C}\sb{\hat H^{\b{L}},\hat \rho}}\\&=\f{i}{\hbar}\Big\langle{\sb{\hat H_{C},\hat H_{SC}^{\b L}}}\Big\rangle =-\f{i}{\hbar}\Big\langle{\sb{\hat H_{S},\hat H_{SC}^{\b L}}}\Big\rangle_D.
    \label{eq:power_a_2}
\end{array}
\end{eqnarray}

In the first equality we utilize Eq. (\ref{eq:Work}), while the  second and third equalities follow from the Liouville-von Neumann equation and the identity 
\begin{equation}
    \textrm{tr}\b{\hat{A}\sb{\hat{B},\hat{\rho}}}=\textrm{tr}\b{\sb{\hat{A},\hat{B}}\hat{\rho}}~~,
    \label{eq:identity}
\end{equation} for any operators $\hat{A}$ and $\hat{B}$. The last equality is a direct consequence of the strict energy conservation condition between the system and control Eq. (\ref{eq:SEC_local}), and the definition of the reduced state $\hat{\rho}_{D} = \textrm{tr}_E\b{\hat{\rho}}$. In these equalities we introduced the notation $\mean{\bullet}_i\equiv\textrm{tr}\b{\bullet \hat{\rho}_i}$.

%The commutation relation in the power highlights that any energy transition in the battery must be generated by non-diagonal terms in the eigen-basis of $\hat{H}_C$, %which suggests that $\sb{\hat{H}_{C},\hat{W}_j}\neq 0$. 
%Alternatively,
 %when the bare battery Hamiltonian and coupling terms commute, there exists a basis $\{\ket{\phi_k}\}$ which simultaneously diagonalizes these operators. Therefore, the battery coupling terms and Hamiltonian acquire the following form: $\sum_k h_k \ket{\phi_k}\bra{\phi_k}$, preventing energy transitions within the battery. 

An alternative expression for the power is obtained by directly employing the device's master equation, Eq. (\ref{eq:local_me_thermo_device}), and identity (\ref{eq:identity})
\begin{eqnarray}
\begin{array}{ll}
    {\cal P}^{a\b L}& = -\textrm{tr}\b{\hat H_{C}\dot{\hat \rho}_{D}} \\&
    =\f{i}{\hbar}\textrm{tr}\b{\hat{H}_{C}\sb{\hat{H}_{D}^{\b L}+\hat{H}_{D,LS}^{\b L},
    \hat \rho_{D}}}-\textrm{tr}\b{\hat H_{C}{\cal D}^{\b L}_D\sb{\hat \rho_{D}}}\\&
    =\f{i}{\hbar}\Big\langle{\sb{\hat H_{C},\hat H_{SC}^{\b L}+\hat{H}_{D,LS}^{\b L}}}\Big\rangle_D-\textrm{tr}\b{\hat H_{C}{\cal D}^{\b L}_D\sb{\hat \rho_{D}}}~~.
    \label{eq:derivation1}
    \end{array}
\end{eqnarray}
Notice that the first term of the final expression includes the autonomous power (Eq. (\ref{eq:power_a_2})).
Moreover, the device's Lamb-shift $\hat{H}_{D,LS}^{\b L}$  is only composed of the device's non-invariant eigenoperators (both $\hat{H}_D$ as well as $\bar{H}_D^{\b L}\b t$ are spanned by $\{\hat{R}_j^{\b L}\}$). In turn, the separable structure of the eigenoperators, Eq. (\ref{eq:eigenoperators_seper}), infers that $\hat{H}_{D,LS}^{\b L}$ commutes with the  primary-system and control bare Hamiltonians, $\hat{H}_S$ and $\hat{H}_C$.
These properties imply  that
\begin{equation}
    \textrm{tr}\b{\hat H_{C}{\cal D}^{\b L}_D\b{\hat \rho_{D}}}=0~~.
    \label{eq:derivation2}
\end{equation}
which means that the control energy is unaffected by the system interaction with the environment, despite the fact that ${\cal L}_D^{\b L}$ constitutes a master equation of the primary system and control. 
This result could have been anticipated from the tandem interaction setup of the control-system and environment. %\footnote{When Lambshift is explicitly included in the derivation one obtains     $\textrm{tr}\b{\hat H_{C}{\cal D}^{\b L}\b{\hat \rho_{D}}}-i\textrm{tr}\b{\sb{\hat{H}_C,\hat{H}_{LS}}\hat{\rho}_{SC}}=0$. This expression complies with the fact that the environment does not influence the work  }. 

To express the heat flux in terms of local primary system operators we first write the energy flux in terms of commutators. Employing the Liouville-von Neumann equation and Eq. (\ref{eq:identity}) we get
\begin{eqnarray}
\begin{array}{ll}
 \dot{E}_S^{a\b L}&=\textrm{tr}\b{\hat{H}_{S}\dot{\hat \rho}}=-\f{i}{\hbar}\textrm{tr}\b{\hat{H}_{S}\sb{\hat H,\hat \rho}}\\&=-\f{i}{\hbar}\Big\langle{{\sb{\hat H_{S},\hat H_{SC}^{\b L}}}}\Big\rangle-\f{i}{\hbar}\Big\langle{\sb{\hat H_{S},\hat H_{SE}^{\b L}}}\Big\rangle\\ &={\cal P}^{a\b L}-\f{i}{\hbar}\Big\langle{\sb{\hat H_{S},\hat H_{SE}^{\b L}}}\Big\rangle~~.
 \label{eq:dot_E_S_23}
 \end{array}
\end{eqnarray}
   This relation complies with the first law of thermodynamics, since 
 \begin{eqnarray}
\begin{array}{ll}
      \dot{\cal{Q}}^{a\b L} &=-\textrm{tr}\b{\hat H_{E}\dot{\hat \rho}}=\f{i}{\hbar}\textrm{tr}\b{H_{E}\sb{\hat H,\hat \rho}}\\&=\f{i}{\hbar}\Big\langle{\sb{\hat H_{E},\hat H_{SE}^{\b{L}}}}\Big\rangle=-\f{i}{\hbar}\Big\langle{\sb{\hat H_{S},\hat H_{SE}^{\b{L}}}}\Big\rangle~~,
  \end{array}
   \end{eqnarray}
   where the last equality follows from the strict energy conservation condition. 
 By utilizing  the first law, strict energy conservation between the system and control, Eq. (\ref{eq:derivation2}) and the commutativity of the Lamb-shift term with $\hat{H}_S$, the heat flux can now be expressed in terms of the system and control operators 
 \begin{equation}
     { \dot{\cal Q}}^{a\b L} =\textrm{tr}\b{\hat{H}_{S}{\cal D}_{S}^{\b L}\sb{\hat{\rho}_{S}}}~~.
     \label{eq:local_heat_flux}
 \end{equation}
This expression highlights that in the local  autonomous setup, heat flux has  a single dissipative contribution. Note, the commutivity of the invariant eigenoperators $\{\hat{P}_i\}$ with $\hat{H}_S$ infers that the pure dephasing term of Eq. (\ref{eq:local_master_eq}) does not contributes to the heat.

In order to compare varying thermodynamic definitions from different paradigms it is be beneficial to express the heat flux in terms of the Lindblad jump operators (Eqs. (\ref{eq:local_master_eq}) and (\ref{eq:local_heat_flux}))
\begin{equation}
    { \dot{\cal Q}}^{a\b L} =-\sum_\alpha \hbar\omega_{\alpha}^{\b L}f_{\alpha\alpha}^{\b L}\mean{\hat{F}_{\alpha}^{\b{L}\dagger}\hat{F}_{\alpha}^{\b L}}~,
    \label{eq:comparison_heat}
\end{equation}
where $\omega_\alpha$ is the Bohr frequency associated with the $\alpha$ transition (if $\hat{F}_\alpha$ transitions between $\ket{m}$ and $\ket{n}$ eigenstates of the primary system, i.e.  $\hat{F}_\alpha^{\b L}\propto\ket{n}\bra{m}$, then $\omega_\alpha^{\b L} = \b{\epsilon_m^{\b{L}}-\epsilon_n^{\b L}}/\hbar$, where $\{\epsilon^{\b L}\}$ are the primary-system eigenenergies). 

\section{Semi-classical regime}
\label{sec:semi_classical}

The semi-classical framework considers a primary system, with a bare Hamiltonian $\hat{H}_S$, in the presence of an external scalar driving fields $\{c_j\b t\}$ and the environment. It is represented by a composite Hamiltonian of the form
\begin{equation}
    \hat{H}^{s.c} \b t=\hat{H}_S +\sum_j\hat{S}_j c_j\b t +\hat{H}_{SE}+\hat{H}_{E}~~.
    \label{eq:s_c_Ham}
\end{equation}
The question arises: what is the relation between $\hat{H}^{s.c}\b t$ and $\hat{H}^{\b{L}}$?
That is, starting from a full quantum description of the total Hamiltonian $\hat{H}^{\b{L}}$, we aim to derive the analogous semi-classical Hamiltonian $\hat{H}^{s.c}\b t$. 

%\subsection{From an autonomous to a semi-classical description }
%\label{subsec:autonomous_to_SC}

The semi-classical Hamiltonian emerges from the autonomous local setting by neglecting quantum correlations between the primary-system and control and tracing over the control degrees of freedom \cite{glauber1963quantum,dann2021quantum}. It can be obtained by the following procedure. We first partition the composite density operator to a separable part and a term containing system-control correlations
\begin{equation}
 \hat{\rho}\b t = \hat{\rho}_{SE}\b t \otimes \hat{\rho}_C\b t +\hat{\chi}\b t
 \label{eq:partition_rho}~~,
\end{equation}
where $\hat{\rho}_{SE}\b t=\textrm{tr}_C\b{\hat \rho\b t}$, $\hat{\rho}_C\b t = \textrm{tr}_{SE}\b{\hat{\rho}\b t}$ are the reduced system-environment and control states. %The partial traces on subsystem $i$ or a combination of subsystems are desiganted by  $\textrm{tr}_{i}$, with $i=S,C,E$. 
Substituting Eqs. (\ref{eq:partition_rho}) and (\ref{eq:local_SE_interaction}) into Eq. (\ref{eq:local_Ham})  leads to \footnote{The correlation term $\hat{\chi}$ can be defined by a tensor product of traceless operators. Therefore, partial trace on the control system, $\tr\b{\hat{X}\hat{\chi}}$, will vanish unless $\hat{X}$ is a suitable operator of the control.}
 \begin{eqnarray}
\begin{array}{ll}
    \textrm{tr}_C\b{\hat{H}^{\b L}\hat{\rho}\b t}=& \sb{\hat{H}_S+\sum_j\hat{S}_j\cdot \Big{\langle}{\hat{C}_j}\Big{\rangle}_C\b{t}+\hat{H}_{SE}^{\b L}+\hat{H}_E} \hat{\rho}_{SE}\b t \\&+\sum_j\hat{S}_j\cdot\textrm{tr}_{C}\sb{\hat{C}_j \hat{\chi}\b t}~~,
    \label{eq:HC}
    \end{array}
\end{eqnarray}
Next, we neglect the system-control correlations, allowing to discard the last term. This results in a term of the form $\hat{H}^{s.c}\b t\hat{\rho}_{SE}\b t$. Since $\hat{\rho}\b t$ (and as a result $\hat{\rho}_{SE}\b t$) is an arbitrary state, this suggests that
\begin{equation}
   \hat{H}^{s.c}\b t= \hat{H}_S+\sum_j\hat{S}_j\cdot \Big{\langle}{\hat{C}_j}\Big{\rangle}_C\b{t}+\hat{H}_{SE}^{\b{L}}+\hat{H}_E
\end{equation}
 with scalar fields
\begin{equation}
 c_j\b t= \big{\langle}{\hat{C}_j}\big{\rangle}_C\b{t}~~.
 \label{eq:w_j}
\end{equation}

As expected, equation (\ref{eq:w_j}) infers that the underlying source of the time-dependent classical scalar fields is the dynamics of the quantum control. If the control is a small quantum system the dynamics of $\hat{\rho}_C \b t$ can be calculated explicitly, however, with increasing system size an exact solution becomes unfeasible. A simplified solution can be attained when the system-control coupling has a negligible affect on $\{c_j\b t\}$.
In this regime Eq. (\ref{eq:w_j}) becomes
\begin{equation}
    c_j\b t \approx \textrm{tr}_C\b{\hat{C}_j\hat{U}_C\hat{\rho}_C\b 0 \hat{U}_C^\dagger}
    \label{eq:w_j_approx}~~,
\end{equation} 
with $\hat{U}_C=e^{-i\hat{H}_Ct/\hbar}$.
As a consequence, the control (or battery) state must be non-stationary in order to generate an explicit time-dependent semi-classical Hamiltonian.  

%\tg{The conditions leading to the semi-classical limit and Eq. \eqref{eq:w_j_approx} are analysed in detail in Appendix \ref{apsec:semicalssical_approx}.}

This example can be formalized in terms of two conditions which define the `{\emph{semi-classical limit}}'\cite{dann2021quantum}:
\begin{enumerate}
%    \item Weak WM-control coupling relative to the Bohr frequencies of the bare control Hamiltonian
    \item  The control state is only slightly affected by the interaction with the system.
    \item The correlations between the primary-system and control can be discarded, allowing to express the system state as a separable state $\hat{\rho}_D\b t=\hat{\rho}_S\b t\otimes\hat{\rho}_C\b t$.
\end{enumerate} 
 
%Overall, the presented procedure associates a semi-classical Hamiltonian with an initial autonomous description.

Condition 1 is satisfied when the control is initialized in a highly excited state and the interaction with the system is negligible relative to the typical control energy scale $||\hat{H}_{SC}||\ll||\hat{H}_C||$. As a result, the control system dynamics are dominated by the bare control Hamiltonian $\hat{\rho}_C\b t\approx\hat{U}_C\hat{\rho}_C\hat{U}_C^\dagger$.
This property also influences the emergence of correlations between the system and control. When the system and control are initialized in separable state, the independent dynamics infers that the states remain approximately separable. Alternatively, if the effective system control coupling in the semi-classical framework is of the order of the system typical energy, then increasing the control energy increases the magnitude of  $||\hat{C}_j\hat{\rho}_C||$  and the coupling strength must be scaled accordingly (decreased). In turn, decreasing the coupling decreases the influence of the primary system on the control dynamics \cite{dann2021quantum}.

\subsection{Thermodynamic variables in the Semi-classical approach}
\label{subsec:s_c_thermo_analysis}
The semi-classical approach identifies work as the average change in the internal energy of the total system, including a time-dependent system and environment, ${\cal W}^{s.c}=\Delta E^{s.c} \equiv \textrm{tr}\b{\hat{H}^{s.c}\b t\hat{\rho}_S\b t-\hat{H}^{s.c}\b 0\hat{\rho}_S\b 0}$. This approach was pioneered by Alicki, Spohn and Lebowitz \cite{alicki1979quantum,spohn1978irreversible} and is ubiquitous in the thermodynamic analysis of quantum heat engines \cite{kosloff1984quantum,quan2007quantum,rossnagel2016single,dann2020quantum}. 
By decomposing the composite density operators into a separable part and a term including the system-environment correlation, the total energy change can also be written in terms of the semi-classical power of the primary system (see  \ref{apsec:s_c_work}). This leads to the prevalent expression
\begin{equation}
    {\cal{W}}^{s.c}=\int_{t_i}^{t_f}{\cal{P}}^{s.c}\b t\,dt
    \label{eq:s_c_work}~~,
\end{equation}
where the semi-classical power is defined in terms of the rate of change of the semi-classical Hamiltonian
\begin{equation}
    {\cal{P}}^{s.c}\b t\equiv \bigg{\langle}{{\pd{\hat{H}^{s.c}\b{t}}{t}}}\bigg{\rangle}~~.
    \label{eq:s_c_power_op}
\end{equation} 
    %~~\textrm{with}~~\hat{P}^{s.c}\b t=\pd{\hat{H}^{s.c}\b{t}}{t}~~.    

The derivation does not include any limitation on the driving, thus, it is apparently valid for adiabatic as well as non-adiabatic driving. In addition, relation (\ref{eq:s_c_work}) is independent of the system-environment coupling, therefore, naturally applies for a closed system scenario. A subtlety emerges in the fact that for non-vanishing system-environment coupling, calculating the work  requires the system reduced dynamics $\hat{\rho}_S\b t$. Such a calculation is generally involved, nevertheless, the exact structure of the master equation, complying with the thermodynamic quantum postulates and the local interaction model, can be obtained (Sec. \ref{sec:local_interaction_model}). We utilize this structure to study the form of the associated thermodynamic variables.

In the identification of the system energy there is a certain ambiguity. One can include the explicit time-dependent term, $\sum_j \hat{S}_j c_j\b t$ within the ``system's'' energy  or exclude it. The common treatment includes the driving within the ``system'' Hamiltonian, which leads to a convenient interpretation of heat and work as a byproduct of the differentiation product rule \cite{alicki1979quantum}. Namely, if we define the system's semi-classical Hamiltonian to be $\hat{H}^{s.c}_S\b t = \hat{H}_S +\sum_j \hat{S}_jc_j\b t$, we get 
that the following energy flux
\begin{equation}
    \f{d}{dt}\mean{\hat{H}^{s.c}_S\b t}={\cal P}^{s.c}\b t+\tr\b{\hat{H}^{s.c}_S\b t{\cal{L}}_S\b t\sb{\hat{\rho}_S}}~~,
\end{equation}
where ${\cal L}_S$ is the exact dynamical generator which reduces to ${\cal L}_S^{\b L}$ under the local interaction model. This relation determines the interpretation of the second term as the quantum heat flux. Such approach is usually justified by the fact that the second term varies the entropy of the system. Nevertheless, this identification is inconsistent with the autonomous definition, Eq. (\ref{eq:dot_E_S_23}), which the semi-classical approach should be a limiting case of. Importantly, it leads to an incorrect expression for the heat flux, which is not associated with the energy transfer to or from the environment.
In order to bypass this discrepancy the explicit time-dependent term should be excluded from the semi-classical system energy flux
\begin{equation}
    \dot{E}_S^{s.c}\b t = \textrm{tr}\b{\hat{H}_S{\cal L}_S\b t\sb{\hat{\rho}_S}}~~.
    \label{eq:sc_E_S_35}
\end{equation}
This identification produces the semi-classical definitions as the semi-classical limit of the local autonomous thermodynamic variables, Sec. \ref{sec:autonomous_s_c_connection}. 
The semi-classical heat is now determined by the first law
\begin{equation}
    \dot{\cal{Q}}^{s.c} = \dot{E}_S^{s.c}-{\cal{P}}^{s.c}~~.
    \label{eq:915}
\end{equation}

\trr{One should note that in the adiabatic limit the fixed point of the map is a Gibbs state of the instantaneous semi-classical Hamiltonian (the Lindblad jump operators of Eq. (\ref{eq:local_master_eq}) are eigenoperators of the primary system time-evolution operator). This is a consequence of the underlying dynamical symmetry and a steady state current from the controller to the environment through the system. As a result, the entropy production rate acquires an additional term beyond the anticipated expression in terms of Spohn's inequality \cite{spohn1978entropy}, see Eq.  (\ref{apeq:local_entropy_prod}) 
\ref{apsec:entropy_prod}.}

\section{Autonomous local model - semi-classical connection}
\label{sec:autonomous_s_c_connection}

%\tg{Forging the connection between the autonomous and semi-classical approaches enables comparing the different thermodynamic definitions on even ground. In addition, the comparison highlights their range of validity.}
The question arises, what is the connection between  ${\cal{W}}^{a}$ and ${\cal{W}}^{s.c}$ (Eqs. (\ref{eq:Work}) and (\ref{eq:s_c_work})) or alternatively between the corresponding powers (Eqs. (\ref{eq:derivation1}) and (\ref{eq:s_c_power_op})) under the local interaction model?
To illuminate the relation between the two we derive the semi-classical power from the autonomous definition. We begin the derivation by applying the semi-classical prescription, Eqs. (\ref{eq:s_c_work}) and (\ref{eq:s_c_power_op}), in the interaction picture relative to the bare control Hamiltonian to calculate the semi-classical power. 
In this picture the effective Hamiltonian $\tilde{H}^{\b{L}}\b t=e^{i\hat{H}_C t/\hbar}\hat{H}^{\b{L}}e^{-i\hat{H}_C t/\hbar}$, which together with Eq. (\ref{eq:power_a_2}) leads to 
\begin{equation}
    \textrm{tr}\b{\pd{\tilde{H}^{\b{L}}}t\tilde{\rho}\b t}=\f i{\hbar}\textrm{tr}\b{\sb{\hat{H}_{C},\hat{H}_{SC}^{\b{L}}}\tilde{\rho}\b t}={\cal{P}}^{a{\b{L}}}\b t~~.
    \label{eq:H_p}
\end{equation}

To complete the connection between the two definitions, the partition of the density operator, Eq. (\ref{eq:partition_rho}) and Eq. (\ref{eq:local_SE_interaction}) are substituted into Eq. (\ref{eq:H_p}), leading to a relation between the autonomous and semi-classical power outputs
\begin{equation}
    {\cal{P}}^{a\b{L}}\b t={\cal{P}}^{s.c}\b t+\sum_j\,\textrm{tr}\b{\pd{\tilde{C}_j\b t}{t}\hat{S}_j\tilde{\chi}}~~.
    \label{eq:40corr}
\end{equation}
When the correlation between the system and battery are negligible (semi-classical limit), the autonomous and semi-classical definitions coincide.

%This relation implies that if the correlations between the system and battery were included in the semi-classical description, the semi-classical and autonomous definitions would coincide.

Note, that this is only true in this particular rotating frame (representation). Since different frames lead to varying effective Hamiltonians, each frame  corresponds to different identification of the power operator and work. This property singles out a preferred interaction picture corresponding to the semi-classical thermodynamic analysis. Meaning that the analysis in the interaction picture relative to $\hat{H}_C$, is the only one which is thermodynamically consistent with the full quantum description $\hat{H}^{\b L}$. According to the ``church of the larger Hilbert space" paradigm, such consistency is crucial for the semi-classical description to constitute a limiting case of a universal quantum description. 

We emphasis that the derivation allows for arbitrarily rapid driving. This confirms that in the semi-classical limit ${\cal P}^{s.c}$ is valid beyond the adiabatic driving regime, in contrast to what has been previously assumed \cite{alicki2018introduction,elouard2020thermodynamics}.  

The system energy flux in the semi-classical approach is given by  Eq. (\ref{eq:sc_E_S_35}). Under the local interaction model the exact system dynamical generator coincides with ${\mathcal L}_S^{\b{L}}$ and the two definitions, Eq. (\ref{eq:sc_E_S_35}) and (\ref{eq:dot_E_S_23}), are identical
\begin{equation}
    \dot{ E}^{\b{L}}_S=\dot{E}_S^{s.c}~~.
    \label{eq:41E}
\end{equation}
%\tb{contemplating if to make the left part bold}
The semi-classical heat flux can be obtained by two equivalent procedures: By utilizing first law and or evaluating the local autonomous heat flux, Eq. (\ref{eq:local_heat_flux}), in the semi-classical regime. 
%For further discussion on the form of the local master equation in the semi-classical regime see discussion in the end of Sec. \ref{subsec:local dynamics}.
%\tb{verify the two derivations coincide}
 
 %\trr{In the local interaction model the transition to the interaction picture (with respect to $\hat{H}_C$) does not modify the Lindblad jump operators, which remain the transition operators between the systems energy states. The change is implicitly manifested in the decay rates $\{\Gamma_{nm},\Upsilon_j\}$, where the semi-classical limit implies that $\hat{\rho}_C\b t$ and therefore $\mean{\hat{W}_k\hat{W}_k\hat{\rho}_C\b t}$ are independent of the system. }

\subsection{Example of the Autonomous- Semi-classical connection: thermodynamic analysis of the Jaynes-Cumming model}
\label{sub_sec:jaynes_cumming}
A basic model in quantum optics includes a full quantum treatment of a two-level system (TLS) coupled linearly to a single mode of the electromagnetic field. Under the rotating wave approximation such a system is described by the Jaynes-Cumming model and represented by the following Hamiltonian \cite{cummings1965stimulated,eberly1980periodic,vogel2006quantum} 
\begin{equation}
    \hat{H}_{JC}=\hat{H}_C+\hat{H}_S+\hat{H}_{SC}
    \label{eq:Ham_JC}
\end{equation}
with 
\begin{eqnarray}
\begin{array}{ll}
    \hat{H}_C = \hbar\omega_c\b{\hat{a}^\dagger \hat{a}+\f{1}{2} }
    \nonumber\\
    \hat{H}_S = \f{\hbar\omega_s}{2}\hat{\sigma}_z \nonumber\\
    \nonumber
    \hat{H}_{SC}^{\b {L}} = \hbar g \b{\hat{\sigma}_-\hat{a}^\dagger+\hat{\sigma}_+ \hat{a} }~~.
\end{array}
\end{eqnarray}
%\begin{equation}
 %   \hat{H}_{JC}=\hbar\omega_w\b{\hat{a}^\dagger \hat{a}+\f{1}{2} }+\f{\hbar\omega_s}{2}\hat{\sigma}_z +\hbar g \b{\hat{\sigma}_-\hat{a}^\dagger+\hat{\sigma}_+ \hat{a} }~~.
%\end{equation}
The first two terms correspond to the bare mode and TLS Hamiltonians, while the third term describes an energy conserving interaction with a coupling strength $g$. We will employ this model to illustrate the proposed theory, describing a working medium (TLS) coupled to a work depository (harmonic mode). %\tb{Do we need to define the Hamiltonian any further or this is enough}
%\tg{where $\hat{a}^\dagger$ ($\sigma_+$) and $\hat{a}$ ($\sigma_-$) are the creation annihilation operators of the mode (TLS), $\hat{\sigma}_i$ are the Pauli operators with $i=x,y,z$, $\omega_c$ and $\omega_s$ are the frequencies of the mode and TLS and $g$ is the coupling strength.} 

In the semi-classical limit, the mode effectively converges to an oscillatory driving field $\v E\b t$, inducing the so-called Rabi oscillations in the TLS population \cite{cohen2006quantum}. This relation demonstrates the quantum-semi-classical correspondence between the two models and serves as a natural framework to analyse the relation between the autonomous and semi-classical definitions of work. An additional advantage is that both models, the Jaynes-Cumming (autonomous) and the Rabi (semi-classical), can be solved analytically in the same validity regime.

The oscillations in population can be viewed as a limiting case of the Jaynes-Cumming model. When the field is prepared in a coherent state,  $\ket{\psi\b 0}=\b{a\ket{g}+b\ket{e}}\otimes \ket{\alpha}$, the population of the excited state evolves according to
\begin{eqnarray}
\begin{array}{ll}
    {P}_{e}\b t=\sum_{n=1}^{\infty}\f{e^{-\mean n}\mean n^{n}}{n!}f\b{n,t}~~,
    \\f\b{n,t}\equiv\bigg|\b{\f{2ag\sqrt{n}+b\Delta}{{\Omega_{n}}}}\sin\b{\f{\Omega_{n}t}2}+ib\cos\b{\f{\Omega_{n}t}2}\bigg|^{2}~,
    \end{array}
\end{eqnarray}
where $\Omega_n = \sqrt{\Delta^2+4g^2n}$ is the generalized Rabi frequency of the $n$'th mode and $\Delta = \omega_s-\omega_c$ is the detuning.

When the mean energy of the coherent state is sufficiently large, the evolution of the TLS populations in the Jaynes-Cumming model reduces to the Rabi oscillations \cite{eberly1980periodic,narozhny1981coherence}. To witness the correspondence we take $\mean{n}=|\alpha|^2\gg 1$, while keeping the coupling constant $g |\alpha|=g\sqrt{\mean n}=\textrm{const}$. This leads to the asymptotic behaviour
\begin{equation}
    P_e \b t \simeq f\b{\mean{n},t}  e^{-\zeta t^{2}/{\mean n}}~~,
    \label{eq:P_e}
\end{equation}
where $\zeta = 2{\b{g\sqrt{\mean n}}^{4}}/\Omega_{\mean{n}}^2$ is constant. For sufficiently large $|\alpha|$, $ \mean{n}\gg t^2 \zeta$, which suppresses the decay of probability. When the TLS is initialized in the ground state  ($a=1$, $b=0$) Eq. (\ref{eq:P_e}) coincides with the familiar Rabi formula $P_e^{Rabi}\b t= \f{4\mean n^{2}g^{2}}{\Omega_{\mean{n}}} \sin^{2}\b{\f{\Omega_{\mean{n}}t}{2}}$ \cite{cohen2006quantum}.

The constraint on the coupling strength, namely, that $g\sqrt{n}$ is bounded, can be viewed as a consequence of the rotating wave approximation, which can be viewed as an average over rapid oscillations over the interface energy and leads to the simplified interaction of $\hat{H}_{SC}$. The approximation is valid under the condition that the coupling strength is small relative to the typical transition frequency, i.e., $g|\alpha|\ll \omega_s,\omega_c$. Therefore, when $\mean{n}$ increases $g$ should be scaled accordingly to comply with the relevant operation regime of both models. We can  further infer that for a set coupling strength the change in $\mean{n}$ is restricted. 

An alternative straightforward derivation, leading to the same result, is obtained by the autonomous-semi-classical transformation given in the beginning of Sec. \ref{sec:semi_classical}. Tracing over the autonomous Hamiltonian $\hat{H}_{JC}$ in the interaction picture relative to the control leads to the analogous semi-classical Hamiltonian 
\begin{equation}
    {\hat{H}}_{JC}^{s.c}\b t= \f{\hbar\omega_s}{2}\hat{\sigma}_z +\hbar g \b{\hat{\sigma}_-\alpha^* e^{i \omega_c t}+\hat{\sigma}_+ \alpha e^{-i \omega_c t} }~~.
    \label{eq:JC_sc}
\end{equation}
This Hamiltonian is equivalent to the Hamiltonian of the Rabi model, producing the Rabi formula $P_e^{Rabi}$.

%The system dynamics can be solved analytically with no approximations, leading to an explicit form for populations. When the field is initialized in a coherent state,  $\ket{\psi\b 0}=\b{a\ket{g}+b\ket{e}}\ket{\alpha}$, population of the excited state evolves as 
%This expression results in an oscillatory behaviour ..... 
%The number state distributes Poissonialy, around the mean value $|\alpha|$. In the limit of large $|\alpha|$ the distribution becomes concentrated around the mean value, where the ratio between the standard deviation and the mean scales as $\f{\mean{\Delta n}}{\mean{n}}\sim \f{1}{\sqrt{\mean{n}}}$. Expanding the frequency around the mean value $\mean{n}=|\alpha|^2$ leads to

\subsection*{Thermodynamic correspondence}
In the context of a thermodynamic model, the single field mode of the Jaynes-Cumming model can be naturally identified as a battery, storing the energy extracted from the TLS. The complete Hamiltonian $\hat{H}_{JC}$ describes an autonomous interaction between the system and battery, and work is identified as the mean energy change of the battery, Eq. (\ref{eq:Work}).

The battery and primary system Hamiltonians $\hat{H}_C$ and $\hat{H}_S$ of Eq. (\ref{eq:local_Ham}) correspond to the harmonic mode and TLS bare Hamiltonians, respectively: $\hat{S}_\pm=\sigma_\pm$ and $\hat{C}_{+/-}=\hat{a}/\hat{a}^\dagger$. Substituting these relation into Eq. (\ref{eq:H_p}), the autonomous power becomes
\begin{equation}
    {\cal{P}}^{a}\b t=i\hbar\omega_{c}g\mean{\hat{a}^{\dagger}\hat{\sigma}_{-}-\hat{a}\hat{\sigma}_{+}}~~.
\end{equation}

\begin{figure}[htb!]
\centering
\includegraphics[width=8cm]{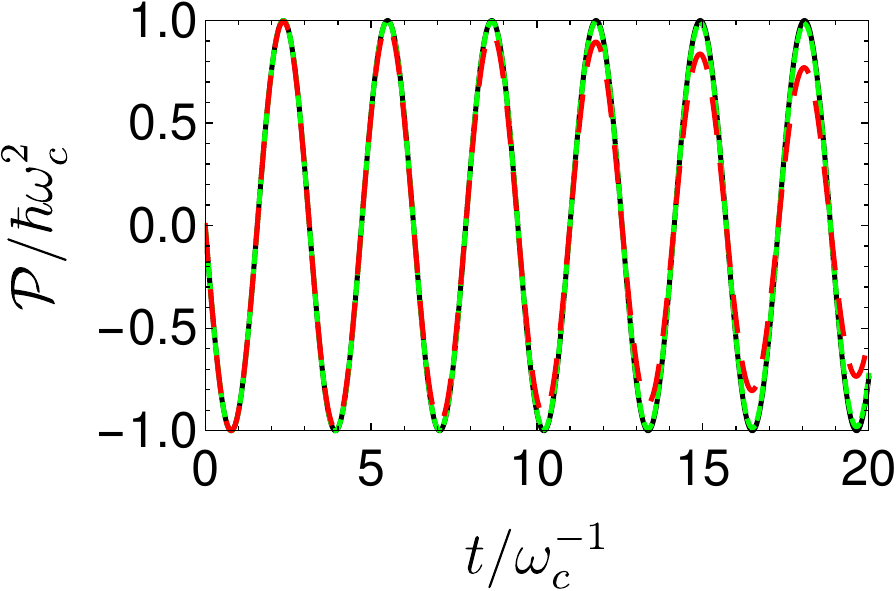}\\
\caption{ Comparison of autonomous and semi-classical  power as function time\label{fig:JC_1} for different values of $\alpha$. $\alpha=100$ - green short-dashed;  $\alpha = 25$ -  red long-dashed, semi-classical power  -black continuous line. With increasing $|\alpha|$ and keeping $g|\alpha|=\textrm{const}$, the amplitude decay is suppressed, eventually leading to the semi-classical power in the limit $\alpha\ra \infty$. The presented model considered equal qubit and mode frequencies, nevertheless similar results are obtained in the non-degenerate case. The model parameters are: $g|\alpha|=1$ with an initial state $\ket{\psi\b 0} = \ket{e}\otimes\ket{\alpha}$, $\omega_c=\omega_s=1$.}   
\end{figure}

In comparison, the semi-classical power is obtained by substituting $\hat{H}^{s.c}_{JC}$ into Eqs. (\ref{eq:s_c_power_op}), giving
\begin{equation}
   P^{s.c}\b t = i\hbar\omega_{c}g\mean{\alpha^{*}\sigma_{-}-\alpha\sigma_{+}}~~.
\end{equation}
In the semi-classical limit, when the correlations between the system and battery are negligible and the battery state dynamics is only slightly perturbed by the two-level system, $\mean{\hat{a}\sigma_+}\approx\mean{\sigma_+} \alpha$ and the autonomous power converges to the semi-classical definition, see Fig. \ref{fig:JC_1}.

\section{Single Shot and Thermodynamic Resource Theory}
\label{sec:single_shot}

The single shot approach formulates quantum thermodynamics as a resource theory (RT) \cite{janzing2000thermodynamic,horodecki2013fundamental,aaberg2013truly,brandao2013resource,halpern2016beyond,lostaglio2019introductory}. Such a theoretical framework, defines a set of free operations that are applied to an initial state (the resource) in an arbitrary combination, and characterizes the possible state transformations. 
The framework is well suited for thermodynamics, as it shares an analogous motivation with the classical theory \cite{callen1998thermodynamics,giles2016mathematical}. Namely, it studies the absolute limits, achievable under certain physical constrains, without imposing any time restriction.

Originally, the thermodynamic RTs were inspired by quantum information studies \cite{werner1989quantum,bennett1996mixed,horodecki2002laws}, such as the RT of entanglement \cite{horodecki2009quantum}. As a consequence, the single shot approach is largely based on a quantum information perspective. It employs information theoretical techniques to provide bounds for thermodynamic variables, such as the maximal extractable work from a certain initial state or the minimal work required to transform a thermal state to the desired quantum state (work of formation).  
 
A number of quantum thermodynamic resource theories have been established, here we focus on the popular framework of the RT of thermal operations \cite{janzing2000thermodynamic,brandao2013resource}. Moreover, the present discussion and results also apply for the RT of extended thermal operations  \cite{cwiklinski2015limitations}, with some slight modifications in our analysis. Importantly, in these RTs the work reservoir is modeled explicitly by a quantum system (a quantum battery), where work is defined in term of a deterministic process involving a transition between the energy eigenstates of the work reservoir. Specifically, Refs. \cite{horodecki2013fundamental} and \cite{aaberg2013truly} employ a work reservoir composed of a `work qubit' (wit) with an energy gap corresponding to the extracted work, however, other possibilities exist \cite{brandao2013resource,gemmer2015single}. In addition, the separable final composite state allows extracting all the work from the quantum battery for further use by means of local unitary operations.   
%\tb{contemplating if to include the red text}

The RT of thermal operations introduces the following set of allowed (free) operations: (i) Adding thermal states at the environment temperature $T$ with an arbitrary Hamiltonian $\hat{H}_E$ (denoted as free states) (ii) Performing energy preserving unitaries, (iii) Discarding subsystems (partial traces). 
This set of operations can be combined and expressed as a quantum channel (CPTP map) acting on a ``device'' system 
\begin{equation}
    \Lambda^{\b{TO}}\sb{\hat{\rho}_D}=\tr\b{\hat{U}^{\b{TO}}\b{\hat{\rho}_D\b 0\otimes \hat{\rho}_E\b 0}\hat{U}^{\b{TO}\dagger}}~~,
    \label{eq:Lambda_TO}
\end{equation} where $\hat{\rho}_D=\hat{\rho}_S\b 0 \otimes \hat{\rho}_C\b 0$, $\hat{\rho}_E\b 0$ is a thermal state, and $\hat{U}^{\b{TO}}$ satisfies
$\sb{\hat{U}^{\b{TO}},\hat{H}_D^{\b{TO}}+\hat{H}_E}=0$.  In the following analysis we modify the original labeling conventions in order to highlight the relation to the other thermodynamic paradigms. In contrast to the local approach Eq. (\ref{eq:11_D}), the thermal operations device Hamiltonian includes only the free primary system  and work reservoir Hamiltonians $\hat{H}_D^{\b{TO}}=\hat{H}_S+\hat{H}_C^{\b{TO}}$, excluding any joint interaction. Such identification then leads to the key commutation relation  
\begin{equation}
    \sb{\hat{H}_S+\hat{H}_C^{\b{TO}}+\hat{H}_E,\hat{U}^{\b{TO}}}=0~~,
    \label{eq:commU_TO}
\end{equation}
which manifests the conservation of energy in basis of the free Hamiltonian. Alternatively, Eq. (\ref{eq:commU_TO}) implies that the subsystems interactions, induced by $\hat{U}^{\b{TO}}$, cannot introduce additional work.
We note that in the original analysis of Ref. \cite{horodecki2013fundamental}, the device also includes an auxiliary thermal system with an arbitrary Hamiltonian $\hat{H}_{S'}$. For the sake of conciseness, we absorb the auxiliary system within the primary-system, as this modification does not affect the final conclusions.

Our present aim is to relate the framework of thermal operations to the dynamical ensemble approaches. The connection is achieved by representing thermal operations in terms of a Hamiltonian description (rather than in terms of dynamical maps). Notice that $\Lambda^{\b{TO}}$ can always be interpreted as a dynamical map of a total system with an Hamiltonian $\hat{H}^{\b{TO}}=\b{i \hbar/t}\textrm{ln}\b{\hat{U}^{\b{TO}}}~~$ \footnote{We can always express the total Hamiltonian as $\hat{H}^{\b{TO}}=\hat{H}_I+\hat{H}_S+\hat{H}_C^{\b{TO}}+\hat{H}_E$, where Eq. (\ref{eq:commut_TO}) implies that the `interaction' term satisfies $\sb{\hat{H}_I,\hat{H}_S+\hat{H}_C+\hat{H}_E}=0$. These relations along with the initial stationary environmental state are sufficient to deduce the form of the dynamical generator.}. This Hamiltonian includes the free Hamiltonians of  environment, control and primary system  and may also include arbitrary coupling terms between all the subsystems, with the constraint (Eq. (\ref{eq:commU_TO})):
\begin{equation}
    \sb{\hat{H}_S+\hat{H}_C^{\b{TO}}+\hat{H}_E,\hat{H}^{\b{TO}}}=0~~.
    \label{eq:commut_TO}
\end{equation}
This familiar relation is analogous to Eq. (\ref{eq:commu2}) which was obtained under the local interaction model. As a consequence, the framework of thermal operations (Eqs. (\ref{eq:Lambda_TO}), (\ref{eq:commut_TO}) and the initial stationary environmental state) provide sufficient conditions for the construction of the local master equation, presented in Sec. \ref{subsec:local dynamics}. This infers that the dynamics of the primary system under thermal operations $\Lambda^{\b{TO}}$, is generated by a generator of local form ${\cal{L}}^{\b{L}}_S$, Eqs. (\ref{eq:local_master_eq}) and (\ref{eq:local_me_thermo_system}).

The operatorial form of the dynamical generator serves as a bridge between the dynamical map description of the single shot paradigm, and the dynamical approaches which are formulated in terms of the dynamical generators and thermodynamic fluxes. In addition, the dynamics of both the single shot and local autonomous approaches leave  $\hat{H}_S+\hat{H}_C+\hat{H}_E$ invariant. Therefore, the identification of work with the energy change of the work reservoir (which is true for both frameworks) implies that the thermodynamics definitions are equivalent. Namely, in both approaches $\Delta E_S=\tr \b{\hat{H}_S\b{\hat{\rho}_S\b{t}-\hat{\rho}_S\b 0}}$, ${\cal{W}}=-\Delta E_C$ 
 and ${\cal Q}=-\Delta E_E$, leading to the same form for the  heat and energy fluxes and the power.

Despite of the similarities, the two approaches address different physical realizations. The single shot approach only considers trajectories for which the control system is in an energy eigenstate at initial and final times. As a result, repetition of the experiment leads to the same work output, and is therefore a deterministic quantity. 
In comparison, the local interaction model defines work in terms of an ensemble average, allowing to analyze trajectories involving mixed states of the control (at initial and final times), such as a generalized Gaussian state. The down-fault of the generalization is that work becomes a stochastic quantity.

Note that Eq. (\ref{eq:commut_TO}) is the only restriction regarding the interactions between subsystems. Hence, $\hat{H}^{\b{TO}}$ may include highly non-local terms which may be extremely difficult (or impossible) to actually realize in the lab. In contrast, in the local autonomous approach the interactions are usually set with a specific physical realization in mind, allowing for an easier realization of the theory.

We emphasis that the transformations induced by thermal operations are not equivalent to the ones generated by the Hamiltonian of the autonomous local interaction model $\hat{H}^{\b{L}}$. Thermal operations constitute a larger set of transformations, since $\hat{H}^{\b{TO}}$ may also include coupling terms between the work reservoir and the environment $\hat{H}^{\b{TO}}_{CE}$, and a global coupling term $\hat{H}
^{\b{TO}}_{SCE}$, involving a combined transitions in the primary system, control and environment. %\footnote{Formally, any global unitary can be concatenated to local and bi-partite unitaries. These are generated by a local Hamiltonian and bi-paratatite interaction Hamiltonians}
Alternatively, the difference in the set of dynamical maps, can also be understood from the fact that the local strict energy conservation relations, Eq. (\ref{eq:SEC_local}), imply Eq. (\ref{eq:commu2}), but the converse is generally false (for instance if $\hat{H}^{\b L}$ includes other interaction terms).  

The local interaction model can be extended to a dynamical framework which corresponds to thermal operations. In the extended local dynamical description the total Hamiltonian includes a general coupling term  which commutes with the total Hamiltonian $\hat{H}^{\b{TOext}}=\hat{H}_S+\hat{H}_C+\hat{H}_E+\hat{H}_I$ \footnote{Without loss of generality the interaction term can be chosen to be of the form $\hat{H}_I=\hat{H}_{SC}^{\b L}+\hat{H}_{SE}^{\b L}+\hat{H}_{CE}+\hat{H}_{SCE}$, where $\hat{H}_{SC}^{\b{L}}$ and $\hat{H}_{SE}^{\b{L}}$ satisfy Eq. (\ref{eq:SEC_local}) and $\sb{\hat{H}_C+\hat{H}_E,\hat{H}_{CE}}=0$.}. The thermal operation framework can be viewed as a special case of the extended local interaction model. It is confined to the cases that involve initial and final control energy eigenstates. 
This restriction introduces a strict distinction between the semi-classical description and the thermal operations paradigm. An energy eigenstate is fundamentally a quantum entity which has no semi-classical analogue. Hence, the semi-classical limit does not lead to the thermodynamical semi-classical definitions.

\section{Global interaction autonomous model}
\label{sec:global_autonomous}
The global interaction autonomous model includes the effect of the environment on both the primary and control quantum systems, cf. Fig. \ref{fig:1-a}. 
Such a physical scenario is represented by the total Hamiltonian
\begin{equation}
    \hat{H}^{\b{G}} = \hat{H}_D^{\b G}+\hat{H}_E+\hat{H}_{DE}^{\b{G}}~~,
    \label{eq:global_Ham}
\end{equation}
where the device interacts with the environment by an energy conserving interaction 
%where $\hat{H}_{DE}^{\b G}$  is the interaction terms between the device and the environment, and the global device Hamiltonian is given by
\begin{equation}
    \sb{\hat{H}_D^{\b G}+\hat{H}_E,\hat{H}_{DE}^{\b G}}=0~~.
    \label{eq:global_sec}
\end{equation}
Similarly to the local scenario, the device is composed of the primary system and control
\begin{equation}
    \hat{H}_D^{\b G} = \hat{H}_S+\hat{H}_C+\hat{H}_{SC}^{\b{G}}~~.
\end{equation}
where,  $\hat{H}_{SC}^{\b G}$  is the internal interaction between the primary system and control in the global setup.  
Contrary to the local setup, strict energy conservation is not imposed between the system and control. 

The global strict energy condition, Eq. (\ref{eq:global_sec}), implies that the interface between the composite system (primary system and control) and environment does not accumulate any energy. Therefore a transition of a quanta of energy to or from the environment is mirrored by a transition in the energy of the device. We stress that such a transition differs from the local scenario Eq. (\ref{eq:SEC_local}),  where a change of energy in the environment is mirrored by an energy change only in the primary system.

In the autonomous framework this modification does not change the notion of autonomous work and heat, Eqs. (\ref{eq:Work}) and (\ref{eq:Heat}), as defined in terms of the total state $\hat{\rho}\b t$. Nevertheless, as will be shown, the cardinal difference in the environmental interaction and dynamical symmetries leads to distinct merits of work and heat in the semi-classical limit.
%This difference arises from the distinct dynamics of the reduced primary system state under the global interaction $\hat{H}_{DE}^{\b{G}}$.

\subsection{Global reduced system dynamics}
The symmetry based axiomatic approach (cf. Sec. \ref{subsec:local dynamics}) enables deducing the precise structure of the reduced dynamics $\hat{\rho}_S\b t$ under the global Hamiltonian $\hat{H}^{\b G}$. Strict energy conservation between the device and environment along with the initial stationary environment state imply that the open system map is time-translation symmetric with respect to the free device dynamics \cite{dann2021quantum}
 \begin{equation}
    {\cal{U}}_{D}^{\b G}\sb{\Lambda^{\b G}\sb{\hat{\rho}_{D}}}={\Lambda^{\b G}\sb{{\cal{U}}_{D}^{\b G}\sb{\hat{\rho}_{D}}}}~~.
    \label{eq:maps_commute_global}
\end{equation}
This condition can be equivalently stated as ${\cal{U}}_{D}^{\b G}\circ\Lambda^{\b G}={\Lambda^{\b G}\circ{\cal{U}}_{D}^{\b G}}$,
where
$\Lambda^{\b G} =\textrm{tr}\b{\hat{U}^{\b G}\b{t,0}\hat{\rho}\b 0\hat{U}^{\b G\dagger}\b{t,0}}$, and  ${\cal U}_{D}^{\b{G}}\sb{\bullet}=\hat{U}_{D}^{\b{G}}\b{t,0}\bullet\hat{U}_{D}^{{\b{G}}\dagger}\b{t,0}$.
 Here, $\hat{U}_{D}^{\b{G}}\b{t,0}=e^{-i \hat{H}_{D}^{\b{G}} t/\hbar}$ is the free propagator of the device and $\hat{U}^{\b G}\b{t,0}=e^{-i\hat{H}^{\b G}t/\hbar}$.

 Under the non-degeneracy condition, the symmetry of the dynamical map infers that the Lindblad operators of the master equation are eigenoperators of the device's free propagator ${\cal U}_D^{\b{G}}$ \cite{dann2021nonmarkovian}. As a consequence, the dynamical symmetric structure of the device is given by 
 \begin{eqnarray}
 \begin{array}{ll}
    {\cal L}_D^{\b{G}}\b{t} \sb{\bullet}& = -i\sb{\bar{H}^{\b{G}}(t),\bullet}\\ &
   +\sum_{\alpha=1}^{N\b{N-1}} g_{\alpha\alpha}^{\b{G}}\b{t} \left( \hat{
   G}_{\alpha}^{\b G} \bullet\hat{G}_{\alpha}^{\b G\dagger} -\f{1}{2}\{\hat{G}_{\alpha}^{\dagger\b G}\hat{G}_{\alpha}^{\b G},\bullet\}\right)\\&
    +\sum_{i,j=1}^{N-1}r_{ij}^{\b G}\b{t}\b{\hat{R}_{i}^{\b G}\bullet\hat{R}_{j}^{\dagger\b G}-\f 12\{\hat{R}_{j}^{\b{G}\dagger}\hat{R}_{i}^{\b G},\bullet\}}~~,
   \label{eq:global_master_eq}
\end{array}
\end{eqnarray}
where $\{\hat{G}_\alpha^{\b{G}}\}$ are the non-invariant eigenoperators of ${\cal U}_D^{\b{G}}$, and $\{\hat{R}_j^{\b G}\}$ constitute an operator basis of the invariant subspace. The kinetic coefficients satisfy analogous properties to the local interaction case as described below Eq. (\ref{eq:local_master_eq_device}). In terms of the thermodynamic ingredients the global master equation can be expressed in the concise form 
 \begin{equation}
    {\cal L}^{\b{G}}_D\b t\sb{\bullet} = -\f{i}{\hbar}\sb{\hat{H}_D^{\b G}+\hat{H}_{D,LS}^{\b G}\b t,\bullet}+{\cal D}^{\b G}_D\b t\sb{\bullet}~~,
    \label{eq:global_me_thermo_device}
\end{equation}
with $\hat{H}_{D,LS}^{\b G}\b t=\bar{H}^{\b{G}}_D\b{t}-\hat{H}_D$.

The global Lamb-shift is composed of the global invariant eigenoperators $\{\hat{R}_j^{\b G}\}$. Unlike the local interaction model these eigenoperators cannot be decomposed to a product of primary-system and control eigenoperator. As a consequence, $\hat{H}_{D,LS}$ does not generally commute with $\hat{H}_S$ and $\hat{H}_C$, and therefore will contribute to the thermodynamic fluxes.

 %, analogous to the set $\{\hat{P}_j\}$ in the local-interaction setup. Similarly, the following analogies exist (satisfying similar mathematical properties as defined below Eq. \eqref{eq:local_master_eq}) between the global and local cases $g_{\alpha\alpha}\leftrightarrow c_{\alpha\alpha}$, $r_{ij}\leftrightarrow d_{ij}$ and $\bar{H}^{\b G}\leftrightarrow\bar{H}$. 
 
  Overall, the master equations of the global and local interaction models (Eq. (\ref{eq:global_master_eq}) and Eq. (\ref{eq:local_master_eq})) share a similar formal structure. The crucial difference concerns the physical  interpretation of the jump operators. 
  The global jump operators $\{\hat{G}_\alpha^{\b{G}}\}$ constitute transition operators between the energy states of the device. Hence, the second term of Eq. (\ref{eq:global_master_eq}) induces energy transitions in the device's energy populations. These correspond to a combined energy transition in the primary system and control. In addition,  $\hat{H}_{SC}^{\b G}$  is not a dynamically invariant quantity, therefore energy transition includes a contribution from the system-control interface. 
 A similar reasoning applies for the invariant eigenoperators $\{\hat{R}_i^{\b G}\}$ and the Hermitian operator $\bar{H}^{\b{G}}$. Consequently, the first and last terms of Eq. (\ref{eq:global_master_eq}) represent the unitary contribution and dephasing in the device's energy basis, correspondingly.

%Such a form will serve us in the thermodynamic analysis.
%This form separates the environment impact to a unitary and dissipative contributions which will serve us in the thermodynamic analysis.

%Comparing the global and local dynamical genrators, Eqs. \eqref{eq:global_master_eq} and \eqref{eq:local_master_eq}, the two superoperators share an identical form. Nevertheless, a crucial distinction concerns the fact that the global transition operators cannot generally be written as a product of primary system and control eigenoperators (Eq. \eqref{eq:G_k_seperable}). That is, $\hat{J}_k\neq \hat{T}_k\otimes\hat{T}_k$. As a consequence, the collective effect of the control and environment on the primary system differs from the local interaction scenario.

 \subsection{Global thermodynamic analysis}
 \label{sec:global_thermo_analysis}

The exact global master equation allows expressing the autonomous energy flux, power and heat in terms of device operators. 
Employing the autonomous work, Eq. (\ref{eq:Work}) and the energy flux definitions, the power and heat flux become 
\begin{equation}
 \dot{E}^{a\b G}_S\b t=\textrm{tr}\b{\hat{H}_S {\cal{L}}_D^{\b{G}}\b t\sb{\hat \rho_D}}
 \end{equation}
 \begin{equation}
     {\cal{P}}^{a\b{G}}\b t=-\textrm{tr}\b{\hat{H}_C {\cal{L}}_D^{\b{G}}\b t\sb{\hat \rho_D}}~~.
 \label{eq:global_power}
\end{equation}
Heat flux is now obtained from the first law and by recalling that strict energy conservation infers the conservation of the total energy of the device and environment (Eqs. (\ref{eq:global_Ham}) and  (\ref{eq:global_sec})),
\begin{equation}
    \dot{\cal{Q}}^{a\b{G}}\b t=\textrm{tr}\b{\b{\hat{H}_D^{\b G}-\hat{H}_{SC}^{\b G}} {\cal{L}}_D^{\b{G}}\b t\sb{\hat \rho_D}}~~.
    \label{eq:Q^aG}
\end{equation}
In terms of the non-invariant eigenoperators of the global master equation this expression becomes
\begin{eqnarray}
\begin{array}{ll}
\dot{\cal{Q}}^{a\b{G}}\b t&=-\sum_\alpha \hbar \omega_\alpha^{\b{G}}g_{\alpha\alpha}^{\b G}\mean{\hat{G}_\alpha^{\b G\dagger}\hat{G}_\alpha^{\b G}}\\&-\textrm{tr}\b{\b{\hat{H}_{SC}^{\b G}} {\cal{L}}_D^{\b{G}}\b t\sb{\hat \rho_D}}~~,    
\label{eq:59_Qa}
\end{array}
\end{eqnarray}
where $\omega_\alpha^{\b{G}}$ is the Bohr frequency, associated with the $\alpha$ energy transition of the device.
Expressing the heat flux in terms of the eigenoperators  will prove to be useful to relate $\dot{\cal{Q}}^{a\b{G}}$ to alternative notions of the heat flux.

In the global model, the heat flux differs from the energy current to the environment $-\tr\b {\hat{H}_E\dot{\hat{\rho}}_E}$ (as obtained in the local interaction model). The difference arises from the second term of Eq. (\ref{eq:59_Qa}), which represents the energy flux into the interface between the primary system and control. Unlike the local interaction model the sum of interaction terms $\hat{H}_{DE}^{\b{G}}+\hat{H}_{SC}^{
\b G}$ does not commute with the total Hamiltonian, thus, energy is accumulated in the interface between the sub-systems. This energy is stored within global correlations of the primary-system and control, which cannot be further consumed by a local operations on the control. Hence, according to the defined framework (Sec. \ref{sec:Framework}) the ``trapped'' energy must be considered as heat. 

Another distinction between the local and global fluxes, is the contribution of the Lamb-shift to the thermodynamics. The device's Lamb-shift $\hat{H}_{D,LS}^{\b G}$, Eq. (\ref{eq:global_me_thermo_device}), may not commute with $\hat{H}_S$ and $\hat{H}_C$. This property leads to terms of the form
$\b{i/\hbar}\bigg{\langle}{\sb{\hat{H}_S,\hat{H}_{D,LS}^{\b G}}}\bigg{\rangle}$ in Eqs. (\ref{eq:Q^aG}) and (\ref{eq:global_power}), and an additional unitary contribution to all the global thermodynamic fluxes. The result highlights the fact that in the global interaction setup, the Lamb-shift should be treated with certain care, and cannot be neglected without justification.

\subsection{Semi-classical limit of the global autonomous approach}
\label{sec:sc_limit_global}

%The effect of the composite dissipator ${\cal D}^{\b G}$ on the primary system can be understood by following the analysis in Ref. \cite{dann2021quantum} and studying the form of the device's propagator $\hat{U}_{D}^{\b G}$.
We next present a concise derivation of  the semi-classical limit of global autonomous approach, for a detailed analysis see \cite{dann2021quantum}. In this limit, the device's time-evolution operator is effectively replaced by the global semi-classical operator  ${\bm{\hat{U}}_S^{\b{G}}}$ operating on the
system's Hilbert space. Such an operator is generated by an explicit time-dependent Hamiltonian, which leads to the reduced system dynamics in terms of the semi-classical eigenoperators.
In the following we denote the semi-classical operators, superoperators and thermodynamic variables by bold letters. For example, ${\bm{\mathcal{L}}}_S^{\b{G}}$ and
$\bm{\mathcal{Q}}^{\b G}$ are the semi-classical limit of the autonomous master equation Eq. (\ref{eq:global_master_eq}), and the autonomous heat in the global interaction model Eq. (\ref{eq:Q^aG}).
 
\subsubsection{Time-evolution operator in the semi-classical limit}
%This operator defines the isolated system map which in turn determines the eigenoperators $\{\hat{J}_k\}$. 
The primary system's reduced dynamics are derived by evaluating the semi-classical limit of the device's time-evolution operator $\hat{U}^{\b G}_D$. In general, the evolution operator can be decomposed as
\begin{equation}
\hat{U}_{D}^{\b{G}}=\hat{U}_C\tilde{U}_D^{\b G}~~,
\label{eq:decomp_global}
\end{equation}
where
$\hat{U}_C\b{t,0}$ is the bare control propagator and
$\tilde{U}_D^{\b G}\b{t,0}={\cal T}\exp\b{-\f{i}{\hbar}\int_0^t\tilde{H}_D^{\b G}\b{\tau}d\tau}$ is generated by the effective composite Hamiltonian in the interaction picture $\tilde{H}_D^{\b G}\b t=\hat{H}_S+\hat{U}_C^\dagger\b{t,0}\hat{H}_{SC}^{\b G}\hat{U}_C\b{t,0}$  and $\cal T$ is the chronological time-ordering operator \cite{dyson1949radiation}.

In the semi-classical limit, the quantum control behaves approximately as a classical system. This is manifested by an independent evolution of the control and the suppression of quantum correlations between the primary system and control (Sec. \ref{sec:semi_classical}). These two properties allow approximating $\tilde{U}^{\b G}_D$ by the {\emph{semi-classical}} propagator
\begin{eqnarray}
\begin{array}{ll}
\tilde{U}_{D}^{\b G}\b{t,0} &\cong \bm{\hat{U}}_S^{\b{G}} \b{t,0}\otimes \hat{I}_C  \\&= {\cal T}\exp\b{-\f{i}{\hbar}\int^{t}_{0}\hat{H}_S^{s.c}\b{\tau}d\tau}\otimes \hat{I}_C~~,
\label{eq:prop_sc_global}
\end{array}
\end{eqnarray}
where  
\begin{eqnarray}
\begin{array}{ll}
\hat{H}_{S}^{s.c}\b{t}& \equiv\textrm{tr}_C\b{\tilde{H}^{\b G}_D\hat{\rho}_C\b 0}\\&= \hat{H}_S+\textrm{tr}_C\b{\b{\hat{U}_C^\dagger\b{t,0}\hat{H}_{SC}^{\b G}\hat{U}_C\b{t,0}}\hat{\rho}_C\b 0}
\label{eq:63scHam}
\end{array}
\end{eqnarray}
is the semi-classical Hamiltonian \cite{dann2021quantum}. The $\cong$ sign designates equalities which are satisfied in the semi-classical limit. Notice, that in the interaction picture relative to the bare control Hamiltonian (in the semi-classical regime)  $\tilde{U}_D^{\b{G}}$ acts trivially on the control system Eq. (\ref{eq:prop_sc_global}), thus, $\bm{\hat{U}}_S^{\b{G}}$ is an operator acting only on the primary system state. While the influence of the control on the primary system is manifested by the explicit time-dependence in $\hat{H}^{s.c}_S$.

In the presence of the environment the composite system is represent by the composite semi-classical Hamiltonian
\begin{equation}
    \hat{H}^{s.c}\b t = \hat{H}^{s.c}_S\b t +\hat{H}_I^{s.c} +\hat{H}_E
    \label{eq:63hsc}
\end{equation}
where $\hat{H}_I^{s.c}$ is the interaction term between the semi-classical system and environment.
The dynamics of the total system is governed by the ${\bm{\mathcal U}}\b{t,0}={\cal T}\exp\b{-\b{i/\hbar}\int_0^t d\tau \hat{H}^{s.c}\b \tau}$, while the reduced dynamical map becomes 
\begin{equation}
    {\bm{\Lambda}}^{\b G }\b{t}\sb{\bullet} = \textrm{tr}_E\b{{\bm{\mathcal U}}\b{t,0}\bullet {\bm{\mathcal U}}^\dagger\b{t,0}}~~.
\end{equation}

\subsubsection{Semi-classical eigenoperators}
\label{sec:s_c_global_eigen_operators}

There is a fundamental difference between the autonomous description and its semi-classical limit, concerning it symmetry properties. This is manifested in a difference in the relation between the dynamical generator and the associated time-evolution operator.  In the autonomous approach, the device free dynamics, governed by the time-evolution operator $\hat{U}_D^{\b{G}}$, are generated by the device Hamiltonian $\hat{H}^{\b{G}}$.
Since the device Hamiltonian does not depend on time, it commutes with $\hat{U}_D^{\b G}$. As a consequence, the global eigenoperators ${\hat{G}_{\alpha}^{\b G}, \hat{R}_i^{\b{G}}}$ constitute eigenoperators of the both the Heisenberg equation and the evolution super operators ${\cal{U}}_D^{\b{G}}$. For instance 
\begin{eqnarray}
    \begin{array}{ll}
    \f{d}{dt}\hat{G}_\alpha^{\b{G}H}\b t&= {\cal H}_D\sb{\hat{G}_\alpha^{\b{G}}} \\&= \hat{U}_D^{\b G \dagger} \b t{\f{i}{\hbar}\sb{\hat{H}_D^{\b G},\hat{G}_\alpha^{\b G}}}\hat{U}_D^{\b G}\b t\\& =-i\omega_\alpha \hat{G}_\alpha^{\b{G}H}\b t
\end{array}
\end{eqnarray}
where the superscript $H$ designates operators in the Heisenberg picture. Similarly,
\begin{equation}
    {\cal U}^{\b G}_D\b{t,0}\sb{\hat{G}^{\b G}_D} = \hat{U}_D^{\b G \dagger} \b{t,0} \hat{G}^{\b G}_D
    \hat{U}_D^{\b G } \b{t,0}= e^{-i\omega_\alpha t} \hat{G}^{\b G}_D~~.
\end{equation}

In comparison, in the semi-classical limit the dynamical generator $\hat{H}^{s.c}_S\b t$ does not commute with itself at different times, therefore it generally does not commute with the semi-classical time-evolution operator $\bm{\hat{U}}_S^{\b{G}}$. This property leads to a different set of eigenoperators, one for the superoperator of the semi-classical Heisenberg equation
\begin{equation}
    \bm{\mathcal H}_D^{\b G} \sb{\bullet}\equiv \hat{U}_D^{\b G \dagger} \b{ t,0}\b{\f{i}{\hbar}\sb{\hat{H}_D^{\b G},\bullet}+\pd{\bullet}{t}}\hat{U}_D^{\b G}\b{ t,0}~~,
    \label{eq:65}
\end{equation}
and the other for the Heisenberg picture evolution superoperator
\begin{equation}
    \bm{\mathcal U}_D^{\b G\ddagger}\sb{\bullet} \equiv \hat{U}_D^{\b G\dagger} \b{ t,0}\bullet \hat{U}_D^{\b G } \b{ t,0}~~.
    \label{eq:66}
\end{equation}
The question arises, which one of these set of eigenoperators is the relevant one which will appear in the master equation?

In Ref. \cite{dann2021quantum} we analysed this issue by building upon the commutation properties of the semi-classical open and isolated system maps. Namely, the global strict energy conservation condition, Eq. (\ref{eq:global_sec}), infers that (cf. Appendix B of Ref. \cite{dann2020quantum}) 
\begin{equation}
    {\bm \Lambda}^{\b G}\circ {\bm{\mathcal U}}_S^{\b G} 
    ={\bm{\mathcal U}}_S^{\b G} \circ{\bm\Lambda}^{\b G}
    \label{eq:comm69}
\end{equation}
where ${\bm{\mathcal U}}_S^{\b G}\b t \sb{\bullet}= {\bm{U}}_S^{\b G}\b{t,0}\bullet {\bm{U}}_S^{\b G \dagger} \b{t,0} $.
Here, we employ a symmetry argument to reach the same final conclusion. Such analysis highlights the relation between thermodynamics in the quantum regime and dynamical symmetries.

To identify the correct set of semi-classical eigenoperators we utilize the dynamical symmetries which are inherent to the present problem. In the autonomous description, When the coupling to the environment vanishes the device Hamiltonian as well as the eigenoperators ${\hat{R}_j^{\b G}}$ constitute a dynamical invariants of the quantum equation of motion. 
In comparison, in the semi-classical description the device Hamiltonian is ill defined as  the semi-classical limit includes tracing over the control degrees of freedom. Nevertheless, we can identify the corresponding invariant eigenoperators as the eigenoperators {of the dynamical generator in Heisenberg picture in the semi-classical limit} $\bm{\mathcal H}^{s.c}$ with zero eigenvalue. These operators are time-dependent constants of motion, characterized by constant expectation values 
\begin{equation}
    \f{d}{dt}\hat{\bm P}_j^H\b t ={\bm{\mathcal{H}}^{s.c}\sb{\hat{\bm P}_j\b t}}= 0\ra \mean{\hat{\bm P}_j^H\b t}=\mean{\hat{\bm P}_j\b 0}~~.
    \label{eq:P_j_67}
\end{equation}
However, this condition does not determine $\hat{\bm P}_j\b t$ uniquely. Any operator of the form
\begin{equation}
\sum_{k=1}^{N^2}c_k \hat{\bm{U}}_S^{\b{G}\dagger}\b{t,0} \hat{O}_k\hat{\bm{U}}_S^{\b{G}}\b{t,0}
\label{eq:67}
\end{equation}
constitutes an invariant eigenoperator of ${\bm{\mathcal{H}}}^{s.c}$, where $c_k$ are time-independent coefficients and $\{\hat{O}_k\}$ is an arbitrary basis of time-independent operators, spanning the system's operator vector space.

In order to set $\{\hat{\bm{P}}_j\b t\}$, we assume that at initial time the control term vanishes, and require that the semi-classical eigenoperators coincide with the system autonomous eigenoperators at initial time. The additional condition on the control does not effectively limit the validity regime of the analysis, as any realistic experiment includes switching on and off control fields within the experimental time duration.
Importantly, the procedure determining the set $\{\hat{\bm{P}}_j\b t\}$ singles out a constant of motion $\hat{X}^{s.c}$ which serves as the relevant operator by which the
eigenstates, $\{\ket{\varphi_j\b t}\}$, and eigenvalues $\{\epsilon_j\}$ constitute a preferred representation of the semi-classical dynamics. $\hat{X}^{s.c}_S\b t$ is chosen such that the initially it coincides with the primary system Hamiltonian
\begin{equation}
\hat{X}^{s.c}_S\b t=\sum_j \epsilon_j \bm{P}_j\b t  ~~.
\label{eq:X^s.c}
\end{equation}
This choice implies that $\{\ket{\varphi_j\b t}\}$ , are stationary states, namely, an isolated semi-classical system initializing in such a state will remain in this state throughout the evolution. Only in the presence of an environment will populations transfer occur between the eigenstates. 
The identification of a preferred basis by which to  study the dynamics (motivated by symmetry considerations), allows identifying the  non-invariant eigenoperators as transition operators between the eigenstates $\{\hat{\bm F}_{ij}\b t\}=\{\ket{\varphi_i\b t}\bra{\varphi_j\b t}\}$ with $i\neq j$. 

Next, we evaluate the dynamics of the eigenstates $\{\ket{\varphi_j\b t}\}$ and eigenoperators in the Heisenberg picture. The dynamics of the invariant eigenoperator are by definition (see Eq. (\ref{eq:P_j_67}))
\begin{equation}
    \hat{\bm{P}}_j^{H}\b t = \hat{\bm{P}}_j\b 0~~.
    \label{eq:70}
\end{equation}
Since unitary dynamics does not modify the rank of an operators, the relation can also be expressed as
\begin{eqnarray}
\begin{array}{ll}
\hat{\bm{P}}_j^{H}\b t &= \hat{\bm{U}}_S^\dagger\b{t,0}\hat{\bm{P}}_j\b 0 \hat{\bm{U}}_S\b{t,0}\\&=
\hat{\bm{U}}_S^\dagger\b{t,0}\ket{\varphi_j\b 0 }\bra{\varphi_j\b 0 } \hat{\bm{U}}_S\b{t,0}\\&=
\ket{\varphi_j^H\b t }\bra{\varphi_j^H\b t }=\ket{\varphi_j\b 0}\bra{\varphi_j\b 0 }  ~~, 
\label{eq:71a}
\end{array}
\end{eqnarray}
where $\ket{\varphi_j^H\b t }$ is an eigenstate of the Heisenberg picture operator $\hat{\bm{P}}_j^{H}\b t$.
Equation (\ref{eq:71a}) and the relations $\b{\bra{\varphi_j\b t }}^\dagger=\ket{\varphi_j\b t }$ infer that the evolution of the stationary states must be of the following form
\begin{equation}
\ket{\varphi_j^H \b t}=e^{i\chi_j\b t}\ket{\varphi_j\b 0}~~,  
\label{eq:71}
\end{equation}
where $\chi_j\b t$ are real scalars.
Utilizing Eq. (\ref{eq:71}) the dynamics of the non-invariant eigenoperators becomes 
\begin{eqnarray}
 \begin{array}{ll}
        \hat{\bm F}^H_{ij}\b t &= \ket{\varphi_i^H \b t}\bra{\varphi_j^H \b t}\\&=e^{i\b{\chi_i\b t-\chi_j\b t}}\ket{\varphi_i \b 0}\bra{\varphi_j \b 0} =e^{-i\theta_{ij}\b t}\hat{\bm F}_{ij}\b 0 ~~,
    \label{eq:72F}
\end{array}
\end{eqnarray}
with $\theta_{ij}=\chi_j\b t-\chi_i\b t$.
In the following analysis we replace the double index $i,j$ by a single index $\alpha$ running over all the possible transitions. This notation leads to the compact expression for the non-invariant eigenoperators
\begin{equation}
\hat{\bm F}_\alpha^H\b t=e^{-i\theta_\alpha\b t}\hat{\bm F}_\alpha\b 0 ~~.
\label{eq:73}
\end{equation}

\subsubsection{Global semi-classical master equation}

The commutation of the open and isolated dynamical maps, Eq. (\ref{eq:comm69}), leads to the general form of the global master equation in the semi-classical limit. This equation governs the reduces system dynamics in the interaction picture relative to the  control free dynamics (see \cite{dann2021quantum} for further details)
\begin{eqnarray}
  \begin{array}{ll}
   \tilde{\bm{\mathcal{L}}}^{\b {G}}_S\b t \sb{\bullet} &= -\f{i}{\hbar}\sb{\bar{\bm H}^{\b G}(t),\bullet} \\&
  + \sum_{\alpha=1}^{N\b{N-1}} \bm{c}_{\alpha\alpha}\b{t}
  \left(
  \hat{\bm F}_{\alpha}\b t \bullet\hat{\bm F}_{\alpha}^{\dagger}\b t
    -\f{1}{2}\{\hat{\bm F}_{\alpha}^{\dagger}\b t\hat{\bm F}_{\alpha}\b t,\bullet\}\right) \\&
  +  \sum_{i,j=1}^{N-1}\bm{d}_{ij}\b{t}\b{\hat{\bm P}_{i}\b t\bullet\hat{\bm P}_{j}^{\dagger}\b t-\f 12\{\hat{\bm P}_{j}^{\dagger}\b t\hat{\bm P}_{i}\b t{,\bullet\}}}~,
   \label{eq:L_time_dependent_sc}
\end{array}
\end{eqnarray}
where $\bar{\bm H}^{\b G}\b{t}$ is defined analogously as below Eq. (\ref{eq:local_master_eq}).
Expressing the semi-classical generator in terms of a dissipator and Lamb-shift enables expressing Eq. (\ref{eq:L_time_dependent_sc}) in a concise form
\begin{equation}
    \tilde{\bm{\mathcal L}}^{\b{G}}_S\b t\sb{\bullet} =
    -\f{i}{\hbar}\sb{\hat{H}_S+\hat{\bm H}_{S,LS}^{\b G}\b t,\bullet}+\bm{\mathcal D}^{\b G}_S\b t\sb{\bullet}~~,
    \label{apeq:local_me_thermo_system}
\end{equation}
with $\hat{\bm H}_{S,LS}^{\b G}\b t=\bar{\bm H}^{\b{G}}_S\b{t}-\hat{H}_S$. 
This master equation generates the primary system's state  dynamics, under the global interaction model in the semi-classical limit. Note that in correspondence with the autonomous case, the Lamb-shift term does not commute $\hat{H}_S$ and therefore will contribute to the primary-system's energy change and to the thermodynamic analysis.

An essential distinction between the local and global dynamical equations arises from the difference in the dynamical symmetries of the map.  In the local setup the symmetry is such that the device eigenoperators constitute a product of system and control dependent eigenoperators, Eq. (\ref{eq:eigenoperators_seper}). Conversely, in the global setup the eigenoperators, operating on the system's Hilbert space, also depend on the control state. As a result, the semi-classical Lindblad jump operators are control dependent.

The dependency of the dissipative dynamics on the coherent evolution constitutes the main feature of Eq. (\ref{eq:L_time_dependent_sc})
This property enables manipulating the dissipation by the application of appropriate control protocols \cite{dann2019shortcut,dann2020fast,kallush2021controlling}.
Such protocols pave the way to control transitions involving entropy changes, such as cooling or the generation of effectively unitary gates under dissipative conditions.

\subsection{Semi-classical limit of the global thermodynamic variables}

The global thermodynamic variables depend on the dynamics of the device, therefore both heat and work generally include combined changes of both the system and control. As a consequence, the semi-classical limit should be taken with certain care.

In the semi-classical limit, the global autonomous dynamical generator in the interaction picture relative control free dynamics, $\tilde{\cal{L}}_S^{\b G}$, is replaced by the semi-classical generator $\tilde{\bm{\mathcal{L}}}_{S}^{\b G}$. As a result, the autonomous primary system energy, Eq. (\ref{eq:global_power}), has a direct semi-classical limit
\begin{equation}
    \dot{\bm{E}}_S^{\b{G}}\b t=\textrm{tr}\b{\hat{H}_S \tilde{\bm{\mathcal{L}}}_S^{\b{G}}\b t\sb{\hat \rho_S}}~~.
    \label{eq:global_E_sc}
\end{equation}
However, the autonomous power ${\cal{P}}^{a\b{G}} $ and heat flux $\dot{\cal{Q}}^{a\b{G}} $  do not have a simple semi-classical limit.
The complication arises from the fact that both $\hat{H}_C$
and $\hat{H}^{\b G}_D$ are ill defined in the semi-classical description. 

By applying similar symmetry considerations as discussed in the previous subsection we can find a replacement for $\hat{H}^{\b G}_D$ in the semi-classical limit. When the device is isolated from the environment, $\hat{H}^{\b G}_D$ is replaced by a time-dependent constant of motion $\hat{X}^{s.c}_S\b t$. This identification enables performing the semi-classical limit to the autonomous thermodynamic definitions Eqs. (\ref{eq:global_power}) and (\ref{eq:Q^aG}). The limit is achieved by replacing the autonomous superoperator and operators by their semi-classical counterparts ${\cal L}_D^{\b G}\ra \tilde{\bm{\mathcal L}}_S^{\b G}$, $\hat{H}^{\b G}_D\ra \hat{X}^{s.c}_S\b t$ and  $\hat{H}_{SC}^{\b G}\ra {\bm{H}}_{SC}^{\b G}=\hat{U}_C^\dagger\b{t,0}\hat{H}_{SC}^{\b G}\hat{U}_C\b{t,0}$
\begin{equation}
\dot{\bm{\mathcal{Q}}}^{\b{G}}\b t=\textrm{tr}\b{\b{\hat{X}_S^{s.c}\b t-\hat{\bm H}_{SC}^{\b G}} \tilde{\bm{\mathcal{L}}}_S^{\b{G}}\b t\sb{\hat \rho_S}}~~,\label{eq:80a}
\end{equation}
The power is now obtained from the first law
\begin{equation}
{\bm{\mathcal P}}^{\b G}\b t = \dot{\bm E}^{\b G}_S\b t-\dot{\bm{\mathcal{Q}}}^{\b{G}}\b t~~.
\label{eq:82_cons}
\end{equation}

By substituting Eq. (\ref{eq:X^s.c}) and (\ref{eq:L_time_dependent_sc}) into Eq. (\ref{eq:80a}) the heat flux can be expresses as 
\begin{equation}
\dot{{\bm{\mathcal Q}}}^{\b{G}}\b t=-\sum_{\alpha}\hbar \omega_\alpha \bm c_{\alpha\alpha}\b t\mean{\hat{\bm F}_{\alpha}^{\dagger}\b t\hat{\bm F}_{\alpha}\b t}-\bm{\Phi}^{\b G}\b t~~,
\label{eq:Q_sc_G_final}
\end{equation}
where  $\omega_\alpha=\b{\epsilon_m-\epsilon_n}/\hbar$ and 
\begin{equation}
    \bm{\Phi}^{\b G}\b t =\textrm{tr}\b{\hat{\bm H}_{SC}^{\b G} \tilde{\bm{\mathcal{L}}}_S^{\b{G}}\b t\sb{\hat \rho_S}}~~.
\end{equation}

The physical meaning of the sum of Eq. (\ref{eq:Q_sc_G_final})  is illuminated by recalling that $\bm{c}_{\alpha\alpha}\b t$ are kinetic coefficients, and that the expectation values correspond to the populations of the eigenstates from which the population is transferred. Altogether, each term in the sum corresponds to the effective energy transfer rate associated with the transition $\alpha$ (where effective energy is defined in terms of the eigenbasis of the semi-classical time-evolution operator). The sum is therefore the accumulated effective (semi-classical) energy transfer to the environment. 

\trr{In adiabatic limit (with in the semi-classical description), the invariant $X_S^{s.c}\b t$ converges to the instantaneous semi-classical Hamiltonian, $H^{s.c}_S\b t$ Eq. (\ref{eq:prop_sc_global}). In addition, the semi-classical limit allows neglecting the contribution of the system control interaction $\hat{\bm{H}}_{SC}^{G}$ (see discussion above Eq. (\ref{eq:P_e})). Equation (\ref{eq:80a}) then converges to  the familiar definition for the heat current 
\begin{equation}
\dot{\bm{\mathcal{Q}}}^{\b{G}}\b t\approx\textrm{tr}\b{{\hat{H}_S^{s.c}\b t} \tilde{\bm{\mathcal{L}}}_S^{\b{G}}\b t\sb{\hat \rho_S}} 
\label{eq:adiabatic_heat}
\end{equation}
\cite{alicki1979quantum} and one obtains a positive entropy production \cite{spohn1978entropy} (see \ref{apsec:entropy_prod} for a detailed derivation)
\begin{equation}
\dot{\Sigma} = \f d{dt}S_{D}+\beta{\rm{tr}}\b{{\cal L}\b{\hat \rho_{D}}{\rm{ln}}\hat \rho_{D}^{th}}~~,
\label{eq:adiabatic_ep}
\end{equation}
where
 $\hat \rho_{D}^{th}=Ze^{-\beta \hat H^{s.c}_S\b t}$.}

%In the semi-classical limit, the global thermodynamic variables can be expressed in terms of local system and control operators. In this regime, the control state propagates independently of the system and the device state is a product of the control and system states $\hat{\rho}_D\b t\cong\hat{\rho}_D^{s.c}\b t\equiv\hat{\rho}_C^{s.c}\b t\otimes\hat{\rho}_S^{s.c}\b t$, where $\hat{\rho}_C^{s.c}\b t = \hat{U}_C\b{t,0}\hat{\rho}_C\b 0\hat{U}_C^\dagger\b{t,0}$. 

%\tg{Here, we introduced the ``s.c" initials to emphasis that dynamics of the system in the interaction picture is generated by the semi-classical master equation Eq. \eqref{eq:D_54}}. \tg{This relation together with Eq. \eqref{eq:P^G} lead to the global semi-classical power 
%\begin{equation}
 %   {\cal P}^{s.c \b{G}} = i\textrm{tr}\b{\sb{\hat{H}_C,\hat{H}_{SC}}\hat{\rho}_D^{s.c}}~~.
%\end{equation}
%and Eq. \eqref{eq:Q^g} gives the global semi-classical heat flux
%\begin{equation}
 %{\cal Q}^{{s.c}\b{G}} =    
  %  \textrm{tr}\b{\hat{H}_S{\cal D}^{\b G s.c}\b{\hat{\rho}_S}}\\-i\textrm{tr}\b{\sb{\hat{H}_C+\hat{H}_{SC}^{\b G},\hat{H}_{DE}}\hat{\rho}_D^{s.c}}~~.
   % \label{eq:Q^g_2}
%\end{equation}}

%\tb{I think we should write it in terms of the semi-classical generator.}

\section{Thermodynamic definitions from an external viewpoint}
\label{sec:external approach}
The external approach identifies the heat flux in terms of the integrated energy flow into the environment: $\dot{E}_E\equiv \f{d}{dt}\textrm{tr}\b{\hat{\rho}\b t\hat{H}_E}$ (with an additional minus sign to get the heat flux with respect to the system) \cite{esposito2010entropy,elouard2020thermodynamics}.
Such identification is equivalent to the local autonomous definition, Eq. (\ref{eq:Heat}). However, since the solution for the environment dynamics is unfeasible, this identification is only a starting point to derive the heat flux in terms of local system operators. In Ref. \cite{elouard2020thermodynamics}  expressions for the ``external heat flux" are derived for qubit undergoing periodic driving, using the Born-Markov-secular approximation scheme (so-called Davies construction).
In \ref{apsec:external_derivation} we generalize the previous results to an arbitrary finite level system and driving (not necessarily periodic). %following the construction of the non-adiabatic master equation (NAME) \cite{dann2018time}. 
Here, we introduce the generalized thermodynamic definitions and compare them  to the alternative definitions.

The construction of the external heat flux follows the standard (Born-Markov-Secular) approximation scheme leading to the reduced dynamics of open quantum systems \cite{breuer2002theory}. The control is treated semi-classically and the coupling between the system and environment must be weak. In addition,  the environment is assumed to be very large compared to the system and its dynamics is much faster then the system's. As a result, the evolution is effectively Markovian and the environment remains in the initial stationary state.
Such a scenario is represented by the total Hamiltonian
\begin{equation}
    \hat{H}^{{e}} \b t=\hat{H}_S^{{e}}\b t +\hat{H}_{I}^{ e}+\hat{H}_{E}~~,
    \label{eq:s_c_Ham_2}
\end{equation}
where the superscript ${e}$ is introduced to emphasis that the semi-classical drive is incorporated within $\hat{H}_S^{{e}}\b t$, and that the interaction term does not necessarily satisfy strict energy conservation.
For simplicity, we consider the case of a single system-reservoir interaction term 
\begin{equation}
    \hat{H}_{I}^{ e}=\hat{S}\otimes \hat{E}~~,
    \label{eq:interaction_floquet}
\end{equation}
where the generalization to the multiple interaction terms and reservoirs follows a similar procedure, see for example \cite{szczygielski2013markovian}. The environment interaction operator satisfies $\langle{\hat{E}}\rangle_E=0$, with out loss of generality \footnote{The condition can be achieved by a suitable transformation of the Hamiltonian if not present from the beginning.}.

The identical structure of $\hat{H}^{e}\b t$ (Eq. (\ref{eq:s_c_Ham_2})) and $\hat{H}^{s.c}\b t$ (Eq. (\ref{eq:63hsc})) enables employing the same symmetry analysis of subsection \ref{sec:s_c_global_eigen_operators} to the framework of the external map. This leads analogous to invariant and non-invariant eigenoperators $\{\hat{P}^{\b{e}}_j\}$ and   $\{\hat{F}^{\b{e}}_\alpha\}$ and associated phases $\{\theta_\alpha\b t\}$, which satisfy
similar relations as Eqs. (\ref{eq:70}) and (\ref{eq:73}). The correspondence between the two descriptions allows replacing the external eigenoperators by $\hat{\bm{P}}_j$ and
$\hat{\bm{F}}_\alpha$ altogether. Here, we keep in mind that the replacement requires that $\hat{H}^{e}\b t$ must be identical to $\hat{H}^{s.c}\b t$. We emphasis that the equivalence of the Hamiltonian does not imply that the thermodynamical analysis are the same.

%These conditions infer strict relations between the typical timescales of the system, decay of environment correlation functions and relaxation process,  see \tb{Appendix \ref{} for further details}.  
Under the weak coupling and Markovian assumptions, the external heat flux is given by (cf.  \ref{apsec:external_derivation})
\begin{equation}
\dot{{\cal Q}}^{e}=-\sum_{\alpha}\hbar{\omega_\alpha^{\b e}}G\b{\omega_\alpha^{\b e}}\mean{\hat{ \bm F}_{\alpha}^{\dagger}\hat{ \bm F}_{\alpha}}~~.
\label{eq:Q_e_final}
\end{equation}
where $\omega^{\b{e}}_\alpha\b t\equiv \f{d}{dt}\theta_{\alpha}\b {t'}\bigg{|}_{t'=t} $

 %Hence, it constitutes the Markovian weak coupling limit of $\bm{c}_{\alpha\alpha}$.

%$\hat{\bm{F}}_\alpha$ are the non-invariant eigenoperators of the semi-classical time-evolution operator $\hat{\bm U}_S\b t$, satisfying \tb{...} and \tb{$\nu_\alpha=...$}

The expression for the external heat flux shares a similar form and interpretation as the first term of Eq. (\ref{eq:Q_sc_G_final}). Thus, the total external heat flux is a sum over all the energy fluxes, each associated with an exchange of a defined quanta of energy with the environment. The positive frequencies correspond to dissipation of heat, while the negative frequencies are associated with absorption of energy from the environment.
The dependence of the external heat flux Eq. (\ref{eq:Q_e_final}) 
on the global Lindblad operators in the semi-classical limit, $\hat{\bm{F}}_\alpha$, suggests that the evolution of the
reduced state is generated by a master equation of the form of $\bm{\mathcal {L}}^{\b G}_S$ Eq. 
(\ref{eq:L_time_dependent_sc}). This identification is crucial for the consistency of the thermodynamic analysis and the open system dynamics. In turn, the operatorial structure of the master equation implies an underlying time-translation symmetry Eq. (\ref{eq:maps_commute_global}). This relation demonstrates the deep connection between the chosen thermodynamic approach and the underlying dynamical symmetry in the autonomous description of the same physical system.   %\tb{In turn, this strongly implies that $\hat{H}_{SE}^{ e}$ satisfies the global strict energy conservation condition.} 

The external energy flux is given by
\begin{equation}
    \dot{E}^{e}_S=\f{d}{dt}\tr \b{\hat{H}^{e}_S\b t \hat{\rho}_S^{ e}\b t}~~
    \label{eq:E^e_1}
\end{equation}
and the work can be determined by the first law.

In the following section, we connect the external thermodynamic fluxes to the global autonomous and semi-classical limit definitions. Comparing the two approaches will highlight an inconsistency in the definition of $\dot{E}_S^e$ (analogously to  $\dot{E}^{s.c}$ in Sec. \ref{sec:semi_classical}), which can be solved by modifying the original definition, Eq. (\ref{eq:E^e_1}).

\subsection{Connection to alternative definitions of heat flux and power}

We begin by focusing on the external energy flux.
To relate Eq. (\ref{eq:E^e_1}) to the autonomous approaches we embed the time dependence of  $\hat{H}^{e}\b t$ within an autonomous description, including  a primary-system, control and environment. In such a description the control is treated as a quantum system and the ``system'' Hamiltonian of the external approach is obtained by tracing over the control in the interaction picture, cf. Sec.  \ref{sec:semi_classical}), 
\begin{equation}
\hat{H}^{e}_S\b t = \hat{H}_S+\tr{\b{\hat{U}_C^\dagger\b{t,0}\hat{H}_{SC}^{{e}}\hat{U}_C\b{t,0}\hat{\rho}_C\b 0}}~~.
\label{eq:H_S_e_aut}
\end{equation}
We emphasis that the system control interaction term $\hat{H}_{SC}^{e}$ (similarly to $\hat{H}_{SC}^{\b{G}}$)  does not satisfy the local strict energy conservation condition, Eq. (\ref{eq:SEC_local}).
Substituting Eq. (\ref{eq:H_S_e_aut}) into Eq. (\ref{eq:E^e_1}) leads to 
\begin{equation}
    \dot{E}^{ e}_S = \dot{\bm{E}}^{\b G }_S+\bm{\Phi}^{\b G}~~
    \label{eq:E^e_deriv}
\end{equation}
where the third term is obtained by noticing that  $\hat{H}^e_{SC}$ is equivalent to $\hat{H}^{\b G}_{SC}$, both represent a general interaction between the primary system and control.

%\tb{For similar reasons as described in the semi-classical approach, Sec \ref{subsec:s_c_thermo_analysis}, we will demonstrate that this definition is inconsistent with the autonomous approaches. Following, we introduce a modification of the definition to restore consistency with the complete quantum description. \tb{this section should be organized it is incoherent}}

%The external approach can be shown to be a limit of the global semi-classical and global autonomous approaches (sections  \ref{sec:sc_limit_global} and \ref{sec:global_thermo_analysis}).  
 The external heat flux can be similarly related to the global semi-classical and global autonomous definitions (section  \ref{sec:sc_limit_global} and \ref{sec:global_thermo_analysis}).
Under Markovian dynamics and in the weak coupling limit the semi-classical kinetic coefficients $\{\bm{c}_{\alpha\alpha}\}$, Eq. (\ref{eq:Q_sc_G_final}), converge to  $G\b{\nu_\alpha}$, Ref. \cite{dann2021nonmarkovian} Sec. VII. This leads to
\begin{equation}
\dot{{\cal Q}}^{e}\b t\eqsim\dot{\bm {\mathcal Q}}^{\b{G}}+\bm \Phi^{\b G}~~,
\label{eq:72}
\end{equation}
where $\eqsim$ designates that the equality holds only in the limit of semi-classical, Markovian dynamics and weak system-environment coupling.
Therefore, the heat fluxes (in the considered regime) differ by a term proportional to the primary system-control coupling strength.

Relation between (\ref{eq:72}), and the autonomous definition of ${\cal Q}^{a\b G}$, Eq. (\ref{eq:Q^aG}) suggests that the  external heat flux emerges from the energy flow from the device in the autonomous description
\begin{equation}
\dot{\cal{Q}}^{\b e}\eqsim\textrm{tr}{\b{\hat{H}_D^{\b G} {\cal{L}}_D^{\b{G}}\sb{\hat \rho_D}}}~~.    
\end{equation}
In the global interaction setup this coincides with the energy flux to the environment, in compliance with the original identification.
%The connection between the external energy flow, Eq. \eqref{eq:E^e_deriv}, and semi-classical limit of the  global autonomous result $\dot{\bm{E}}_S^{\b G}$, Eq. \eqref{eq:global_E_sc}, is more subtle.  This can be seen by comparing the powers of the two approaches. 
A relation between the powers are derived by substituting Eq. (\ref{eq:72}) into Eq. (\ref{eq:E^e_deriv}) and utilizing the first law, Eq. (\ref{eq:82_cons}), we get
\begin{equation}
    {\cal P}^e \eqsim \bm{\mathcal{P}}^{\b G}.
    \label{eq:75}
\end{equation}

These relations imply that the external and global approached differ by the way they treat the interface energy between the system and control. The external approach includes this energy as a part of the primarly system's energy, while the the global approach includes it as a part of the environment.
This discrepancy can be fixed in two ways. One can redefining  $E^e_S$ and identify it with the energy of the time-independent part of $\hat{H}^e_S\b t$, i.e, $\hat{H}_S$ of Eq. (\ref{eq:H_S_e_aut}). This definition leads to 
\begin{equation}
\dot{E}^e_S = \dot{\bm{E}}^{\b G}_S~~.
\label{eq:E^e_S}
\end{equation}
Alternatively, in the limit of negligible primary system control coupling strength, $\bm{\Phi}^{\b G}$ can be discarded and the two approaches converge.

%Equations \eqref{eq:E^e_deriv} and \eqref{eq:72} highlight the distinction between the two sets of thermodynamic definitions.
%These infer an inconsistency in the thermodynamic analysis, as both approaches are associated with the same autonomous setup, the global interaction model.  
%\tg{This discrepancy between the external and global approaches could have been foreseen from the beginning. The external heat flux was originally defined as the negative of the energy flux to the environment, $-\dot{E}_E$. This definition complies with the local autonomous approach, Eq. \eqref{eq:Heat}, but differs from the global one \eqref{eq:Q^aG}. Therefore, if one employees a master equation which constitutes a semi-classical limit of the dynamics of the global approach (e.g. Flouqet, adiabatic, or the non-adiabatic master equations), the thermodynamic analysis will have an inherent inconsistency. To overcome this issue energy accumulated in the primary system  control coupling should be included in the external heat, restoring the consistency with the global interaction model.}
%\tb{I DON'T THINK THE LAST PART IS CORRECT, DISCUSS THIS}

%If the coupling strength is negligible throughout the evolution, the two heat fluxes coincide.

\section{Dynamical map approach: First law from Entropy production}
\label{sec:dynamical map}
The dynamical map approach identifies the heat flux from the entropy production rate of the primary system. This proposal relies on the `strict' Markovianity of the map, which is a limiting case of the general dynamics. In the strict Markovian regime, memory effects are neglected and the system dynamics are governed by 
a semi-group dynamical map which is contracting. This mathematical property can be interpreted  as a monotonic decrease in distinguishability between different initial states. This effectively leads to Spohn's inequality \cite{spohn1978entropy}:
\begin{equation}
 \Sigma\sb{\hat{\sigma}}=k_B\b{\dot{\cal{S}}_{V.N}\sb{\hat \sigma}+\textrm{tr}\b{{\cal{L}}_S^M\sb{\hat{\sigma}}\textrm{ln} \hat{\sigma}_{i.a}}}\geq 0~~,
 \label{eq:spohn_ineq}
\end{equation}
where $k_B$ is the Boltzmann factor, $\dot{{\cal S}}_{V.N}\sb{\hat \sigma} = -\textrm{tr}\b{\hat{\sigma}\textrm{ln}\hat \sigma}$ is the  von-Neumann entropy of the reduced state and  ${\cal{L}}_S^M$ is the Markovian generator of the dynamical map which may be time-dependent. $\hat{\sigma}_{i.a}\b t$ is the instantaneous attractor, which constitutes the fixed point of the dynamical map at time $t$, ${\cal{L}}_S^M\b t\sb{\hat{\sigma}_{i.a}\b t} =0$. 

The positivity of $\Sigma\sb{\hat{\rho}_{S}}$ allows interpreting it as the entropy production and Eq. (\ref{eq:spohn_ineq}) as the microscopic differential version of the second law of thermodynamics \cite{alicki1979quantum}. In the presence of a thermal bath, the system's change in entropy corresponds to $k_B \dot{{\cal S}}_{V.N}\sb{\hat{\rho}_{S}}$, which motivates identifying the heat flux as
\begin{equation}
\dot{{\cal Q}}^{d.m}=- k_B T\textrm{tr}\b{{\cal{L}}_S^M\sb{\hat{\rho}_{S}}\textrm{ln} \b{\hat{\rho}_{S}}_{i.a}}~~,
\label{eq:Q_dm}
\end{equation}
in order to guarantee consistency with the second law.
In turn, the dynamical map power ${\cal{P}}^{d.m}$ is evaluated from the first law. 

The dynamical map approach has been previously employed to analyse the thermodynamics of time-independent systems \cite{alicki1979quantum} and then generalized for periodic driving  \cite{alicki2012periodically,kolavr2012quantum,levy2012quantum,szczygielski2013markovian} as well as for inertial driving protocols \cite{dann2020quantum,dann2021inertial}.

The Markovian generator ${\cal L}_S^M$ and instantaneous attractor both depend on the specific physical scenario under study and the nature of the primary system environment interaction. Specifically, in the local interaction setup under Markovian dynamics, ${\cal L}_S^M$  coincides with ${\mathcal L}_S^{\b L}$ Eq. (\ref{eq:local_me_thermo_device}), while for a global interaction it coincides with ${\cal L}_S^{\b G}$ (\ref{eq:global_me_thermo_device}), or in the semi-classical limit it converges to the semi-classical counterpart $\bm{\mathcal L}_S^{\b{G}}$. Therefore, in order to compare $\dot{\cal Q}^{d.m}$ to the other definitions we assume a general form for ${\cal L}_S^M$ and derive the  heat flux in terms of general Lindblad jump operators. In the following section, we consider specific instances, where ${\cal L}_S^M$ constitutes the master equation of a certain description (local/global or autonomous/semi-classical) and compare the thermodynamic fluxes of the dynamical map approach to the other approaches.

\subsection{General form of the dynamical heat flux}

%\tb{\begin{itemize}
%    \item Calculate the general form for $Q^{d.m}$
 %   \item describe what it becomes in the autonomous global and local approaches and semi-classical limits.
  %  \item Write the energy flux in each of the regimes.
   % \item Compare to each of the definitions in various regimes
%\end{itemize}}

The heat flux of the dynamical map approach is described in terms of the dynamical generator and the instantaneous attractor. 
In order to derive  the dynamical heat flux, we first define a general Markovaian generator ${\cal{L}}_S^M$ and calculate the associated instantaneous attractor. Formally, we introduce the generalized invariant and non-invariant Lindblad operators $\{\hat{P}_i\}$ and $\{\hat{F}_{\alpha}\}$, respectively, which coincide with the suitable jump operators of the chosen description (in the Markovian limit). These eigenoperators allow expressing the general form of the primary system Markovian master equation
\begin{eqnarray}
    \begin{array}{ll}
    {\cal L}_S^M\b{t} \sb{\bullet}& = -i\sb{\bar{H}\b t,\bullet}\\& 
   +\sum_{\alpha=1}^{N\b{N-1}} G_{\alpha}\b{t} \left( \hat{
   F}_{\alpha} \bullet\hat{F}_{\alpha}^{\dagger} -\f{1}{2}\{\hat{F}_{\alpha}^{\dagger}\hat{F}_{\alpha},\bullet\}\right)\\&
    +\sum_{i,j=1}^{N-1}k_{ij}\b{t}\b{\hat{P}_{i}\bullet\hat{P}_{j}^{\dagger}-\f 12\{\hat{P}_{j}^{\dagger}\hat{P}_{i},\bullet\}}~~,
   \label{eq:79}
\end{array}
\end{eqnarray}
where $\{{G}_\alpha\}$ and $\{k_{ij}\}$ are generalized kinetic coefficients, which coincide with the kinetic coefficients of the chosen description. The strict Markovianity implies that they must be non-negative at all times. 

The instantaneous attractor effectively constitutes an invariant operator of both $ {\cal L}_S^M\b{t}$ and the free dynamics. This implies it is composed of the invariant operators of the free dynamics $\{\hat{P}_i\}$. As a consequence,
the first and third terms, which are also composed of the invariant eigenoperators, do not affect the form of the instantaneous attractor. This property allows us to focus only on the second term, denoted as ${\cal G}_S\b t$, which generates energy transitions within the primary system.
The crucial term constitutes a sum of local generators ${\cal {G}}_S\b t=\sum_{\alpha}'{\cal G}_{\alpha}\b t$, with
\begin{eqnarray}
\begin{array}{ll}
   {\cal G}_{\alpha}\b t\sb{\bullet} &=G_\alpha \b t\Big(\hat{F}_{\alpha}\bullet\hat{F}_{\alpha}^{\dagger}-\f 12\{\hat{F}_{\alpha}^{\dagger}\hat{F}_{\alpha},\bullet\}\Big)\\&+G_{-\alpha} \b t\Big(\hat{F}_{\alpha}^\dagger\bullet\hat{F}_{\alpha}-\f 12\{\hat{F}_{\alpha}\hat{F}_{\alpha}^\dagger,\bullet\}\Big)~~,
    \label{eq:floquet_local generator}
\end{array}
\end{eqnarray}
where $G_{-\alpha}$ designates the kinetic coefficient of the reverse  transition and the prime sum indicates a sum only over half of  the values of $\alpha$ (counting each pair of energy levels once). 

%and \tb{the kinetic coefficients satisfy a skewed detailed balance with respect to the effective frequencies $\{\alpha_k\}$, $G\b{-\alpha_k}=e^{-\hbar\alpha/k_B T}G\b{\alpha_k}$.}

Each local generator is associated with an independent interaction channel leading to  an energy transition between two distinct states. The ${\cal{G}}_{\alpha}$ generator induces an energy transfer of a possibly time-dependent quanta $\hbar{\omega_\alpha}$ to or from the reservoir at rates $G_\alpha$ and $G_{-\alpha}$, accordingly.
%In addition, the positively of the kinetic coefficients motivates expressing them in terms of a detailed balance style relation $G_\alpha =e^{-\Lambda_\alpha}G_{-\alpha}$, for some positive scalar $\Lambda_\alpha$. 
As a consequence, each local channel has an associated heat flux ${\cal J}_\alpha$ of the form of Eq. (\ref{eq:Q_dm}), defined by the system dynamical generator and the associated instantaneous attractor
\begin{equation}
    \b{\hat{\rho}_{S}}_{i.a}^\alpha\b t =Z^{-1}_\alpha\exp\b{-\b{\epsilon_\alpha\b t \hat{F}_\alpha^\dagger \hat{F}_\alpha+\epsilon_{-\alpha}\b t \hat F_\alpha \hat F_\alpha^\dagger}}
    ~~.
    \label{eq:80}
\end{equation}
Here $\epsilon_\alpha-\epsilon_{-\alpha}=\textrm{ln}\b{G_{\alpha}/G_{-\alpha}}$ and $Z_\alpha$ is the associated partition function. %\footnote{To show that $\b{\hat{\rho}_S}_{i.a}^\alpha$, Eq. \eqref{eq:80}, is invariant under the action of ${\cal G}_\alpha$, Eq. \eqref{eq:floquet_local generator}, we express $\hat{F}_\alpha$ in terms of the instantaneous eigenstates of time-evolution operators (similarly to Eq. \eqref{eq:F_alpha_sc}) and utilize the orthogonality condition.}.
$F_\alpha^\dagger F_\alpha$ and $F_\alpha F^\dagger_\alpha$ are  projection operators and therefore commute, therefore, the total instantaneous attractor can be written as a product of the channel attractors $\b{\hat{\rho}_S}_{i.a}=\Pi_\alpha' \b{\hat{\rho}_{S}}_{i.a}^\alpha$. Finally, by gathering together Eqs. (\ref{eq:Q_dm}),  (\ref{eq:floquet_local generator}) and (\ref{eq:80}) we obtain
\begin{equation}
    \dot{\cal{Q}}^{d.m}=\sum_\alpha{\cal J}_\alpha = k_{B}T\sum_{\alpha}G_{\alpha} \textrm{ln}\b{\f{G_{-\alpha}}{G_{\alpha}}}\mean{\hat F_{\alpha}^{\dagger}\hat F_{\alpha}}~~,
    \label{eq:85}
\end{equation}
where both the kinetic coefficient and eigeoperators may be time-dependent.

Finally, when the environment remains in a thermal state throughout the dynamics, the kinetic coefficients satisfy the detailed balance relation $G_{-\alpha} =e^{-\hbar \omega_\alpha/k_B T}G_{\alpha}$. This simplifies Eq. (\ref{eq:85}) to 
\begin{equation}
    \dot{\cal{Q}}^{d.m}=-\sum_{\alpha}\hbar \omega_\alpha G_{\alpha} \mean{\hat F_{\alpha}^{\dagger}\hat F_{\alpha}}~~.
    \label{eq:90_Qdm}
\end{equation}

\subsection{Comparison to alternative approaches}
The dynamical map definitions of heat flux and power serve as a general framework, which can be employed whenever the dynamics are Markovian. Therefore, the paradigm is applicable under various dynamical settings (global or local), and under autonomous or semi-classical descriptions. As a result, depending on the underlying open system dynamics, the dynamical map definitions are related to different thermodynamic setups. We next analyse the connections to the ensemble average based thermodynamic definitions of heat and work.

Under the local interaction model, the generalized Lindblad jump operators $\{\hat{F}_\alpha\}$ coincide with local operators $\{\hat{F}_\alpha^{\b L} \}$, the local kinetic coefficients $f_{\alpha \alpha}^{\b{L}}$ converge to Markovian coefficients $\{G_\alpha\}$, and $\textrm{ln}\b{G_{\alpha}/G_{-\alpha}}\ra\omega_\alpha^{\b L}$. Therefore, in this regime $\dot{\cal Q}^{d.m}$ is equivalent to $\dot{{\mathcal{Q}}}^{a\b L}$, Eq. (\ref{eq:comparison_heat}) (see also Table \ref{table:thermodfluxes} for a summary of the thermodynamic fluxes).
%The difference between the definitions involves a term proportional to the Lamb-shift. Since the energy fluxes of both approaches coincide, this correction is included within the power of the dynamical map approach. When the Lamb-shift is negligible and the dynamics are Markovian the two approaches become equivalent. 

Under a global dynamical description, an analogous identification  leads to $\dot{\mathcal{Q}}^{d.m}=-\sum_{\alpha}\hbar \omega_\alpha^{\b G} G_{\alpha} \mean{\hat G_{\alpha}^{\dagger \b{G}}\hat G_{\alpha}^{\b{G}}}$. This expression is identical to the first term of $\dot{{\mathcal Q}}^{a\b G}$ (Eq. (\ref{eq:59_Qa})) in the Markovian limit, with a correction term proportional to the primary-system control interaction $-\textrm{tr}\b{\b{\hat{H}_{SC}^{\b G}} {\cal{L}}_D^{\b{G}}\b t\sb{\hat \rho_D}}$. In addition, the energy fluxes of both approaches are identical, which implies that the energy accumulated in the interface between the primary-system and control is considered as useful work in the dynamical map definitions.

In the semi-classical limit, an equivalent correspondence is obtained for $\dot{\bm{\mathcal Q}}^{\b G}$, Eq. (\ref{eq:Q_sc_G_final}).
Finally, the external and dynamical map heat flux 
$\dot{\cal Q}^e$ (\ref{eq:Q_e_final}) are clearly identical in the semi-classical regime.

%This expression was first introduced in [31] and has be-
%come an established method by which to analyse period-
%ically driven and steady state autonomous systems [31–
%34].

\section{Summary: comparison between alternative definition of heat and work}
\label{sec:summary}

\begin{figure*}[ht]
\centering
\includegraphics[width=15 cm]{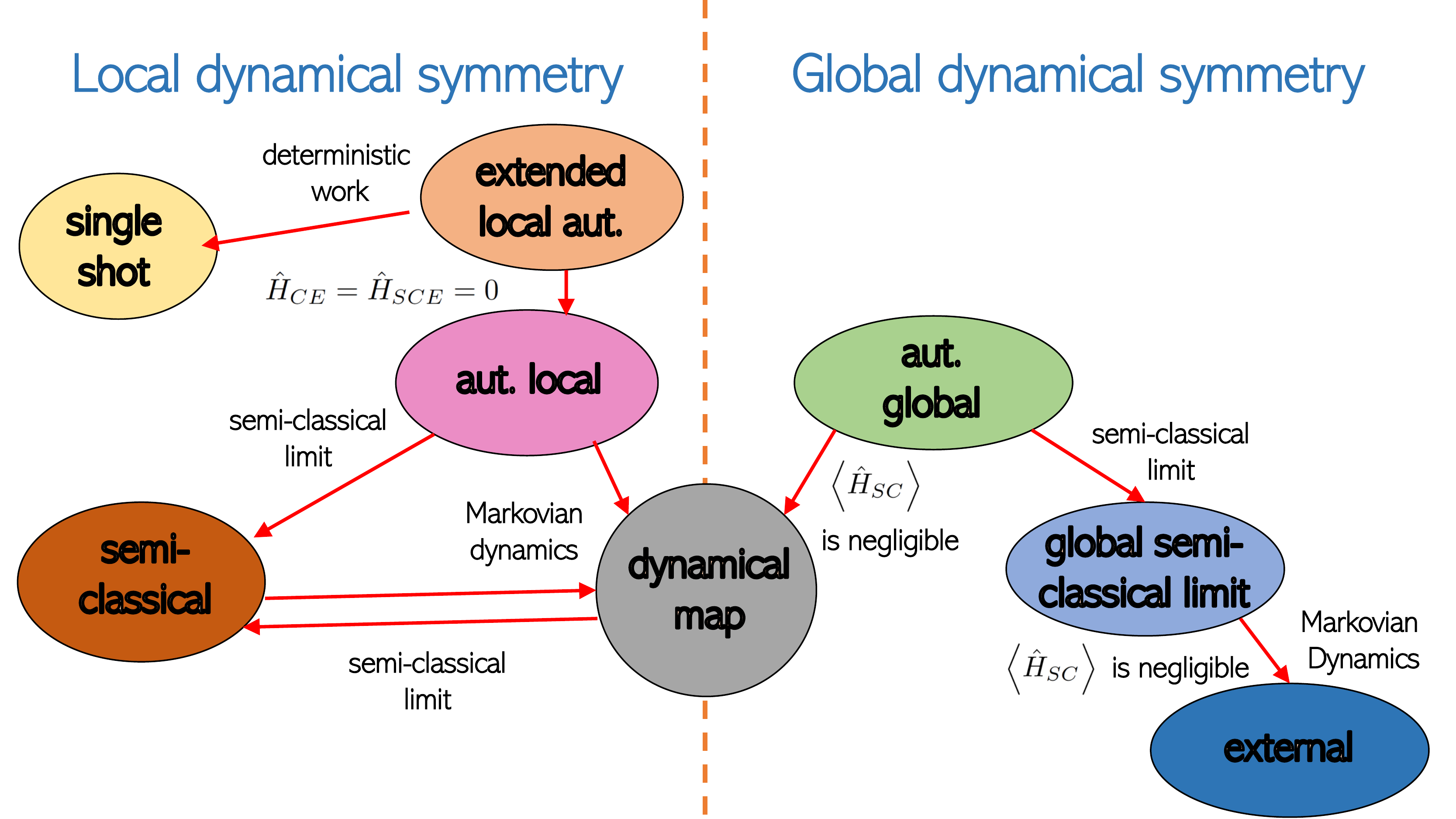}\\
\caption{Interrelations between different thermodynamic approaches. The different approaches can be categorized according to the underlying dynamical symmetry of the associated autonomous description, local or global. The building blocks of the various approaches are the autonomous approaches which describe the work reservoir or control as quantum systems. The connections between the different definitions presented by red arrows.}
\label{fig:interrelations}
\end{figure*}

Traditionally, thermodynamics identified work and heat in terms of local system observables. 
%Conservation of energy is then employed in order to associate heat to entropy change.
For practical purposes, it is desirable to have an analogous construction in the quantum theory. However, in quantum mechanics two system observables which do not commute are simultaneously ill defined. As a result, internal definitions of heat, work or energy change may be incompatible. The present solution to this problem is to identify (and measure) work in terms of  local work reservoir operators which are always compatible with the energy change of the primary system. In turn, when the sum of the subsystems' local energies is conserved, heat and work are then simultaneously well defined. Such a theoretical structure motivates the assumption of strict energy conservation on the interface.

The local nature of the thermodynamic definitions implies that work can only be extracted by means of local operations on the control. As a consequence, any energy stored in the interface (correlations) between the work reservoir and other subsystems is inaccessible, and therefore degrades the amount of extractable work. Such correlations are then accounted as heat, and are associated with an increase in the local subsystems entropy.
In the ideal limiting case the global correlations can be ignored and work coincides with the local energy change of the work reservoir.

These considerations lead to the autonomous thermodynamic definitions, where work, heat and the energy change are naturally identified in terms of energy changes of the associated quantum system Eqs. (\ref{eq:Work}) and (\ref{eq:heat_pre}). Moreover, strict energy conservation introduces a well defined partition between the subsystems, allowing to ignore the interface energy, and infers an associated dynamical symmetry.

\begin{figure}[htb!]
\centering
\includegraphics[width=8.4cm]{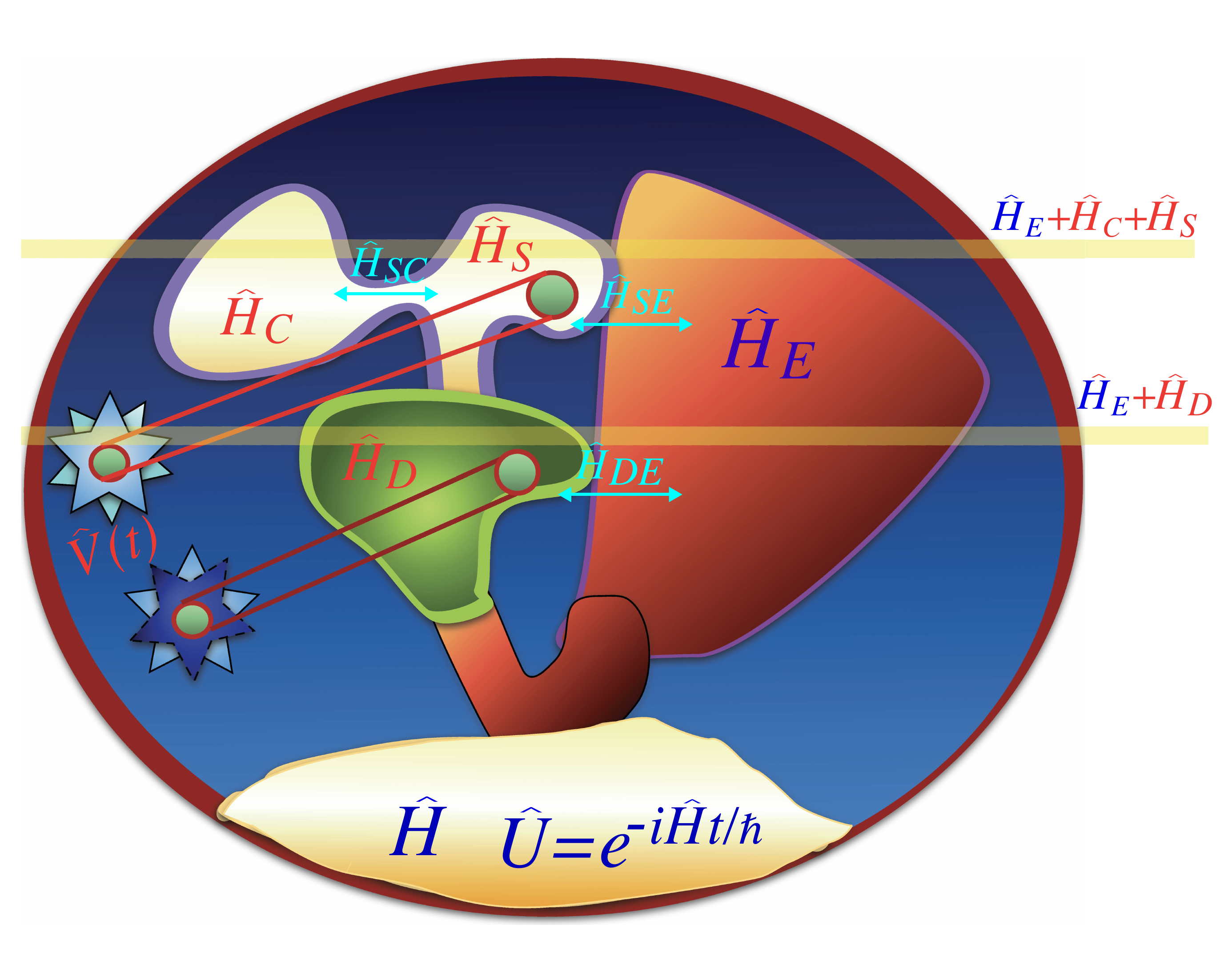}\\
\caption{ Schematic representation of energy conservation and possible partitions. The total Hamiltonian $\hat H$ is an conserved quantity, which implies that the total energy is a constant of motion. This property is represented by the surrounding (red) ellipsoid which confines the combined system.  It is then partitioned in to device $\hat H_D$ (green boundary) and environment $\hat H_E$ (in orange). Under the ``global'' strict energy conservation the sum:
$\hat H_D+\hat H_E$ is a conserved quantity.
A further partition to control and primary-system defines $\hat H_C+\hat H_S+\hat H_E$ as a conserved quantity under the ``local'' strict energy conservation (purple boundary). The control in the semi-classical limit can be expressed as an external time dependent drive $\hat{V}\b t =\sum_j \hat{S}_j c_j\b{t}$, indicated by cog-wheels. These are depicted as cog wheals.} 
\label{fig:blob}
\end{figure}

In the present study, we focused on two possible partitions, local and global, cf. Fig. \ref{fig:1-a}
and Fig. \ref{fig:blob}. The local setup considers a primary-system independently coupled to both the work reservoir (control) and environment. While the global autonomous approach includes a device system, encompassing the work reservoir and primary system, coupled to the environment. 
These two thermodynamic approaches serve as the building blocks of the popular thermodynamic paradigms that rely on a semi-classical description of the work-reservoir (or control). These are the semi-classical, external and dynamical map approaches. 
In addition, an extension of the local autonomous approach serves as the underlying dynamical description of the single shot framework of quantum thermodynamics. The different definitions are summarized in Table \ref{table:thermodfluxes}, the relations between them are presented graphically in Fig. \ref{fig:interrelations}.

The semi-classical approach defines power as the expectation value of the time-derivative of the total Hamiltonian. Heat is then obtained from the first law.  By studying the semi-classical limit of the local autonomous thermodynamic power a connection is obtained between the two approaches. 
Specifically, the local autonomous and semi-classical definitions %Eqs. \eqref{eq:Heat} and \eqref{eq:power_a_2}, semi-classical definitions, Eq. \eqref{eq:s_c_work} Eq. \eqref{eq:sc_E_S_35}, 
converge  under two conditions: The work reservoir evolves independently and the correlations between the primary system and control are negligible (Eqs. (\ref{eq:40corr}) and Eq. (\ref{eq:41E})). These conditions define the limiting case of the semi-classical regime, which occurs when the control resides in a highly excited ``classical" state.  The relation between the two powers highlights an inconsistency in the common identification of the ``system'' energy. Namely, to insure consistency of both thermodynamic approaches the energy of the primary system should only include the time-independent part of the system Hamiltonian $\hat{H}_S$. Excluding the time-dependent scalar terms, such as $\hat{S}_j c_j\b t$, which originate from the primary-system control interaction terms in the autonomous description. Under this identification the two approaches coincide in the semi-classical regime.

The connection between the local autonomous and semi-classical paradigms highlight two important points. Primarily, the work reservoir can be equivalently regarded as a quantum controller, inducing changes in the primary system state. Crucially, the choice of a  preferred interpretation depends on the specific physical scenario and does not affect the thermodynamic analysis. 
In addition, the connection also proves that the semi-classical definitions are valid for arbitrarily rapid driving, in contrast to what has been previously conjectured.

The dynamical map approach can be also related to the local interaction setup. The former approach is a thermodynamic ``template'' which is applicable to whenever the open system dynamics are Markovian (local or global, autonomous or semi-classical). It defines the heat flux in terms of the entropy production rate and the change of von-Neumann entropy, by relying on the well-known Spohn's inequality \cite{spohn1978entropy}. Since the inequality holds for any contracting (Markovian) map, the dynamical map approach  serves as a general structure which one can employ to analyse the thermodynamics.

When the open system dynamics obeys the local dynamical symmetry Eq. (\ref{eq:commu2}) and the dynamics are Markovian, the dynamical map and local interaction model thermodynamic definitions coincide (cf. Table \ref{table:thermodfluxes} and Eqs. (\ref{eq:40corr}) and \ref{eq:41E}). These relations carry over to the semi-classical regime, and therefore establish a relation between the thermodynamic definitions of the dynamical map and the semi-classical approach.

A similar correspondence exists between the semi-classical limit of the global autonomous approach and the external approach. The external approach identifies heat flux in terms of the energy current to the environment. It then employs a microscopic approximation scheme to express the thermodynamic fluxes in terms of the primary-system observables.  Comparing the two approaches, the powers coincide, Eq. (\ref{eq:75}), while the heat and energy fluxes vary by a term proportional to the primary-system control coupling, cf. Eqs. (\ref{eq:E^e_deriv}) and (\ref{eq:72}). The external approach includes the primary-system control interface energy within the system's energy. 
Similarly, the definitions of the dynamical map and external approaches differ by the same term. This discrepancy can be resolved by redefining what one considers as the primary system's energy Eq. (\ref{eq:E^e_1}). 

\trr{Finally, in the adiabatic limit, the global setting results in the standard definition of heat flux Eq. (\ref{eq:adiabatic_heat}) and entropy production rate (\ref{eq:adiabatic_ep}). While the local setting expressions deviate from the standard definitions due to a non-vanishing energy current from the control system through the system to the environment (see \ref{apsec:entropy_prod}). Importantly, in the global setting, the energy flux is associated only with the primary system. This definition differs from the standard approach which also includes the energy change in the control \cite{alicki1979quantum}.  }
%when the open system dynamics is generated by $\bm{\mathcal{L}}_S^{\b G}$, the dynamics are Markovian, and the environment remains in a thermal state throughout the evolution, the external and dynamical map approaches coincide. This highlights the underlying assumptions of  the dynamical map approach in this regime.  \tb{add to the Fig. 3 that the environment is in a thermal state in the arrow between the external and dynamical map}

%\begin{figure}[htb!]
%\centering
%\includegraphics[width=9cm]{}\\
%\caption{The generation of partitions between subsystem and their influence on definitions of heat and work.  }
%\label{fig:tree}
%\end{figure}

%  =-\textrm{tr}\b{\hat H_{C}\dot{\hat\rho}} =\f{i}{\hbar}\textrm{tr}\b{\hat H_{C}\sb{\hat H^{\b{L}},\hat \rho}}\\=\f{i}{\hbar}\Big\langle{\sb{\hat H_{C},\hat H_{SC}^{\b L}}}\Big\rangle =-\f{i}{\hbar}\Big\langle{\sb{\hat H_{S},\hat H_{SC}^{\b L}}}\Big\rangle_D.

%\section{Equivalence between approaches}
%We previously showed, that the autonomous and semi-classical thermodynamic approaches are equivalent in the semi-classical limit, Sec. \tb{\ref{}}. Next, we turn to focus and the various approaches, including a semi-classical description of the external driving. We show that the external and dynamical map approaches are identical under the customary assumptions that lead to the reduced dynamical description (\tb{stated in.....}). 

\section{Discussion}
\label{sec:discussion}
Quantum mechanics is considered as a comprehensive theory, incorporating the physics of the microscopic as well as macroscopic objects. Such a belief is corroborated by many experimental verifications on ever increasing physical systems \cite{martinis1987experimental,monroe1996schrodinger,brune1996observing,arndt1999wave,arute2019quantum,van2000quantum,friedman2000quantum}. As a comprehensive theory, it overlaps with the thermodynamics.  Therefore the well established thermodynamic principles should emerge from an underlying quantum description. The present study employs this philosophy (or paradigm ) to elucidate the nature of different notions of work and heat in the quantum regime.

Since the basic theory is quantum all notions of work and heat must emerge as limiting cases of a complete quantum description, which we denoted as the autonomous framework. 
Derivations of the known definitions from an underlying theoretical framework  illuminates the basic assumptions and idealizations.
Notably, certain idealizations in one description, e.g. semi-classical or classical, are only implicit in another. This may potentially lead to inconsistencies and controversy between accepted quantum thermodynamic approaches.

In the autonomous description we adopt the thermodynamic idealization of isothermal partition in order to construct a simple and basic theory. This idealization suggests strict conditions on the coupling terms between the subsystems of the autonomous description. In turn, such strict partition infer a dynamical symmetry of the open system. 

We emphasis that in the microscopic regime isothermal partition is not always satisfied, as the interface between microscopic systems does not generally satisfy strict energy conservation. In the microscopic world, this may lead to accumulation of energy on the boundary which is comparable with the subsystems energy. In such a case, the boundaries cannot be neglected and should be viewed as distinct thermodynamic constituents.  When work can be extracted only by means of local operations, the energy trapped in the interfaces between subsystems should be regarded as heat. 
In the present study we focused on the idealized case, a further extension to general interactions is the prospect of future studies. Crucially, such an extension of the theory first requires an exact form of the dynamical equations of motion  under arbitrary interactions. 

In the present work we considered two types of isopartitions or interaction models, and analysed the emerging thermodynamic relation. A recent work suggests that the distinction between the global and local autonomous interaction models is related to the dynamical timescale regime of interest \cite{winczewski2021bypassing}. Namely, the results of Winczewski et al. suggest that in the short timescale regime the dynamics are governed by a master equation emerging from the local interaction model, while for long times the dynamical equations converge to the master equation of the global interaction model. This result hints of the possibility of time-dependent thermodynamics. The observation is in line with the time-energy uncertainty relation. 

A natural extension of the current analysis is to consider other dynamical symmetries, conservation of the number of particles (excitations). This will lead to analogous equations of motion and thermodynamic relations. Such a framework will serve as a dynamical extension of  thermodynamics in the grand-canonical ensemble.

Beyond basic structure of the underlying theoretical model, a guiding principle asserts that the division of energy change between heat and work should be an element of reality. Meaning that it must be possible to confirm the notions of work and heat by experiment. It is therefore, important to define how they can be measured. Moreover, for a meaningful interpretations of the first law, the balance between heat and work should be inferrable simultaneously. This is possible if we can (even in principle) deduce work, heat and energy change by measurements on different subsystems. Such an approach is inline with Refs. \cite{skrzypczyk2014work,horodecki2013fundamental,beyer2020work,beyer2020work}. An alternative viewpoint deduces the work and energy change by measurements on the primary-system. Within this framework, there is a dispute between two possibilities: defining work as an observable \cite{allahverdyan2005fluctuations,silva2021quantum} or in terms of a two point measurement \cite{kurchan2000quantum,tasaki2000jarzynski,piechocinska2000information,talkner2007fluctuation,talkner2016aspects}. The difficulty in defining work, heat and energy in terms system observables arises from the fact that the associated operators may not commute, for example see Refs. \cite{allahverdyan2005fluctuations,silva2021quantum}. Therefore, they do not have a theoretical simultaneous existence \cite{bohr1937causality}. 
Alternatively, the two point measurement scheme is by nature evasive, eliminating the initial and final coherences. Therefore, the measurement procedure becomes inherent within the definition of work and heat. This excludes initial coherence from the thermodynamic description.  In addition, the inherent role of the measurement in this notion of work may require taking into account the energetic cost of measurement \cite{guryanova2020ideal}. Even an initial local energy measurement on the control is evasive, since it illuminates the local coherence which is the main source of transient dynamics. Within the local semi-classical framework and when system and bath are initially uncorrelated and diagonal in the free energy representation, the two-point measurement protocol coincides with the local semi-classical approach.

In comparison, the various approaches  considered and the hierarchy between them suggests an alternative experimental validation scheme. Initially, prepare the system and control in a known state %(or measure them in the eigenbasis of the corresponding density operator)
and measure the final energies of the control and primary-system. Assuming the initial preparation step is well-defined the energy partition can be inferred. By repeating the experimental protocol many time, the initial preparation step can be verified. An illustration of this type of inference process is exhibited in the preparation of a coherent field state in an optical cavity \cite{brune1996observing,guerlin2007progressive}. In this scheme, the initial control energy is inferred from the preparation process and not by measurement.

To conclude, the present work harnesses dynamical symmetry to establish a unified approach to the first law of thermodynamics. We found two relevant symmetries and extracted from them the corresponding definitions of heat and work.
As a result, two distinct partitions emerge which encompass the popular thermodynamic definitions. Such insight is essential for the modeling of open quantum systems and their thermodynamical analysis. The dynamical equation of the open system can be classified according to their symmetry class and the consistent thermodynamic definitions of heat and work can be identified.  
Overall, the extension of traditional thermodynamics to dynamical processes, underlines a deep relation between the thermodynamic principles and dynamical symmetries.

\begin{table*}[h]
\begin{center}
\small{
\begin{tabular} {|l|l|l|l|}
\hline
Description & Variable & Expression & Equation \\ 
\hline
\hline
Autonomous energy flux & $\dot{E}_S^{a\b L}$ & $\textrm{tr}\b{\hat{H}_{S}\dot{\hat \rho}}$ & (\ref{eq:dot_E_S_23})\\
\hline
Autonomous local power &   ${\cal P}^{a\b L}$   &$-\textrm{tr}\b{\hat H_{C}\dot{\hat\rho}}=\f{i}{\hbar}\Big\langle{\sb{\hat H_{C},\hat H_{SC}^{\b L}}}\Big\rangle$ & (\ref{eq:power_a_2}) \\ 
\hline
Autonomous local heat flux &${ \dot{\cal Q}}^{a\b L}$ &$-\sum_\alpha \hbar\omega_{\alpha}^{\b L}f_{\alpha\alpha}^{\b L}\mean{\hat{F}_{\alpha}^{\b{L}\dagger}\hat{F}_{\alpha}^{\b L}}$ & (\ref{eq:comparison_heat})\\
\hline
Autonomous energy flux & $ \dot{E}^{a\b G}_S$& $\textrm{tr}\b{\hat{H}_S {\cal{L}}_D^{\b{G}}\b t\sb{\hat \rho_D}}$ & (\ref{eq:global_power}) \\
\hline
Autonomous global power &   ${\cal P}^{a\b G}$   &  $-\textrm{tr}\b{\hat{H}_C {\cal{L}}_D^{\b{G}}\b t\sb{\hat \rho_D}}$ & (\ref{eq:global_power})\\ 
\hline
Autonomous global heat flux &${ \dot{\cal Q}}^{a\b G}$ & $\textrm{tr}\b{\b{\hat{H}_D^{\b G}-\hat{H}_{SC}^{\b G}} {\cal{L}}_D^{\b{G}}\b t\sb{\hat \rho_D}}$& (\ref{eq:Q^aG}), (\ref{eq:59_Qa}) \\
~& ~& $= -\sum_\alpha g_{\alpha\alpha}^{\b G}\hbar \omega_\alpha^{\b{B}}\mean{\hat{G}_\alpha^{\b G\dagger}\hat{G}_\alpha^{\b G}}$& \\  
& & $-\textrm{tr}\b{\b{\hat{H}_{SC}^{\b G}} {\cal{L}}_D^{\b{G}}\b t\sb{\hat \rho_D}}$&\\
\hline
\hline
Semi-classical energy flux (after modification)& $\dot{E}_S^{s.c}$ & $\tr\b{\hat{H}_S{\cal L}_S\b t\sb{\hat{\rho}_S}}$ & (\ref{eq:sc_E_S_35}) \\
\hline
Semi-classical power & ${\cal{P}}^{s.c}$ & $\bigg{\langle}{{{\partial\hat{H}^{s.c}\b{t}}/{\partial t}}}\bigg{\rangle}$ & (\ref{eq:s_c_power_op})  \\
\hline
Semi-classical heat flux & $\dot{\cal{Q}}^{s.c}$  &  $\dot{E}_S^{s.c}-{\cal{P}}^{s.c}$& ~\\
\hline 
\hline
External energy flux before modification & $\dot{E}_S^e$ & $\f{d}{dt}\tr \b{\hat{H}^{e}_S\b t \hat{\rho}_S^{ e}\b t}$ & (\ref{eq:E^e_1}) \\
\hline
External energy flux after modification & $\dot{E}_S^e$& $\textrm{tr}\b{\hat{H}_S {\cal{L}}_S^{\b{G}}\b t\sb{\hat \rho_S}}$   & (\ref{eq:E^e_S}) \\
\hline
External power & ${\cal P}^e$ & $\dot{E}_S^e-\dot{{\cal Q}}^{e}$ &~\\
\hline
External heat flux & $\dot{{\cal Q}}^{e}$ & $-\sum_{\alpha}\hbar{\omega_\alpha^{\b{G}}}G\b{\omega_\alpha^{\b{G}}}\mean{\hat{\bm F}_{\alpha}^{\dagger}\b t\hat{\bm F}_{\alpha}\b t}$ & (\ref{eq:Q_e_final}) \\
\hline
\hline
Dynamical map energy flux (after modification)  & $\dot{E}_S^{d.m}$ & $\tr\b{\hat{H}_S{\cal L}_S\sb{\hat{\rho}_S}}$& ~\\
\hline
Dynamical map power &  ${\cal P}^{d.m}$ & $\dot{E}_S^{d.m}-\dot{{\cal Q}}^{d.m}$ & ~\\
\hline
Dynamical map heat flux & $\dot{\cal{Q}}^{d.m}$& $- k_B T\textrm{tr}\b{{\cal{L}}_S^M\sb{\hat{\rho}_{S}}\textrm{ln} \b{\hat{\rho}_{S}}_{i.a}}$ & \\ ~& ~& $=k_{B}T\sum_{\alpha}G_{\alpha} \textrm{ln}\b{\f{G_{-\alpha}}{G_{\alpha}}}\mean{\hat F_{\alpha}^{\dagger}\hat F_{\alpha}}$ & (\ref{eq:90_Qdm}) \\
\hline
\end{tabular}
}
\end{center}
    \caption{\normalsize{Summary of the various definitions of thermodynamic energy fluxes}}
    \label{table:thermodfluxes}
\end{table*}

\begin{table*}[h]
\begin{center}
\small{
\begin{tabular} {|l|l|}
\hline
Variable & Symbol \\ 
\hline
\hline
Total state & $\hat{\rho}$  \\ 
\hline
Subsystem states & $\hat{\rho}_A$ with $A=E,D,C,S$ \\
\hline
Total Hamiltonian & $\hat{H}^{\b B}$ with $B=L,G$   \\ 
\hline
Subsystem free Hamiltonians & $\hat{H}_A$ with $A=E,D,C,S$ \\
\hline
Total time-evolution operator & $\hat{U}$  \\ 
\hline
Subsystem (isolated) time-evolution operator & $\hat{U}_A$ with $A=E,D,C,S$ \\ 
\hline
von-Neumann/energy entropy & ${\cal S}_{VN}/{\cal S}_{E}$ \\
\hline
Autonomous approach local thermodynamic fluxes & $\dot{E}_S^{a\b{L}}, {\cal P}^{a\b L}, \dot{\mathcal{Q}}^{a\b{L}}$\\
\hline
Autonomous approach global thermodynamic fluxes & $\dot{E}_S^{a\b{G}}, {\cal P}^{a\b G}, \dot{\mathcal{Q}}^{a\b{G}}$\\
\hline
Semi-classical limit of  autonomous local/global variables & $\dot{\bm E}_S^{a\b{B}}, \bm{\mathcal P}^{a\b B}, \dot{\bm{\mathcal{Q}}}^{a\b{B}}$\\ ~&  with $B=L,G$\\
\hline
Semi-classical approach thermodynamic fluxes & $\dot{E}_S^{s.c}, {\cal P}^{s.c}, \dot{\mathcal{Q}}^{s.c}$
\\
\hline
External approach thermodynamic fluxes & $\dot{E}_S^{e}, {\cal P}^{e}, \dot{\mathcal{Q}}^{e}$
\\
\hline
Dynamical map approach thermodynamic fluxes & $\dot{E}_S^{d.m}, {\cal P}^{d.m}, \dot{\mathcal{Q}}^{d.m}$
\\
\hline
Autonomous non-invariant&\\ eigenoperators of the primary system & $\hat{F}_\alpha^{\b B}$ with $B=L,G$ 
\\
\hline
Autonomous invariant eigenoperators  of the primary system & $\hat{P}_i^{\b B}$ with $B=L,G$ 
\\
\hline
Dynamical invariant of the semi-classical & \\limit of the global autonomous approach &  $\hat{X}^{s.c}\b t$ (\ref{eq:X^s.c})
\\
\hline
Semi-classical limit of global non-invariant eigenoperators & $\hat{\bm F}_\alpha^{\b G}$
\\
\hline
Semi-classical limit of global invariant eigenoperators & $\hat{\bm P}_j^{\b G}$ 
\\
\hline
Eigenstates and eigenvalues of $\hat{X}^{s.c}\b t$ & $\ket{\varphi_j}$, $\epsilon_j$ 
\\
\hline
Operator in the interaction picture relative to $\hat{H}_C$  & $\tilde{X}$ 
\\
\hline
Operator in the interaction picture relative to $\hat{H}_S+\hat{H}_C$  & $\widetilde{X}$  
\\
\hline 
Primary system  energy eigenstates ($\hat{H}_S$) & $\ket{n}$ 
\\
\hline
Device energy eigenstates ($\hat{H}_D$) & $\ket{\phi_i}$
\\
\hline
Eigenstates of the semi-classical evolution & \\ operator $\hat{\bm{U}}_S\b{t,0}$ & $\ket{\psi_j}$
\\
\hline
Thermal operations dynamical map/& \\ time-evolution operator/ & \\ total Hamiltonian  & $\hat{\Lambda}^{\b{TO}}/\hat{U}^{\b{TO}}/\hat{H}^{\b{TO}}$ \\
\hline
\end{tabular}}
\end{center}
    \caption{\normalsize{Notation table: The capital letters correspond to $E$-environment, $D$-device, $C$-control, $S$-primary system, $L$-local interaction model, $G$-global interaction model.}}
    \label{table:notations}
\end{table*}

\clearpage

 \ack
We thank Cyril Elouard and Juliette Monsel for an insightful correspondence. This research was supported by the Adams Fellowship  Program of the Israel Academy of Sciences and Humanities and The Israel Science Foundation Grant No.  2244/14.

\appendix
\section{derivation of the autonomous local interaction master equation}
 \label{apsec:appendix_A}

The local interaction commutation relations, Eq. (\ref{eq:SEC_local}), infer the conservation of the interaction energy
 \begin{equation}
     \sb{\hat{H}^{\b{L}},\hat{H}_S+\hat{H}_C+\hat{H}_E}=0~~.
     \label{eqap:commu2}
 \end{equation}
This relation along with the initial stationary state of the environment imply a crucial dynamical symmetry. Under these postulates, 
the device's dynamical map $\Lambda^{\b{L}}_{D}\b t$ commutes with the free propagator of the total system ${\cal{U}}_0$ (lacking interactions between the subsystems, cf. \ref{apsec:cummut}).
\begin{equation}
    {\cal{U}}_0\sb{\Lambda^{\b{L}}_{D}\sb{\hat{\rho}_{D}}}={\Lambda^{\b{L}}_{D}\sb{{\cal{U}}_0\sb{\hat{\rho}_{D}}}}~~,
    \label{eq:maps_commute}
\end{equation}
or alternatively ${\cal{U}}_0\circ \Lambda_D^{\b{L}}=\Lambda_D^{\b{L}}\circ{\cal{U}}_0$.
Here,  ${\cal U}_{0}\sb{\hat{\rho}_{D}\b 0}=\hat{U}_{0}\b{t,0}\hat{\rho}_{D}\b 0\hat{U}_{0}^{\dagger}\b{t,0}$, with  $\hat{U}_0\b{t,0}=e^{-i\b{\hat{H}_S+\hat{H}_C+\hat{H}_E}t/\hbar}$. 
and the dynamical map is given by (Eq. (\ref{eq:13_lambda_D}))
\begin{equation}
\Lambda^{\b{L}}_{D}\b t\sb{\hat{\rho}_{D}\b{0}}=\textrm{tr}_{E}\b{\hat{U}^{\b{L}}\b{t,0}\hat{\rho}\b 0{\hat{U}^{\b{L}}\b{t,0}}}~~,
\end{equation}
where $\hat{U}^{\b{L}}\b{t,0}=e^{-i \hat{H}^{\b{L}} t/\hbar}$, and $\hat{H}^{\b{L}}$ is defined in Eq. (\ref{eq:local_Ham}).
Relation (\ref{eq:maps_commute}) manifests the so-called time-translation symmetry \cite{marvian2014extending}.

The time-translation symmetry allows employing a spectral analysis, which infers that the dynamics of the device energy populations and coherences (with respect to $\hat{H}_D$) are decoupled.

In the Heisenberg picture, this symmetry is manifested by the decoupling of the dynamics of the device's non-invariant $\{\hat{G}_\kappa^{\b{L}}\}$ and invariant $\{\hat{R}_j^{\b{L}}\}$ eigenoperators. These eigenoperator obey the eigenvalue-type relation ${\cal{U}}_0\sb{\hat{G}_\kappa^{\b{L}}}=\lambda_\kappa \hat{G}_\kappa^{\b{L}}$  and ${\cal{U}}_0\sb{\hat{R}_\kappa^{\b{L}}}= \hat{R}_\kappa$, where $\lambda_k\in\mathbb{C}$ and generally differs from unity.
Since the device Hamiltonian has no explicit time-dependence,  $\{\hat{G}_{\kappa}=\ket{\phi_n}\bra{\phi_m}\}$ constitute transition operators between the energy states of $\hat{H}_D^{\b L}$ ($\{\ket{\phi_n}\}$ are the device eigenstates). The invariant operator subspace of ${\cal U}_0$, is spanned by the device's invariant eigenoperators, these are generally linear combinations of the device's energy projection operators ${\ket{\phi_n}\bra{\phi_n}}$.

%In the Heisenberg picture, this symmetry is manifested by a decoupling of the non-invariant eigenoperators $\{\hat{F}_{\alpha}\}$, with $\alpha=1,\dots,N\b{N-1}$ and the invariant eigenoperators  
%\footnote{An operator $\hat{S}$ is an eigenoperator with respect to a map ${\cal{V}}\sb{\bullet}$ if it obeys the eigenvalue-type relation ${\cal{V}}\sb{\hat{S}}=\lam \hat{S}$, where $\lam\in\mathbb{C}$. Invariant eigenoperators or have unity eigenvalues $\lam=1$, while for non-invariant eigenoperators $\lam\neq 1$}.
%For a time-independent system Hamiltonian,  $\{\hat{F}_{\alpha}=\ket{n}\bra{m}\}$ constitute transition operators between the energy states of $\hat{H}_S$. Therefore, $\alpha$ implicitly designates a double index $nm$, referring to the two specific energy levels $\ket{n}$ and $\ket{m}$. The invariant operator subspace of ${\cal U}_0$, is spanned by the invariant eigenoperators, which can be chosen as the energy projection operators $\{\hat{\Pi}_i\}$.

When the spectrum of ${\cal U}_0$ is non-degenerate, the associated non-invariant eigenoperators $\{\hat{G}_{\kappa}\}$ also evolve independently. This condition along with the Hermitiacy and trace preserving properties impose strict constrains on the linear form of the dynamical generator. Following a similar analysis as presented in Ref. \cite{dann2021open}, the outcome of these constraints is a  dynamical symmetric structure of the following form (Eq. (\ref{eq:local_master_eq_device}) in the main text)
\begin{eqnarray}
    \begin{array}{ll}
    {\cal L}^{\b{L}}_D\b{t} \sb{\bullet} &= -\f{i}{\hbar}\sb{\bar{H}^{\b{L}}_D\b{t},\bullet}\\&
   +\sum_{\kappa=1}^{N_{D}\b{N_{D}-1}} g_{\kappa\kappa}\b{t} \left( \hat{G}_{\kappa}^{\b{L}} \bullet\hat{G}_{\kappa}^{\b L\dagger} -\f{1}{2}\{\hat{G}_{\kappa}^{\b L\dagger}\hat{G}_{\kappa}^{\b{L}},\bullet\}\right)\\&
    +\sum_{i,j=1}^{N_D}r_{ij}\b{t}\b{\hat{R}_{i}^{\b{L}}\bullet\hat{R}_{j}^{\b L\dagger}-\f 12\{\hat{R}_{j}^{\b L\dagger}\hat{R}_{i}^{\b{L}},\bullet\}}~~,
   \label{eqap:local_master_eq_device}
\end{array}
\end{eqnarray}
where $\bar{H}^{\b{L}}_D\b{t}=\f 1{2i}\b{\hat{R}^{\dagger}\b{t}-\hat{R}\b{t}}$ is a Hermitian operator, with $\hat{R}\b{t}=\b{\f {1}{N_D}}^{1/2}\sum_{i=1}^{N_{D}-1}r_{iN_D}\b{t}\hat{R}_{i}$ and $N_D=\textrm{dim}\b{H_D^{\b{L}}}=N_S\cdot N_C$. Here, the set $\{\hat{R}_i\}$ constitutes an operator basis for the invariant subspace, satisfying $\hat{R}_{N_D}=\hat{I}/\sqrt{N_D}$, while the rest of the operators are traceless. 
A possible choice for such a set is the diagonal matrices of the $SU\b{N}$ generalized Gell-Mann matrices $\hat{R}_j=\sqrt{\f{2}{j\b{j+1}}}\b{\sum_{k=1}^{j}\ket{\phi_k}\bra{\phi_k}-j\ket{\phi_{j+1}}\bra{\phi_{j+1}}}$, for $j=1,\dots,N_D-1$ \cite{bertlmann2008bloch}. Crucially,
the kinetic coefficients $g_{\kappa \kappa}\b{t}$ must be real and possibly negative, while $r_{ij}\b{t}$ are generally complex and satisfy $r_{ij}\b{t}=r_{ji}^*\b{t}$.

From the thermodynamic point of view it is natural to decompose the local master equation, Eq. (\ref{eq:local_master_eq}), into three significant terms: (i) Isolated unitary dynamics, associated with the device bare Hamiltonian $\hat{H}_D^{\b L}$ (ii) The environment's unitary contribution to the reduced dynamics (a Lamb-shift type term), related to the Hermitian operator $\hat{H}_{LS}$ (iii) Dissipative term, generated by the superoperator  ${\cal D}^{\b{L}}_D$. This decomposition reads (Eq. (\ref{eq:local_me_thermo_device}))
\begin{equation}
    {\cal L}^{\b{L}}_D\b t\sb{\bullet} = -\f{i}{\hbar}\sb{\hat{H}_D^{\b L}+\hat{H}_{D,LS}^{\b L}\b t,\bullet}+{\cal D}^{\b L}_D\b t\sb{\bullet}~~,
    \label{eqap:local_me_thermo_device}
\end{equation}
where the Lamb-shift term may be time-dependent $\hat{H}_{D,LS}^{\b L}\b t=\bar{H}^{\b{L}}_D\b{t}-\hat{H}_D^{\b L}$. The definition of $\hat{H}_{D,LS}^{\b L}\b t$ together Eqs. (\ref{eqap:local_master_eq_device}) and (\ref{eqap:local_me_thermo_device}) define the three significant terms. 
The decomposition simplifies the thermodynamic analysis, as it compresses the complex form Eq. (\ref{eq:local_master_eq_device}) into the basic three thermodynamic ingredients.

\section{Commutivity of maps in the local interaction model}
\label{apsec:cummut}

We set out to prove the following claim:\\
Let $$\hat{H}^{\b{L}} =\hat{H}_S +\hat{H}_C+\hat{H}_{SC}^{\b{L}}+\hat{H}_E+\hat{H}_{SE}^{\b L}~~,$$ be the time-independent Hamiltonian of the composite system, satisfying the relations
\begin{equation} 
    \sb{\hat{H}_S+\hat{H}_{C},\hat{H}_{SC}^{\b L}}=0~~;~~\sb{\hat{H}_S+\hat{H}_{E},\hat{H}_{SE}^{\b L}}=0~~,
    \label{eqap:local_comm}
\end{equation}
and let the initial state $\hat{\rho}_E\b 0$ be a stationary state of  $\hat{H}_E$.
 Then  the dynamical maps $\Lambda^{\b{L}}\sb{\hat{\rho}_S\b{0}}=\textrm{tr}_{E,C}\b{\hat{U}^{\b{L}}\b{t,0}\hat{\rho}\b 0{\hat{U}^{\b{L}}\b{t,0}}}$ and  ${\cal U}_{0}\sb{\hat{\rho}_S\b 0}=\hat{U}_{0}\b{t,0}\hat{\rho}_{S}\b 0\hat{U}_{0}^{\dagger}\b{t,0}$ commute:
 \begin{equation}
    {\cal{U}}_0\sb{\Lambda^{\b{L}}\sb{\hat{\rho}_S}}={\Lambda^{\b{L}}\sb{{\cal{U}}_0\sb{\hat{\rho}_S}}}~~,
    \label{eq:maps_commute_2}
\end{equation}
 where $\hat{U}^{\b{L}}\b{t,0}=e^{-i \hat{H}^{\b{L}} t/\hbar}$ and $\hat{U}_0\b{t,0}=e^{-i\b{\hat{H}_S+\hat{H}_C+\hat{H}_E}t/\hbar}$.
 
 \paragraph*{Proof}  First, we express the local strict energy conservation conditions, Eq. (\ref{eqap:local_comm}), in an alternative form 
 \begin{equation}
     \sb{\hat{H}^{\b{L}},\hat{H}_{SC}^{\b 
     L}+\hat{H}_{SE}^{\b L}}=0\\
     \sb{\hat{H}^{\b{L}},\hat{H}_S+\hat{H}_C+\hat{H}_E}=0~~.
     \label{apeq:commu2}
 \end{equation}
 These will serve us in the proof. Working in the Heisenberg picture, the open system CPTP dynamical map can be expressed in terms of a Kraus decomposition \cite{kraus1971general}. This is achieved by expressing the initial environment state in terms of the eigenstates $\hat{\rho}_E \b 0 =\sum_i\lambda_i\ket{\chi_i}\bra{\chi_i}$. Since the $\hat{\rho}_E\b 0$ is stationary with respect to $\hat{H}_E$, $\{\ket{\chi_i}\}$ can be chosen as energy eigenstates and $\{\lambda_i\}$ are then identified as the environment energy population. Utilizing this form the dynamical map can be expressed as  
 \begin{equation}
     \Lambda^{\ddagger\b{L}}\sb{\hat{O}_{D}} = \sum_{ij} \hat{K}_{ij}^{\dagger}\hat{O}_{D}\hat{K}_{ij}~~,
 \end{equation} 
 where $\hat{K}_{ij}=\sqrt{\lambda_{i}}\bra{\chi_{j}}\hat{U}^{\b L}\ket{\chi_{i}}$ are the called the Kraus operators, and $\hat{O}_{D}$ is an arbitrary device operator (composite system of the primary system and control). The Kraus decomposition allows simplifying the product of dynamical maps 
 \begin{eqnarray}
     \begin{array}{ll}
     {\mathcal U}_{0}^{\ddagger}\sb{\Lambda^{\ddagger\b{L}}\sb{\hat{O}_{D}}}&=\hat U_{0}^{\dagger}\b{\sum_{ij}\hat K_{ij}^{\dagger}\hat O_{D}\hat K_{ij}}\hat U_{0}\\&=\hat U_{0}^{\dagger}\b{\sum_{ij}\lambda_{i}\bra{\chi_{i}}{\hat U^{\dagger\b{L}}\ket{\chi_{j}}\hat O_{D}\bra{\chi_{j}}\hat U^{\b{L}}\ket{\chi_{i}}}}\hat U_{0}\\&=\hat U_{0}^{\dagger}\b{\sum_{j}\lambda_{i}\bra{\chi_{i}}\hat U^{\dagger\b L}\hat O_{D}\hat U^{\b{L}}\ket{\chi_{i}}}\hat U_{0}~~.
     \label{apeq:U_prod_1}
 \end{array}
 \end{eqnarray}
 
The free propagator is a product of three local propagators
\begin{equation}
    \hat{U}_0 = \hat{U}_S\hat{U}_C\hat{U}_E~~.
\end{equation}
Substituting this relation into Eq. (\ref{apeq:U_prod_1}) and utilizing the eigenvalue relation $\hat{U}_E \ket{\chi_i}=e^{-i\nu_i t}$ we get
\begin{eqnarray}
    \begin{array}{ll}
    {\mathcal U}_{0}^{\ddagger}\sb{\Lambda^{\ddagger\b{L}}\sb{\hat O_{D}}}
    &= \hat U_{E}^{\dagger}\sum_{i}\lambda_{i}\bra{\chi_{i}}\hat U_{E}\hat U_{C}^{\dagger}\hat U_{S}^{\dagger}\hat U^{\dagger\b L}\hat O_{D}\hat U^{\b{L}}\hat U_{S}\hat U_{C}\hat U_{E}\ket{\chi_{i}}\hat U_{E}^\dagger\\&
    =\sum_{i}\lambda_{i}\bra{\chi_{i}}\hat U_{0}^{\dagger}\hat U^{\dagger\b L}\hat O_{D}\hat U^{\b{L}}\hat U_{0}\ket{\chi_{i}}\otimes\hat{I}_E~~.
\end{array}
\end{eqnarray}
where $\hat{I}_E$ is the identity operator of the environment's Hilbert space.
Equation (\ref{apeq:commu2}) now implies that $\sb{\hat{U}^{\b{L}},\hat{U}_0} = 0$, which leads to
\begin{equation}
     {\cal U}_{0}^{\ddagger}\sb{\Lambda^{\ddagger\b{L}}\sb{\hat O_{D}}} =\sum_{i}\lambda_{i}\bra{\chi_{i}}\hat U^{\dagger\b L}\hat U_{0}^{\dagger}\hat O_{D}\hat U_{0}\hat U^{\b{L}}\ket{\chi_{i}}\otimes\hat{I}_E~~.
     \label{apeq:first_prod_final}
\end{equation}

Following a similar procedure the reversed order product can be expressed as 
\begin{eqnarray}
\begin{array}{ll}
   \Lambda^{\ddagger\b{L}}\sb{{\cal U}_{0}^{\ddagger}\sb{\hat O_{D}}}&=\sum_{ij}\hat K_{ij}^{\dagger}\hat U_{0}^{\dagger}\hat O_{D}\hat U_{0}\hat K_{ij}\\&
=\sum_{ij}\lambda_{i}\bra{\chi_{i}}{\hat U^{\dagger\b L}\ket{\chi_{j}}}\hat U_{0}^{\dagger}\hat O_{D}\hat U_{0}\bra{\chi_{j}}{\hat U^{\b{L}}\ket{\chi_{i}}}\\&=\sum_{ij}\lambda_{i}\bra{\chi_{i}}\hat U^{\dagger\b L}\hat U_{S}^{\dagger}\hat U_{C}^{\dagger}\ket{\chi_{j}}\hat U_{E}^{\dagger}\hat O_{D}\hat U_{E}\bra{\chi_{j}}\hat U_{C}\hat U_{S}\hat U^{\b{L}}\ket{\chi_{i}}
\end{array}
\end{eqnarray}

since $\sb{\hat{U}_{E},\hat{O}_{D}}=0$
\begin{equation}
    =\sum_{i}\lambda_{i}\bra{\chi_{i}}\hat U^{\dagger\b L}\hat U_{S}^{\dagger}U_{C}^{\dagger}\hat O_{D}\hat U_{C}\hat U_{S}\hat U^{\b{L}}\ket{\chi_{i}}\otimes \hat{I}_E~~.\\
\end{equation}
Inserting the identity $\hat U_{E}^{\dagger}\hat U_{E}=\hat I_{E}$ within the brakets and utilizing the commutation of environment operators with device operators, we obtain
\begin{eqnarray}
    \begin{array}{ll}
   \Lambda^{\ddagger\b{L}}\sb{{\cal U}_{0}^{\ddagger}\sb{\hat O_{D}}}&
    =\sum_{i}\lambda_{i}\bra{\chi_{i}}\hat U^{\dagger\b L}\hat U_{S}^{\dagger}\hat U_{C}^{\dagger}\hat U_{E}^{\dagger}\hat O_{D}\hat U_{E}\hat U_{C}\hat U_{S}\hat U^{\b{L}}\ket{\chi_{i}}\otimes \hat I_E\\&=\sum_{i}\lambda_{i}\bra{\chi_{i}}\hat U^{\dagger\b L}\hat U_{0}^{\dagger}\hat O_{D}\hat U_{0}\hat U^{\b{L}}\ket{\chi_{i}}\otimes \hat I_E ~~.
    \end{array}
\end{eqnarray}
Comparing this result to Eq. (\ref{apeq:first_prod_final}) completes the proof 
\begin{equation}
    {\mathcal U}_{0}^{\ddagger}\sb{\Lambda^{\ddagger\b{L}}\sb{\hat O_{D}}}=\Lambda^{\ddagger\b{L}}\sb{{\cal U}_{0}^{\ddagger}\sb{\hat O_{D}}}~~_\blacksquare
\end{equation}
In the Schr\"odinger picture the relation becomes 
\begin{equation}
    {\cal U}_{0}\sb{\Lambda^{\b{L}}\sb{\hat{\rho}_{D}}}=\Lambda^{\b{L}}\sb{{\cal U}_{0}\sb{\hat{\rho}_{D}}}~~.
\end{equation}

\section{Semi-classical work}
\label{apsec:s_c_work}
The semi-classical composite system is represented by $\hat{H}^{s.c}\b t$, Eq. (\ref{eq:s_c_Ham}).
%, where the system-environment interaction term is given in Eq. \eqref{eq:local_SE_interaction} with out loss of generality.
By transforming to Heisenberg representation and utilizing the linearity of both the trace and integration operations, the semi-classical work is expressed as 
\begin{eqnarray}
\begin{array}{ll}    {\mathcal W}^{s.c} &= \textrm{tr}\b{\hat{H}^{s.c}\b{t_f}\hat{\rho}\b{t_f}-\hat{H}^{s.c}\b{t_i}\hat{\rho}\b{t_i}} \\&= \textrm{tr}\b{\sb{\int_{t_i}^{t_f}\pd{\hat{H}^{s.c,H}\b t}{t}dt}\hat{\rho}\b 0 }\\&= \int_{t_i}^{t_f} \textrm{tr}\b{\pd{\hat{H}^{s.c}\b t}{t}\hat{\rho}\b t}\,dt~~.
    \label{eq:delta_E_sc}
\end{array}
    \end{eqnarray}
Here, $\hat{\rho}\b t$ is the composite density operator and $\hat{H}^{s.c,H}\b t$ is the composite Hamiltonian in the Heisenberg representation.

The semi-classical procedure identifies the power operator in terms of the rate of change of the Hamiltonian
\begin{equation}
    \hat{P}^{s.c}\b t=\pd{\hat{H}^{s.c}\b{t}}{t}= \pd{\hat{H}_{S}^{s.c}\b{t}}{t}\otimes \hat{I}_E~~,
    \label{eq:s_c_power_op_ap}
\end{equation} 
where
\begin{equation}
 \hat{H}_{S}^{s.c}\b t=\hat{H}_S +\sum_j\hat{S}_j c_j\b t~~.
 \label{eq:WM_control_SC_ham}
\end{equation}
We next partition the composite state (similarly as in Eq. (\ref{eq:partition_rho}))
to a term including the system-environment correlations, and a separable term: 
\begin{equation}
    \hat{\rho}\b t = \hat{\rho}_{S}\b t\otimes\hat{\rho}_E\b t+\hat{\chi}\b t
    \label{eqap:rho}~~,
\end{equation}
where $\hat{\rho}_{S}\b t=\textrm{tr}_E\b{\hat{\rho}\b t}$ and $\hat{\rho}_E\b t=\textrm{tr}_S\b{\hat{\rho}\b t}$ are the reduced system and environment density operators. Substituting these relations into the average change of internal energy, Eq. (\ref{eq:delta_E_sc}), leads to the {\emph{semi-classical work}}
\begin{eqnarray}
\begin{array}{ll}
{\mathcal W}^{s.c}&= \int_{t_i}^{t_f}\textrm{tr}_S\b{\pd{\hat{H}_{S}^{s.c}\b{t}}{t}\hat{\rho}_S\b{t}}dt\\&+\int_{t_i}^{t_f}\textrm{tr}\b{\pd{\hat{H}^{s.c}\b{t}}{t}\hat{\chi}\b{t}}dt~~.
\end{array}
\end{eqnarray}
The first term on the r.h.s. is the integrated semi-classical power, while the second term contains the accompanied correction that stems from global correlations between the system and environment.
The second term vanishes under the dynamics of $\hat{H}^{s.c}\b t$, for arbitrary driving (see derivation below), which implies that the semi-classical work
coincides with the integrated semi-classical power \cite{campisi2009fluctuation}:
\begin{equation}
    {\cal{W}}^{s.c}=\int_{t_i}^{t_f}{\cal{P}}^{s.c}\b t\,dt
    \label{eq:s_c_work_ap}~~,
\end{equation}
with ${\cal{P}}^{s.c}\b t\equiv \Big{\langle}{{\hat{P}^{s.c}\b t}}\Big{\rangle}_S$.

\subsection*
{Calculation of the semi-classical correlation term}
We calculate the correction term of ${\cal{W}}^{s.c}$, Eq. (\ref{eq:delta_E_sc}), and show it vanishes for arbitrary driving and total state $\hat{\rho}$. 
The correction term is obtained by integrating over
\begin{equation}
{\cal C}\b t=\textrm{tr}\b{\pd{\hat{H}^{s.c}\b{t}}{t}\hat{\chi}\b{t}}~~,   
\label{eqap:C}
\end{equation}
%where $\hat{\chi}_{SW}$ is defined in terms of the composite state, and reduced density operators $\hat{\rho}_{SW}\b t=\textrm{tr}_R\b{\hat{\rho}\b t}$ and $\hat{\rho}_R\b t = \textrm{tr}_S\b{\hat{\rho}\b t}$,
%\begin{equation}
%\hat{\rho}\b t=\hat{\rho}_{SW}\b t\otimes \hat{\rho}_R\b t+ \hat{\chi}_{SW}~~.    
%\label{eqap:rho}
%\end{equation}
Performing partial traces over Eq. (\ref{eqap:rho}), leads to
\begin{equation}
 \textrm{tr}_E\b{\hat{\chi}}=0\,\,\,\, \textrm{and} \,\,\,\,\, \textrm{tr}_S\b{\hat{\chi}}=0~~.
 \label{eq:partial trace}
\end{equation}
%where $\hat{0}_{SW}$ and $\hat{0}_R$ are the zero operators of the system (working medium and control) and reservoir. 

Let $\{{\ket{s_i}}\}$ and $\{{\ket{r_j}}\}$ be orthonormal bases of the system and environment Hilbert spaces ${\cal H}_{S}$ and ${\cal H}_E$. Then the set $\{\ket{s_i }\otimes\ket{r_j} \}$ forms a basis of the composite Hilbert space ${\cal H}={\cal H}_{S}\otimes{\cal H}_E$. Since,  only the system is explicitly time-dependent, the total power operators is separable $\pd{\hat{H}^{s.c}}{t}=\pd{\hat{H}_{S}^{s.c}}{t}\otimes \hat{I}_E$. The system power operator can be expressed in terms of the basis set $\pd{\hat{H}_{S}^{s.c}}{t}=\sum_{n,m,r}c_{nm}\ket{s_n}\bra{s_m}$, leading to 
\begin{equation}
    \pd{\hat{H}}{t}=\sum_{n,m,r}c_{nm}\ket{s_n}\bra{s_m}\otimes\ket{r_k}\bra{r_k}~~.
    \label{eqap:H_dot}
\end{equation}
Next, we insert Eq. (\ref{eqap:H_dot}) into Eq. (\ref{eqap:C}) to obtain (applying the notation $\ket{s_i b_j}\equiv\ket{s_i}\otimes\ket{b_j}$)
\begin{eqnarray}
\begin{array}{ll}
{\cal{C}} \b t& =\sum_{i,j}\bra{s_{i}r_{j}}\b{\sum_{n,m}c_{n,m}{\ket{s_{n}}}\bra{s_{m}}\otimes\sum_{r}\ket{r_{k}}\bra{r_{k}}}\hat{\chi}\ket{s_{i}r_{j}}\\&
    =\sum_{i}\bra{s_{i}}\sum_{n,m}c_{n,m}\ket{s_{n}}\bra{s_{m}}\sum_{j}\bra{r_{j}}\hat{\chi}_{C}\ket{r_{j}}\ket{s_{i}}\\&
=\textrm{tr}_{S}\b{\pd{\hat{H}_{S}^{s.c}}t  \textrm{tr}_{E}\b{\hat{\chi}\b t}}=0~~.
\end{array}
\end{eqnarray}
The third equality follows from Eq. (\ref{eq:partial trace}). Overall, the correction term, resulting from system-environment correlations, vanishes when all the explicit time-dependence is absorbed in the system Hamiltonian.

\section{External heat flux derivation}
\label{apsec:external_derivation}

We consider a system coupled weakly to a thermal reservoir and a semi-classical drive. The representing  Hamiltonian is given by
\begin{equation}
    \hat{H}^{e} \b t=\hat{H}_S^{e}\b t +\hat{H}_{SE}^{e}+\hat{H}_{E}~~,
    \label{apeq:s_c_Ham_2}
\end{equation}
where the superscript ${e}$ is introduced to emphasis that there is no further restriction on the system environment interaction $\hat{H}_{SE}^{e}$ (no strict energy conservation condition). The external system Hamiltonian $\hat{H}_S^e\b t$ describes an arbitrary time-dependent Hermitian operator on the primary system's Hilbert space. It is therefore, equivalent to the semi-classical Hamiltonian $\hat{H}_S^{s.c}\b t$ (\ref{eq:63scHam}).

In the considered regime, the heat flux can be derived from a generalization of the so-called Davis construction \cite{alicki2012periodically,skrzypczyk2014work,breuer2002theory}. This procedure employs the Born-Markov approximation  to obtain a form similar to the Markovian quantum master equation
\begin{equation}
    \dot{E}_E\approx -\dot{\cal Q}^{e}=\\-
    \int^{\infty}_0\textrm{tr}\b{\sb{\widetilde{H}_{SE}^{ e}\b{t},\sb{\widetilde{H}_{SE}^{ e}\b{t-s},\widetilde{\rho}_S\b t\otimes \hat{\rho}_E}}\hat{H}_E}~~,
    \label{eq:heat_flux_derivation}
\end{equation}
where the wide-tilde designate operators in the interaction picture relative to the free dynamics ($\hat{H}_S^{s.c}\b t +\hat{H}_{E}$).
For simplicity we consider the case of a single interaction term 
\begin{equation}
    \hat{H}_{SE}^{ e}=\hat{S}\otimes \hat{E}~~,
    \label{apeq:interaction_floquet}
\end{equation}
where the generalization to the multiple interaction terms and reservoirs follows a similar derivation. The environment interaction operator satisfies $\langle{\hat{E}}\rangle_E=0$, with out loss of generality \footnote{The condition can be achieved by a suitable transformation of the Hamiltonian if not present from the beginning.}, and the reservoir state $\hat{\rho}_E$ is assumed  to remain in a  Gibbs state with a temperature $T$ throughout the dynamics.

We now wish to decompose the interaction operators in the interaction picture relative to the free dynamics in terms of the dynamical eigenoperators.
The similarity between $\hat{H}^e_{S}\b t$ and $\hat{H}^{s.c}_S\b t$ allows performing the same symmetry motivated procedure presented in subsection \ref{sec:s_c_global_eigen_operators}. This leads to analogous expressions as Eq.  (\ref{eq:70}) and (\ref{eq:73}) to the eigenoperators of $\hat{H}^e_{S}\b t$. Namely the invariant and non-invariant  eigenoperators satisfy
\begin{equation}
    \hat{P}_j^{\b{e}H}\b t=\hat{P}_j^{\b{e}}\b 0
\end{equation}
\begin{equation}
    \hat{F}_\alpha^{\b{e}H}\b t=e^{-i\theta_\alpha\b t}\hat{ F}_\alpha^{\b{e}}\b 0 ~~.
\end{equation}
correspondingly. 

Next, we decompose the system's coupling operator in the interaction picture, in terms of the driven system eigenoperators.
\begin{equation}
    \widetilde{S}\b t = \sum_k \hat{{F}}_\alpha^{\b{e}}\b 0 e^{i\theta_\alpha\b t}+\sum_j \hat{{P}}_j^{\b{e}}\b 0
    \label{apeq:S_int}
\end{equation}
Substituting Eqs. (\ref{apeq:S_int}) and (\ref{apeq:interaction_floquet}) into Eq. (\ref{eq:heat_flux_derivation}) and expressing the system state in the Schr\"odinger picture leads to

\begin{eqnarray}
\begin{array}{ll}
    \dot{E}_E&=\f 1{\hbar^{2}}\sum_{\alpha,\alpha'}\int_0^{\infty}{d\tau} e^{-i\b{\theta_{\alpha}\b{t-\tau}-\theta_{\alpha'}\b t}}\times\\&
    \textrm{tr}_S\bigg{(}
    \hat F_{\alpha}^{\b{e}}\widetilde{\rho}_{S}\b t\hat F_{\alpha'}^{{\b{e}}\dagger}
    \textrm{tr}_{E}\b{\widetilde{E}\b{t-\tau}\hat \rho_{E}\widetilde{E}^{\dagger}\b t\hat H_{E}}
    \\&-\hat F_{\alpha'}^{{\b{e}}\dagger}\hat F_{\alpha}^{\b{e}}\widetilde{\rho}_{S}\b t\textrm{tr}_{E}\b{\widetilde{E}^{\dagger}\b t\widetilde{E}\b{t-\tau}\hat{\rho}_{E}\hat H_{E}}\bigg{)}+\textrm{h.c}\\&
    =    \f 1{\hbar^{2}}\sum_{\alpha,\alpha'}\int_0^{\infty}{d\tau} e^{-i\b{\theta_{\alpha}\b{t-\tau}-\theta_{\alpha'}\b t}}\times\\&\textrm{tr}_S\bigg{(}
    \hat F_{\alpha}^{\b{e}}\widetilde{\rho}_{S}\b t\hat F_{\alpha'}^{{\b{e}}\dagger}
    \mean{\widetilde{E}^{\dagger}\b t{\hat H_{E}\widetilde{E}\b{t-\tau}}}
   \\& -\hat F_{\alpha'}^{{\b{e}}\dagger}\hat F_{\alpha}^{\b{e}}\widetilde{\rho}_{S}\b t\mean{\hat H_{E}\widetilde{E}^{\dagger}\b t\widetilde{E}\b{t-\tau}}\bigg{)}+\textrm{h.c}
   \label{eqap:D7}
\end{array}
\end{eqnarray}

For a Markovian environment, the integral of Eq. (\ref{eqap:D7}) is dominated by times which are smaller then the typical timescale characterizing the decay of the environmental correlation functions $\tau_E$.
Assuming the change in $\theta_\alpha\b t$ is slow relative to $\tau_E$, we can expand the phases around the time $t$, in the relevant time regime $0<\tau\sim\tau_E$
\begin{equation}
\theta_{\alpha}\b{t-\tau}\approx\theta_{\alpha}\b t-\tau\b{\f{d}{dt}\theta_{\alpha}\b {t'}\bigg{|}_{t'=t}}~~.
\label{apeq:D8}
\end{equation}

Next we define $\omega^{\b{e}}_\alpha\b t\equiv {\f{d}{dt}\theta_{\alpha}\b {t'}\bigg{|}_{t'=t}} $ and substitute  Eq. (\ref{apeq:D8}) into Eq. (\ref{eqap:D7}). We obtain
\begin{eqnarray}
\begin{array}{ll}
\dot{E}_E=-\sum_{\alpha \alpha'}e^{-i\b{\theta_{\alpha}\b t-\theta_{\alpha'}\b t}}\\\times\textrm{tr}\b{\hat{ F}_{\alpha'}^{{\b{e}}\dagger}\hat{F }_{\alpha}^{\b{e}}\widetilde{\rho}_{S}\b t}{ \cal K}\b{\omega^{\b{e}}_\alpha\b t}+\textrm{c.c}+~\textrm{similar terms},
    \label{apeq:heat_flux}
\end{array}
\end{eqnarray}
where ${\cal {K}}\b{\nu}=\b{1/{\hbar^{2}}}\int_ 0^{\infty}{d\tau}e^{i\nu\tau}\mean{\sb{\widetilde{E}^{\dagger}\b{\tau},\hat H_{E}}\hat{E}\b 0}_E$. This expression is further simplified by substituting the relation $\Re\sb{\cal{K\b{\nu}}}=\hbar\nu G\b{\nu}/2$, \ref{apsec:KGrelation}, where 
\begin{equation}
    G\b{\nu } = \f 1{\hbar^{2}}\int_{-\infty}^\infty{e^{i\nu \tau}\mean{\widetilde{E}^\dagger\b t\hat E\b 0}}_{E}d\tau~~, 
    \label{apeq:reservoir_correlation_func}
\end{equation}
is the Fourier transform of the reservoir correlation function.
At this stage of the derivation, a coarse-graining procedure on a timescale of $\Delta t_{c.g}\gg\textrm{min}\sb{1/\b{{\theta}_\alpha-{\theta}_{\alpha'}}}$ is employed, leading to elimination of the rapidly oscillating terms. Assuming no degeneracies, the coarse-graining eliminates all the terms for which $\alpha\neq \alpha'$. The final expression for the external heat flux then becomes
%
%\tb{not sure if this part is needed}
%Moreover, the spectrum of any realistic reservoirs are bounded from bellow by some positive frequency. This property implies that the density of states at zero frequency vanishes, which leads to vanishing rates $G\b{0}=0$.  
%As a result, all the terms dependent on the non-invariant eigen operators vanish and 

\begin{equation}
\dot{{\cal Q}}^{e}=-\sum_{\alpha} \hbar\omega^{\b{e}}_\alpha\b t G\b{\omega^{\b{e}}_\alpha\b t}\mean{\hat{F}_{\alpha}^{{\b{e}}\dagger}\hat{F}_{\alpha}^{\b{e}}}~~.
\label{apeq:Q_e_final_1}
\end{equation}

\section{\trr{Entropy production in the local and global interaction settings}}
\label{apsec:entropy_prod}

In quantum thermodynamics the positivity of entropy production is a consequence of the initial product state between the various thermodynamic constituents (system, environment and control)
and the introduction of partitions. Any global unitary evolution can generate quantum correlations between the constituents and modify the state of the environment, leading to positive entropy production. In the present study we adopt the conventional definition, highlighted by Esposito et al. \cite{esposito2010entropy} and analyse the entropy production in the local and global interaction settings. The analysis highlights fundamental differences between the two approaches, not only in the first law, but also in connection with the second law of thermodynamics. 

The quantum entropy production can be expressed as \cite{landi2021irreversible}
\begin{equation}
    \Sigma=D\b{\hat \rho_{SE}'||\hat{\rho}_S'\hat{\rho}_E} = I\b{S:E}+D\b{\hat \rho'_{E}||\hat \rho_{E}}~~.
    \label{apeq:entropy_prod}
\end{equation}
where $I\b{S:B}=S\b{\hat \rho'_{S}}+S\b{\hat \rho'_{E}}-S\b{\hat \rho'_{SE}}$ is the quantum mutual information, $S$ is the von-Neumann entropy, and 
$D\b{\hat \rho||\hat \sigma}=\rm{tr}\b{\hat \rho \rm{ln}{\hat \rho}-\hat \rho \rm{ln}\hat \sigma}$ 
is the quantum relative entropy, $\hat \rho_E'={\rm{tr}}_{S,C}\b{\hat \rho\b t}={\rm{tr}}_{S,C}\b{\hat U\b{t,0}\hat \rho_S\b 0\otimes \hat{\rho}_C\b 0\otimes \rho_E \b 0\hat U^\dagger \b{t,0}}$, 
$\hat \rho_S'={\rm{tr}}_{E,C}\b{\hat \rho\b t}$, $\hat{\rho}_{SE}'={\rm{tr}}_C\b{\hat \rho \b t}$ and $\hat \rho_{S/E}\equiv\hat  \rho_{S/E}\b 0$. The first term of Eq. 
(\ref{apeq:entropy_prod}) is associated with the loss of information captured in the correlation between the studied system and environment, while the second term is associated with the lack of information regarding the environment's state change. 

Considering the two physical sources included in the entropy production, in the semi-classical limit the control system does not contribute to the entropy production. 
 In this limit, the correlations between the
 system and control are negligible and it's dynamics is known and approximately unitary,
 $\hat \rho'_{C}\approx \hat U_{C}^{\dagger}\b t\hat \rho_{C}\b 0\hat U_{C}\b t$ (cf. Sec. \ref{sec:semi_classical}
 ). This condition together with the unitarity of the composite system dynamics implies that the dynamics of $\hat{\rho}_{SC}\b t $ is unitary as well.  Such property along with the initial separable state allows simplifying the expression for the entropy production
\begin{eqnarray}
\begin{array}{ll}
   \Sigma&
   =S\b{\hat \rho'_{S}}+S\b{\hat \rho'_{E}}-S\b{\hat \rho_{SE}}+{\rm{tr}}\b{\hat \rho'_{E}{\rm{ln}}\hat \rho'_{E}}-{\rm{tr}}\b{\hat \rho'_{E} {\rm ln}\hat \rho_{E}}\\&=    S\b{\hat \rho'_{S}}-S\b{\hat \rho{}_{S}}-S\b{\hat \rho_{E}}-{\rm{tr}}\b{\hat \rho'_{E}{\rm{ln}}\hat \rho_{E}} \\&
    =\Delta S_{S}-{\rm{tr}}\b{\sb{\hat \rho'_{E}-\hat \rho_{E}}{\rm{ln}}\hat \rho_{E}}~~.
\end{array}
\end{eqnarray}
Assuming the environment initially resides in a thermal state, $\hat \rho_E = \exp\b{-\beta \hat{H}_E}/Z_E$, one obtains the familiar form
\begin{equation}
    \Sigma=\Delta S_{S}+\beta{\rm{tr}}\b{\sb{\hat \rho'_{E}-\hat \rho_{E}}\hat H_{E}}=\Delta S_{S}+\beta\Delta E_{E}~~.
\end{equation}

\paragraph{Local interaction analysis}

We now focus on the the local interaction framework in the Markovian regime, demonstrating that the entropy production rate differs from the one obtained from the so-called Spohn's inequality. By utilizing the local strict energy conservation  
\begin{equation}
    \sb{\hat H_{S}+\hat H_{E}+\hat H_{C},\hat H}=0
\end{equation} 
which implies that    $\Delta E_{S}+\Delta E_{C}+\Delta E_{E}=0$, 
with 
 $\Delta E_{i}={\rm{tr}}\b{\b{\hat \rho'-\hat \rho}\hat H_{i}}$,
one can express the system entropy production in terms of system and control observables:
\begin{equation}
   \Sigma=\Delta S_{S}+\beta{\rm{tr}}\,\b{\b{\hat \rho'_{S}-\hat \rho_{S}}{\rm{ln}}\hat \rho_{S}^{th}}-\beta\, \Delta E_{C}~~,
\end{equation}
where $\rho_{S}^{th}=\exp\b{-\beta\hat{H}_S}/Z_S$.
Next, we consider the entropy production in small time interval $\tau$ which is of the order of the memory time of the environment (after each time step $\tau$, the state of the environment already relaxes back to a thermal state). Assuming Markovian dynamics, the changes of the reduced system occur in timescales much larger than $\tau$. Hence, the entropy production rate can be can approximated as
\begin{eqnarray}
\begin{array}{ll}
    \sb{{\Sigma\b{t+\tau}-\b{\Sigma\b t}}}/{\tau}&\approx\f d{dt}S_{S}+\beta\, {\rm{tr}}\b{{\cal L}\sb{\hat \rho_{S}'}{\rm{ln}}\hat \rho_{S}^{th}}-\f d{dt}E_{C}\\&=-\f d{dt}D\b{\hat \rho_{S}'||\hat \rho_{S}^{th}}-\f d{dt}E_{C}~~,
    \label{apeq:local_entropy_prod}
\end{array}
\end{eqnarray}
where ${\cal{L}}\sb{\hat{\rho}_S'}=\f{d}{dt}\hat{\rho}_S\b t$.
The first term gives the so-called Spohn's inequality (Eq. (\ref{eq:spohn_ineq})) \cite{spohn1978entropy}  (the inequality is obtained from the contractive property of dynamical maps, satisfying the semi-group property). The last term of Eq. (\ref{apeq:local_entropy_prod}) can be interpreted as a non-vanishing energy current from the control system through the system to the environment.

\paragraph{Global interaction analysis}

In the global analysis, we partition the composite system to a device system and the environment.
For an initial separable state $\hat \rho\b 0=\hat \rho_{D}\b 0\otimes\hat \rho_{E}\b 0$, following a similar derivation as in the local analysis, for an environment initially in a thermal state, we have
\begin{equation}
   \Sigma=\Delta S_{D}+\beta\,{\rm{tr}}\b{\sb{\hat \rho'_{E}-\hat \rho_{E}}\hat H_{E}}=\Delta S_{D}+\beta\Delta E_{E} ~~.
   \label{apeq:e8}
\end{equation}
The global strict energy condition (Eq. (\ref{eq:global_sec})) implies the commutation relation $\sb{\hat H_{D}^{\b {G}}+\hat H_{E},\hat H}=0$ 
and as a consequence leads to the conservation of the sum of device and environment bare energies, $\Delta E_{D}+\Delta E_{E}=0 $. Substituting this relation into Eq. (\ref{apeq:e8}), we obtain
\begin{equation}
  \Sigma=\Delta S_{D}+\beta\, {\rm{tr}}\b{\b{\hat \rho'_{D}-\hat \rho_{D}}{\rm{ln}}\hat \rho_{D}^{th}}~~.  
\end{equation}
Next, we consider Markovian dynamics which results in an entropy production rate which coincides with Spohn's inequality 
\begin{equation}
   \f{\Sigma\b{t+\tau}-\b{\Sigma\b t}}{\tau}\approx\f d{dt}S_{D}+\beta{\rm{tr}}\b{{\cal L}\b{\hat \rho_{D}}{\rm{ln}}\hat \rho_{D}^{th}}=-\f d{dt}D\b{\hat \rho_{D}||\hat \rho_{D}^{th}} 
\end{equation}
In the semi-classical limit and adiabatic regime we have $\hat \rho_{D}^{th}=Ze^{-\beta \hat H_{s.c}\b t}$,
 which produces the familiar expression for the entropy production rate in the adiabatic limit.
 
Beyond the adiabatic limit, the entropy production rate in global and local interaction settings,  in the semi-classical description is obtained by replacing the dynamical generators with their semi-classical analogues.

\section{Relation between ${\cal{K}}\b{\nu}$ and the bath correlation function $G\b{\nu}$}
\label{apsec:KGrelation}
The contribution of the reservoir to the heat flux is encapsulated by a single correlation term, Eq. (\ref{apeq:heat_flux}):
\begin{equation}
    {\cal K}\b{\nu}=\f 1{\hbar^{2}}\int_0^{\infty}{d\tau}e^{i\nu\tau}\Big{\langle}\sb{\tilde{E}^{\dagger}\b{\tau},\hat{H}_{E}}\hat{E}\b 0\Big{\rangle}_E~~.
    \label{eqap:K}
\end{equation}
This term is related to the reservoir correlation functions $G$, Eq. (\ref{apeq:reservoir_correlation_func}) by a simple relation.

The reservoir Hamiltonian is a constant, therefore, the interaction operators at the initial time can be easily decomposed as a sum over eigenoperators of $\hat{H}_E=\sum_\epsilon\,\epsilon \ket{\epsilon}\bra{\epsilon}$:
\begin{equation}
   \hat{E}\b 0=\sum_{\omega}\hat{E}\b{\omega} 
   \label{apeq:R}
\end{equation}
 where $\hat{E}\b{\omega}=\sum_{\epsilon'-\epsilon=\hbar\omega}\hat{\Pi}\b{\epsilon}\hat{E}\b 0\hat{\Pi}\b{\epsilon'}$ and $\hat{\Pi}\b{\epsilon}$ is the projection on the eigenspace of $\epsilon$.
 These definitions imply the relation $\sb{\hat{H}_{E},\hat{E}\b{\omega}}=-\hbar\omega \hat{E}\b{\omega}$.  In turn, Eq. (\ref{apeq:R}) and the commutation relation then lead to
 \begin{equation}
     \sb{\tilde{E}^{\dagger}\b{\tau},\hat{H}_{E}}=-\sum_{\omega}\hbar\omega \hat{E}^{\dagger}\b{\omega}e^{i\omega\tau}~~.
 \end{equation}
Substituting this relation into Eq. (\ref{eqap:K}) yields
 \begin{eqnarray}
 \begin{array}{ll}
     {\cal{K}}\b{\nu}&=-\sum_{\omega}\hbar\omega\f 1{\hbar^{2}}\mean{\hat{E}^{\dagger}\b{\omega}\widetilde{E}\b 0}\int_ 0^{\infty}{d\tau}\,e^{i\b{\nu+\omega}\tau}\\&
     =\hbar\nu 
     \sb{\f 1{\hbar^{2}}\mean{\hat{E}^{\dagger}\b{-\nu}\widetilde{E}\b 0}}
     +i\sum_{\omega}\textrm{P}\f{\mean{\hat{E}^{\dagger}\b{\omega}\widetilde{E}\b 0}}{\nu+\omega}~~,
     \label{apeq:D5}
  \end{array}
     \end{eqnarray}
utilizing the identity $\int_ 0^{\infty}{d\tau e^{-i\alpha \tau}=\pi\delta\b{\alpha}-i\textrm{P}\f 1{\alpha}}$, where $\textrm{P}$ denotes the Cauchy principal value.
 
Following the same derivation, the reservoir correlation functions are expressed as 
 \begin{equation}
     G\b{\nu}=\f{1}{\hbar^2 }\int_{-\infty}^{\infty}{e^{i\nu \tau}\mean{\widetilde{E}\b{\tau}\hat{E}\b 0}}_{E}d\tau=\f{2}{\hbar^2}\mean{\hat{E}^{\dagger}\b{-\nu}\hat{E}\b 0}_{E}~~.
      \label{apeq:D6}
 \end{equation}
Equations (\ref{apeq:D5}) and (\ref{apeq:D6}) now imply the relation
\begin{equation}
    \Re\sb{{\cal{K}\b{\nu}}}=\hbar \nu \f{G\b{\nu}}{2}~~.
\label{apeq:K_G_relation}
\end{equation}
%\bibliography{references.bib}


\clearpage
\section*{References}
%\bibliography{references.bib}
\bibliographystyle{unsrt}

\end{document}